\documentclass[lettersize,journal]{IEEEtran}
\usepackage{amsmath,amsfonts}
\usepackage{algorithmic}
\usepackage{algorithm}
\usepackage{array}
\usepackage[caption=false,font=normalsize,labelfont=sf,textfont=sf]{subfig}
\usepackage{textcomp}
\usepackage{stfloats}
\usepackage{url}
\usepackage{verbatim}
\usepackage{graphicx}
\usepackage{xcolor}
\usepackage{cite}
\usepackage[hidelinks]{hyperref}
\usepackage{multirow}
\usepackage{textcomp}
\usepackage{tabularx}
\usepackage{placeins}
\usepackage{pifont}
\newcommand{\cmark}{\ding{51}}%
\newcommand{\xmark}{\ding{55}}%

\begin{document}


\title{\huge A Contemporary Survey on Semantic Communications: Theory of Mind, Generative AI, and Deep Joint Source-Channel Coding}

\author{Loc X. Nguyen, Avi Deb Raha, Pyae Sone Aung, Dusit Niyato~\IEEEmembership{Fellow,~IEEE,}\\
Zhu Han~\IEEEmembership{Fellow,~IEEE,} Choong Seon Hong~\IEEEmembership{Fellow,~IEEE}
\thanks{Loc X. Nguyen, Avi Deb Raha, Pyae Sone Aung, and Choong Seon Hong are with the Department of Computer Science and Engineering, Kyung Hee University,  Yongin-si, Gyeonggi-do 17104, Rep. of Korea, e-mails:{\{xuanloc088, avi, pyaesoneaung,  cshong\}@khu.ac.kr}.}

\thanks{Dusit Niyato is with the College of Computing and Data Science, Nanyang Technological University, Singapore. e-mails:\{dniyato@ntu.edu.sg\}.}
\thanks{Zhu Han is with the Department of Electrical and Computer Engineering at the University of Houston, Houston, TX 77004 USA, and also with the Department of Computer Science and Engineering, Kyung Hee University, Seoul, South Korea, 446-701. email:{\{hanzhu22\}}@gmail.com.}}

\markboth{Journal of \LaTeX\ Class Files,~Vol.~14, No.~8, August~2021}%
{Shell \MakeLowercase{\textit{et al.}}: A Sample Article Using IEEEtran.cls for IEEE Journals}


\maketitle

\begin{abstract}
Semantic communication is emerging as the next pillar in wireless communication technology due to its transformative capabilities in reducing communication overhead, enhancing robustness, and enabling intelligent information exchange. Despite its promising potential, semantic communication still faces several critical challenges that must be addressed before real-world deployment. The most significant obstacle lies in the lack of standardization across various research directions, leading to inconsistencies in interpretation, objectives, and evaluation. In this survey, we provide an in-depth overview of three leading directions in semantic communication, namely Theory of Mind-based semantic communication, Generative AI-driven semantic communication, and Deep Joint Source-Channel Coding (DJSCC)-based semantic communication. These directions have been extensively studied and developed by research institutes worldwide, and their effectiveness continues to improve alongside advances in communication and computing technologies. We first introduce the fundamental concepts and theoretical background of each direction. The Theory of Mind-based semantic communication enables communication agents to interact intelligently, infer each other's intentions, and gradually form a shared understanding or common language. The Generative AI-based semantic communication leverages generative models to create and interpret content beyond traditional compression, allowing flexible semantic encoding and decoding tailored to specific tasks. The DJSCC-based semantic communication direction integrates deep learning models to jointly optimize the source and channel coding processes for efficient semantic information transfer. Next, we present a detailed survey of existing works under each direction, providing a structured overview of past developments, key methodologies, and open research problems in semantic communication. Furthermore, we identify and analyze critical challenges-such as scalability, adaptability, interoperability, and computational efficiency-that currently hinder the deployment of semantic communication systems in practical, dynamic environments. Finally, we discuss potential research opportunities and future directions, including the exploration of emerging technologies such as quantum computing to further enhance the capabilities of semantic communication. This survey aims to serve as a comprehensive reference for researchers and practitioners, offering insights into current progress and guiding future advancements in semantic communication.
\end{abstract}

\vspace{-0.1in}
\section{Introduction}
\subsection{Current Wireless Communication System}

We live in the age of an information society, and the desire to connect to the grid at all times is undeniable. With the development of social applications and mobile games, using mobile devices has become an indispensable activity in our daily lives. With the introduction of virtual reality, metaverse, online gaming, and other bandwidth-consuming applications, the current communication system has already encountered various difficulties in fulfilling their requirement for fast transmission, low latency, and high reliability. In addition, the current increasing pace of connected devices in the network even makes it more challenging for traditional communication to catch up with those demands. The current traditional communication is developed based on the information theory proposed by Claude Shannon in 1948 \cite{Shannon}, whose core idea is reproducing the message at the receiver exactly the same or approximately to the message at the transmitter. This is the first level of communication, also referred to as an engineering problem in his paper. A large number of techniques have been developed to improve its transmission rate by compressing the data \cite{Huffman,Golomb,Rissanen,DeVore}, as well as algorithms \cite{Hamming,Reed,Gallager,Elias,Berrou,Arikan} to protect the message from a noisy transmission channel. Along with other bandwidth resource management, traditional communication has reached its maximum data transmission rate - known as Shannon Capacity, which cannot make any massive improvement by following the information theory. Moreover, the channel capacity depends both on the allocated bandwidth and the noise level from the environment, which is impossible to control or estimate accurately. 
\vspace{-0.1in}
\subsection{Semantic Communication}
Semantic communication is the second level of communication, which was introduced by both Shannon and Weaver\cite{Weaver}, where the objective of the communication system is that the receiver can successfully interpret the received message as the transmitter originally intended. Unlike traditional communication, semantic communication does not care about the difference between the transmitted sequence bits and the received sequence as long as the receiver can understand the message. Therefore, instead of allocating resources for bit-checking errors such as low-density parity-check (LDPC) code, computing resources can be dedicated to the semantic encoding/decoding process. The most important property of semantic communication is the ability to understand the data from both communication parties, which can improve communication efficiency significantly. For example, by understanding the data, the semantic transmitter can acknowledge the essential information from the raw data and the redundant information. This ability allows it to transmit important features to the receiver; thus, it can substantially decrease the overall size of the transmit signal without compromising the receiver's ability to interpret the message. Such a reduction in data volume not only improves the efficiency of the system but also optimizes the use of communication resources such as bandwidth and energy. With these advancements, semantic communication is expected to be the pillar of the next wireless communication generation (Sixth Generation - 6G). In this survey, we aim to cover all the concepts, innovative ideas, breakthroughs, and open challenges of three leading semantic communication systems: \textit{Direction I: Theory of Mind}, \textit{Direction II: Generative AI}, and \textit{Direction III: Deep Joint Source-Channel Coding}.
\begin{table*}[t]
\centering
\caption{A Comparision of Contribution between Existing Survey and our survey}
\label{ComparisionSurvey}
\begin{tabular}{|p{1.1cm}|p{1.1cm}|p{1.2cm}|p{1.1cm}|p{1.1cm}|p{10cm}|}
\hline
\centering Ref.                                      & \centering Theory of  Mind & \centering Generative AI  & \centering Deep-JSCC & \centering Quantum  Semantic &  Contributions                                                                                                        \\ \hline
\centering \cite{SemanticSurvey0}   & \centering  \xmark        & \centering \checkmark & \centering \checkmark                           & \centering \checkmark                                    & Focus on the architectures, the security and privacy problems for multi-agents scenario. Later on, authors considered a novel three-layer architecture for multi-agent communication and exposed the security and privacy threats, and finally, the defense mechanism.                                                     \\ \hline

\centering\cite{ChristinaLessdata} & \centering \checkmark                                 & \centering \checkmark & \centering \xmark                                &\centering  \checkmark                                   & Analyze the shortcomings of the current data-driven semantic communication and the obtained benefits when shifting the focus to improve and design a system with the ability to provide causal-reasoning relationships, akin to human logical thinking.             \\ \hline

\centering\cite{SemanticSurvey1}   & \centering\xmark                                     & \centering\xmark     &\centering\checkmark                            &\centering \xmark                                       & The survey mainly summarized the improvement in the architecture model of the Deep Joint Source-Channel Coding direction. Additionally, the potential applications are provided along with opening the major challenges for further research scientists.                                                  \\ \hline

\centering\cite{SemanticSurvey2}   & \centering\xmark                                     & \centering\xmark     & \centering\checkmark                            &\centering \xmark                                       & A comprehensive discussion is provided to raise underscore the security and privacy issues of deploying the semantic communication system. Then, various techniques have been provided to prevent these threads, such as adversarial training, and blockchain.                                             \\ \hline

\centering\cite{SemanticSurvey3}   & \centering\xmark                                     & \centering\checkmark &\centering \checkmark                            & \centering\xmark                                       & Introduce the concept of intellicise (intelligent and concise) wireless networks to catch up the massive growth in demands of emerging applications. The framework had various intelligence components for signal processing, information transmission, network organization, and finally, service bearing.                \\ \hline

\centering\cite{SemanticSurvey4}   & \centering\checkmark                                 & \centering\xmark     &\centering \checkmark                            & \centering\xmark                                       &  Mainly discuss works follow the Deep-JSCC direction. Additionally, they divided the existing works into three categories: semantic-aware, semantic-oriented, and goal-oriented. Then, the design associated with each category is provided for further details.        \\ \hline

\centering\cite{SemanticSurvey5}   & \centering\xmark                                     & \centering\xmark     &\centering \checkmark                            & \centering\xmark                                       & The survey delves into the research related to goal-oriented semantic communication for the next generation communication. They offered readers insights into the SOTA research landscape, emerging trends, practical use cases, and frameworks.                        \\ \hline

\centering\cite{SemanticSurvey6}   & \centering\xmark                                     & \centering\xmark     &\centering \checkmark                            & \centering\checkmark                                       & It primarily focuses on the lack of unified/universal metrics for the semantic communication system, and its goal to provide a comprehensive survey of current existing metrics.                       \\ \hline

\centering\cite{ChengsiAIGC}       & \centering\xmark                                     & \centering\checkmark & \centering\xmark                                & \centering\xmark                                       & Explore the potential advantages of AIGC models in network management and the architecture of semantic communication. Additionally, they introduced GAI-driven SemCom networks, which can advance the communication infrastructure to next level. \\ \hline

\centering\cite{WangSurvey}       & \centering\xmark                                     & \centering\xmark & \centering\checkmark                                & \centering\xmark                                       & Explore different concepts of semantic information: Feature Vector Representations, Knowledge Graph, and Hierarchical Semantic Tree. And the different processes on the semantic information such as: Extraction, Encoding, Fusion, Transmission, and finally, the architecture of the semantic communication system.  \\ \hline

\centering\cite{MengSurvey}       & \centering\xmark                                     & \centering\xmark & \centering\checkmark                                & \centering\xmark                                       & The survey highlights the benefits of deploying the semantic communication in 6G areas to tackle the problem of seamless connectivity in Space-Air-Ground-Sea integrated networks (SAGSIN) and review some current state-of-the-art works.\\ \hline

\centering\cite{DWonSurvey}       & \centering\checkmark                                     & \centering\xmark & \centering\checkmark                                & \centering\checkmark                                     & It underscore the need for resource management, security, and privacy for efficient and secure semantic communication systems, and later provide a review of the current research, key challenges, and open questions for further exploration.\\ \hline

\centering\cite{ZLuSurvey}       & \centering\checkmark                                     & \centering\xmark & \centering\checkmark                                & \centering\xmark                                       & This survey emphasizes the explicit and implicit reasoning capabilities of communication agents while highlighting the significance of DeepJSCC-based semantic communication techniques. It also provides the background, and taxonomy of current literature.\\ \hline

\centering This survey                                      & \centering\checkmark                                 & \centering \checkmark & \centering\checkmark                            & \centering\checkmark                                   & We are the first to completely discuss the different semantic communications directions in one survey. Without favoring any direction, we illustrate in detail the base theory of each direction, the prospects, and along with their challenges. Finally, we discuss the potential application of quantum technology in semantic communication, which, if feasible, could lead to unprecedented and unforeseen applications.  \\ \hline
\end{tabular}
\end{table*}

\subsection{Contributions}
The existing survey on semantic communication mainly focuses on a single direction or one sub-problem within the direction, and some works even try to deny the existence of others. In contrast to this, in our survey, we are open to any possibility and every promising idea, as shown in Table~\ref{ComparisionSurvey}. Therefore, our survey contains a comprehensive understanding of various approaches to semantic communication and does not favor the approach. Specifically, we delve into detailed explanations of the foundation theory and design modules associated with each approach. Specifically, three prominent research directions are primarily discussed as follows: semantic communication based on \textit{1) Theory of Mind}, \textit{2) Generative AI}, and \textit{3) Deep Joint Source-Channel Coding}. Studies in the field of Theory of Mind (ToM) are built on the foundation of knowledge, where communicative agents can predict the behavior of their counterparts upon receiving a specific signal based on the knowledge they have acquired \cite{ThomasReason1,ThomasPragmatic}. This enables the formation of concise and efficient communication. The studies following the second direction leverage the contextual understanding, cross-modal conversion of the AI model to encode data at extreme compression rate, and its creativity in creating content at the receiver side \cite{GenerativeAISem,GenerativeAISem1}. Finally, the DeepJSCC research has demonstrated the ability of semantic information extraction from raw data and the robustness against channel noise by deploying the advancement of new deep learning architectures and methods. It is one of the most popular directions due to its alignment with the modules of current traditional wireless communication systems \cite{Bourtsoulatze2019DeepJSCC}. Each direction has its own strategy for reducing the transmitted signal while guaranteeing the performance of the system, all of which are developing at a fast pace and becoming pillars of the next communication generation. Therefore, in this survey, we provided a comprehensive survey on the existing works, promising aspects, and finally, the open problems of the directions. The main contributions can be summarized as follows:
\begin{figure*}[t]
\centering
\includegraphics[width=0.9\textwidth]{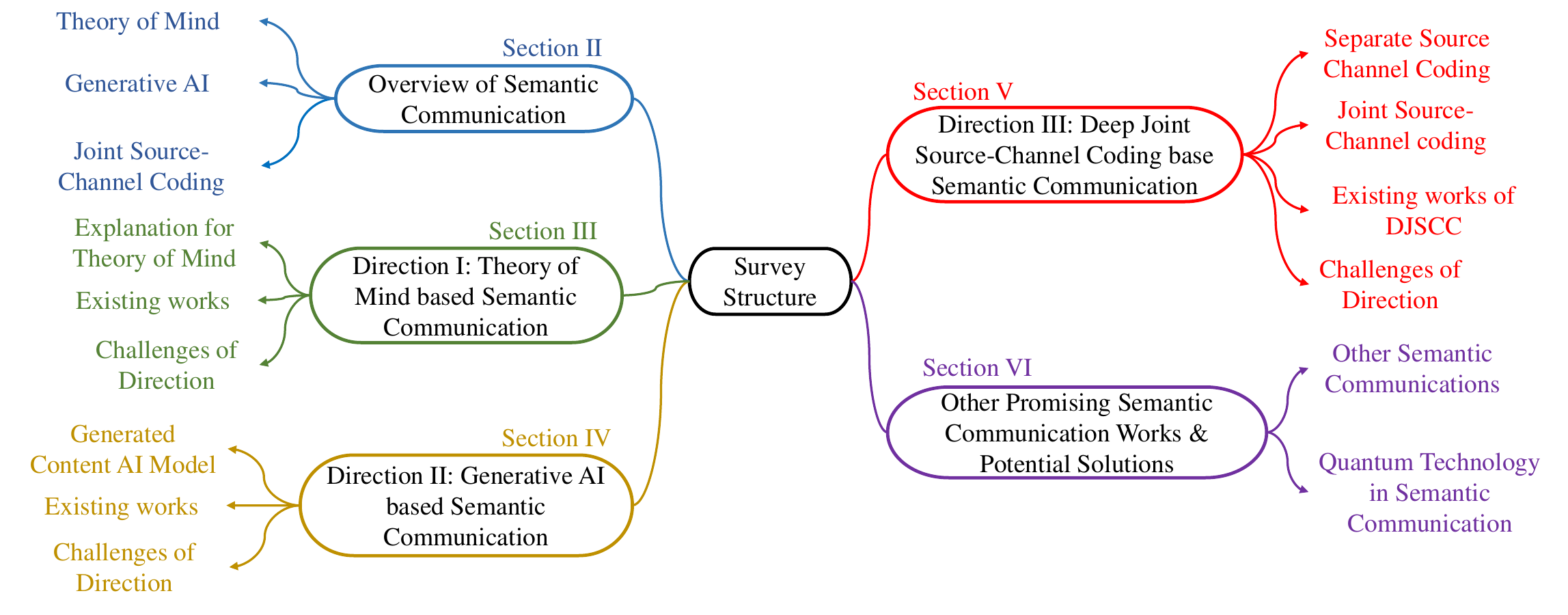}
\caption{The structure and scope of the survey}
\label{StructureFigure}
\end{figure*}
\begin{itemize}
    \item To the best of our knowledge, we are the first to conduct an in-depth exploration of three emerging research directions in semantic communication. For each direction, we discuss its fundamental theory, key components, potential benefits, and the current progress in related research.  
    \item Through detailed explanations of each proposed approach, we demonstrate how the designed scenario effectively eliminates redundant data, ensures the transmission of only meaningful information, and thereby meets the stringent demands of 6G communication systems. Specifically, the first approach establishes a concise communication language among agents, while the second leverages the encoding and generative capabilities of advanced AI models. Finally, the third strategy seamlessly integrates semantic information directly into the encoding process.
    
    \item Finally, we highlight the common and specific challenges associated with each of the three proposed directions: the substantial storage demands of the knowledge base for the Theory of Mind, the computational requirements of devices for deploying AI models, and the scalability constraints of the network in the DJSCC direction. Based on these challenges, we outline future research tasks aimed at addressing these issues. Additionally, we discuss the potential of leveraging quantum technology to further advance semantic communication systems.
\end{itemize}

\begin{figure*}[t]
\centering
\includegraphics[width=0.92\textwidth]{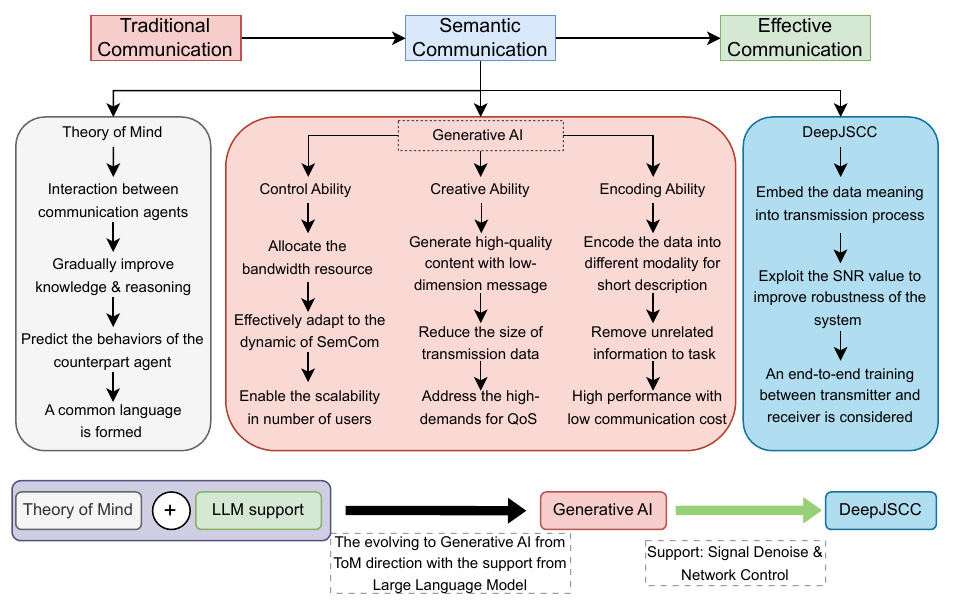}
\caption{The visualization of different directions for semantic communication.}
\label{Visualization}
\end{figure*}
\vspace{-0.15in}
\subsection{Structure of the Survey}
The organization of the survey, as shown in Fig.~\ref{StructureFigure}, is as follows: brief discussions of three leading semantic communication directions are provided in Section~\ref{SectionII}. Then, the comprehensive explanations of the base theory, the existing works, and the challenges of \textit{Theory of Mind, generative AI}, and \textit{Deep Joint Source-Channel Coding} are presented in Sections \ref{DirectionI}, \ref{DirectionII}, and \ref{DirectionIII}, respectively. Additionally, some innovative works, which cannot be categorized into any directions, have been discussed in Section \ref{OtherDirections} along with the potential benefit of applying quantum technology in semantic communication. Finally, we conclude our survey in Section \ref{Conclusion}.
\vspace{-0.1in}
\section{Overview of Semantic Communication}
\label{SectionII}
As discussed earlier, the three primary directions in semantic communication systems are \textit{Theory of Mind, generative AI, and deep joint source-channel coding}, which is later referred to as Direction I, Direction II, and Direction III, respectively, for convenience. The commonality among these approaches is that they are built based on the advancement of machine learning/deep learning models. In essence, they are adopted as a solution method for the proposed semantic communication scenario. The deep learning model is especially capable of understanding the semantic meaning within the data, thus becoming a fundamental component of semantic communication. In the following, we illustrate the definition of semantic communication for each direction and differentiate the scenarios. Fig.~\ref{Visualization} provides an overview scheme related to the abilities, the principles behind each direction, the close relationship between the ToM and Generative AI.
\subsection{Theory of Mind-Based Semantic Communication}
As suggested by its name, the Theory of Mind-based semantic communication is an emerging system that stays at the intersection of cognitive science, artificial intelligence, and communication theory. The Theory of Mind can be referred to as the ability to understand/attribute mental states, such as beliefs, desires, and intentions, to oneself and others \cite{Frith}. Under the context of semantic communication, it is employed to enhance communication efficiency by empowering the transmitter to anticipate or model the outcome of the receivers when acquiring a message. The receiver's behavior can be effectively estimated by collecting a sufficient set of knowledge about the receiver, such as current state, beliefs, logic, and finally, its reasoning \cite{ThomasReason1}. This process mimics the learning behaviors of humans; for example, if the same message is delivered to different listeners, they will react dissimilarly due to their knowledge and personality. Therefore, the speaker needs to take into consideration the listener's knowledge related to that particular topic and sometimes even social status is factored in. With the framework being described, the problem is how to understand the receiver from the perspective of the transmitter. Just like how humans communicate with others, if the transmitter wants to understand the receiver, it has to proactively engage in interaction with the listener. By observing the receiver's behavior against the transmitted message and using the feedback as the supervised signal, we can step by step formulate the knowledge base (KB) for each receiver. Additionally, the knowledge base for the direction is normally described in the form of a graph.

\subsection{Generative AI-Based Semantic Communication}
Artificial Intelligence-Generated Content (AIGC) has acquired various achievements in the field of natural language processing, image modality, and even cross-modality tasks, which inspires wireless communication researchers to adopt it into semantic communication \cite{GenerativeAI_Ref}. Different from the first direction, where the learning model of the transmitter and receiver is designed based on the considered task and trained from scratch. The generative AI-based semantic communication leverages the enormous capabilities of the foundation models for various tasks and employs them as the communication agent. The foundation model, also known as the large AI model, has been trained on vast datasets so it can be applied across a wide range of use cases, from medical application \cite{GenerativeAI_Ref2} to semantic extraction \cite{GenerativeAI_Ref3}. Therefore, employing them as a module of the communication agents can effectively enhance successful semantic extractions with the lowest compression rate. For example, considering the task where the transmitter wants to describe a general description of one observed picture to a receiver, instead of encoding the whole image by JPEG and LDPC techniques, the foundation model can create the caption for the image and transmit the output to the receiver, in which another generative AI model is utilized to create the image follow the description as in \cite{GenerativeAI_Ref4}. In the example, the generative AI-based semantic communication system significantly reduces the volume of transmitted data, allowing for increased user capacity and shorter communication times. With more advanced generative AI models, we can achieve macro-level control over the generated content, aligning it more effectively with specific intentions. 
\subsection{Deep Joint Source-Channel Coding-Based Semantic Communication}
Last but not least is the deep joint source-channel coding-based semantic communication direction, where its framework resembles traditional communication the most. In a traditional communication framework, the source encoder and channel encoder are independently designed with each other. On the other hand, as its name suggests, DJSCC optimizes both modules simultaneously, which can effectively embed the semantic information of the data into the transmission process and also more robust to the channel noise from the physical environment. Most of the works under this direction follow an auto-encoder architecture, where the transmitter employs a deep learning encoder to extract and compress the semantic features, while the receiver has a deep learning-based decoder to decode the message with the objective of reconstructing the original data or conducting certain downstream tasks. The performance of the direction closely attached to the advancement of new learning neural network architectures and techniques, from Convolutional Neural Network (CNN) \cite{LeCun}, Residual Neural Network (ResNet) \cite{Kaiming}, Attention mechanism \cite{Vaswani}, Vision Transformer (ViT) \cite{Dosovitskiy}. With the flexibility offered by the DL models, it can proactively adjust the coding length to adapt to receiver requirements \cite{SemanticDJSCCLength} or the network traffic conditions. The direction is getting the most attention among the three schemes.

However, the most popular way to do semantic communication now does not necessarily imply it will become the definitive way for the future semantic communication system. As we dedicate more effort to the field and it becomes more mature, the most standardized semantic communication system will be shaped by various factors such as continuous experimentation, debate from different research groups, and the performance of different methodologies. In the future, a universally accepted definition of semantic communication may emerge, either from the dominance of a single approach or, more likely, through the integration of multiple perspectives. Each research direction- whether it is based on the Theory of Mind, AI-generated content, or joint source-channel coding- offers unique contributions that address different facets of semantic communication. 
\vspace{-0.15in}
\subsection{The Relations among Directions}
Although each direction is grounded in distinct theoretical foundations, they do not entirely exclude one another but instead demonstrate a strong interrelation. First of all, they all operate on the idea that the transmitter and receiver can collaborate with each other to finish the task with minimal transmission cost. The difference lies in the design of the transmitter,  receiver models. ToM relies on the sender understanding the receiver's knowledge and behavior. This allows the sender to predict how the receiver will respond and send only the minimal signal needed to achieve the expected behavior. Meanwhile, the generative AI approach equips both the transmitter and receiver with advanced AI models, pre-trained on extensive datasets, which can be fine-tuned to minimize the transmitted signal while effectively addressing the specific task at hand. In a way, we can say the generative AI direction is the improved version of the ToM, where the communication agents have amassed extensive knowledge across diverse domains, enabling them to generate more informed and contextually relevant responses. Lastly, the DJSCC achieves the communication efficiency of semantic communication by constructing the communication participant with the deep neural networks to aggressively compress the data and protect it against channel noise.
\section{Direction I: Theory of Mind-based Semantic Communication}
\label{DirectionI}
First, we provide a brief introduction related to the Theory of Mind in this section, explain the benefit of the theory, and explain how it can be integrated into semantic communication scenarios. Then, comprehensive works related to the Theory of Mind are presented and discussed in detail with the aim of providing the reader with an understanding of the contributions of each work.
\vspace{-0.15in}
\begin{figure*}[t]
\centering
\includegraphics[width=0.85\textwidth]{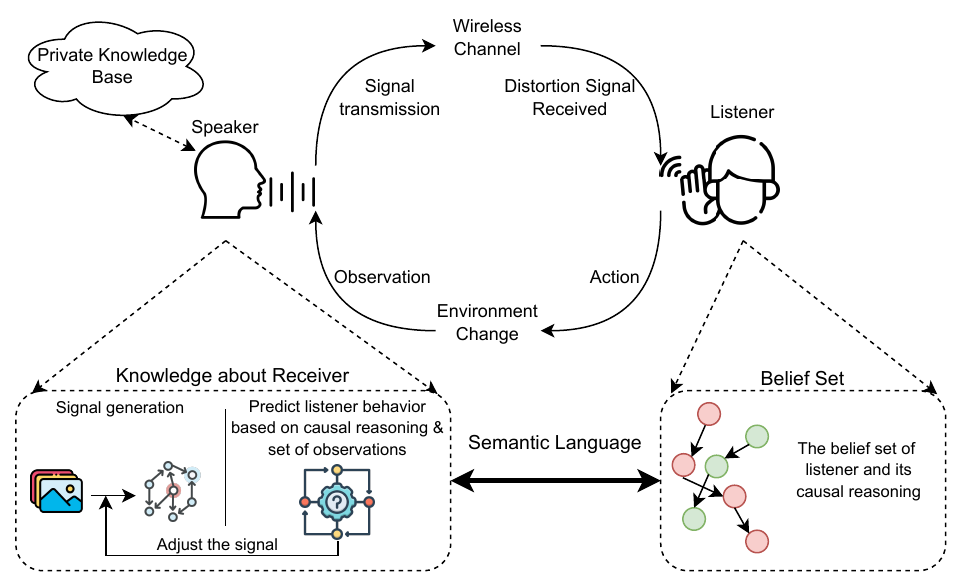}
\caption{The visualization of a general Theory of Mind-based Semantic Communication.}
\label{TheoryofMind}
\end{figure*}

\subsection{Explanation for Theory of Mind}
In order to explain the Theory of Mind, we first briefly provide a very interesting example in \cite{Frith}. A boy named Maxi has a chocolate bar and eats half of it, then stores the remaining half in the kitchen cupboard and goes play outside. The mother sees the chocolate bar in the kitchen cupboard and moves it to the fridge without letting Maxi know; the question is, when Maxi comes back and wants to eat his chocolate, where will he look for it? If your answer is the kitchen cupboard, congratulations! You possess a ``Theory of Mind," the ability to intuitively interpret and explain people's behavior by considering their knowledge, beliefs, and desires. Even though we know that the chocolate was actually in the fridge, that reality does not affect the boy's behavior; only his belief at the time does. The ``Theory of Mind" allows us to manipulate other's behaviors by altering their beliefs. 
\vspace{-0.15in}
\subsection{Novel Works under the ToM-Based Semantic Communication Direction}
Semantic communication works under the Theory of Mind direction, which considers the interactions among communication parties, in which they gradually improve their knowledge, form relations with other communication parties, and acknowledge the condition of the wireless channel. After sufficiently collecting and improving their knowledge, the transmitter and receiver collaboratively create a language of their own to further enhance communication efficiency. Later on, with the improvement of the reasoning ability of each communication party, the transmitter can briefly deliver a set of words extracted from a text sentence, then the received using their knowledge and its logic and reasoning ability to understand the intent meaning of the original message. Fig.~\ref{TheoryofMind} illustrates a general case for the works following the Theory of Mind direction. The considered system mimics the learning behavior of humans; from the birth stage, babies know absolutely nothing about how to communicate with adults except for accumulating the observations of adult's behavior and language. They start developing the cognitive perspective through a twelve to eighteen-month period and are able to talk simple words. After that, they learn how to communicate in the languages of their parents and continue to develop their logic and reasoning abilities later on. As we describe, the similarity between the development of the Theory of Mind-based semantic communication and the human learning process of the human is clearly observed. Therefore, this direction heavily depends on the learning ability of communication agents, which is normally assumed to be unlimited. Next, we provide the literature review in this direction to help readers gain a deeper understanding of the proposed scenarios, as well as their challenges and potential solutions.

Inspired by Daniel Kahneman's book \textit{Thinking Fast and Slow} \cite{Kahneman}, which demonstrates there are two systems of thoughts: one is irrational, and the other is logical that drive human behavior, the authors in \cite{Hyowoon} proposed two levels of Semantic-Native Communication (SNC), named System 1 SNC and System 2 SNC, respectively. In the proposed system, they examine a scenario wherein two agents perform tasks requiring communication or at least coordination; one is the speaker, and the other one is the listener. The speaker transforms the message from the observation space to the concept (meaning) space before symbolizing it to representation space. These three spaces are defined as the triangle of meaning in human cognition in \cite{Ogden}. The author models the transforming and symbolizing process with mathematical modeling. With the received message, the listener reverses the encoding process using the same mathematical modeling as the speaker. However, the authors recognized the shortcomings of System 1, and infused the contextual reasoning ability into each agent, where the speaker not only selectively encodes the obverse but also pays attention to the context of the involved listener. Specifically, the agent mimics its listener using contextual reasoning by conducting self-semantic native communication with a virtual agent, where the process can sort out the redundant information and thus improve the system's efficiency. This process runs internally and iteratively, which gradually structures the contextual reasoning ability.

\begin{figure*}[t]
\centering
\includegraphics[width=0.9\textwidth]{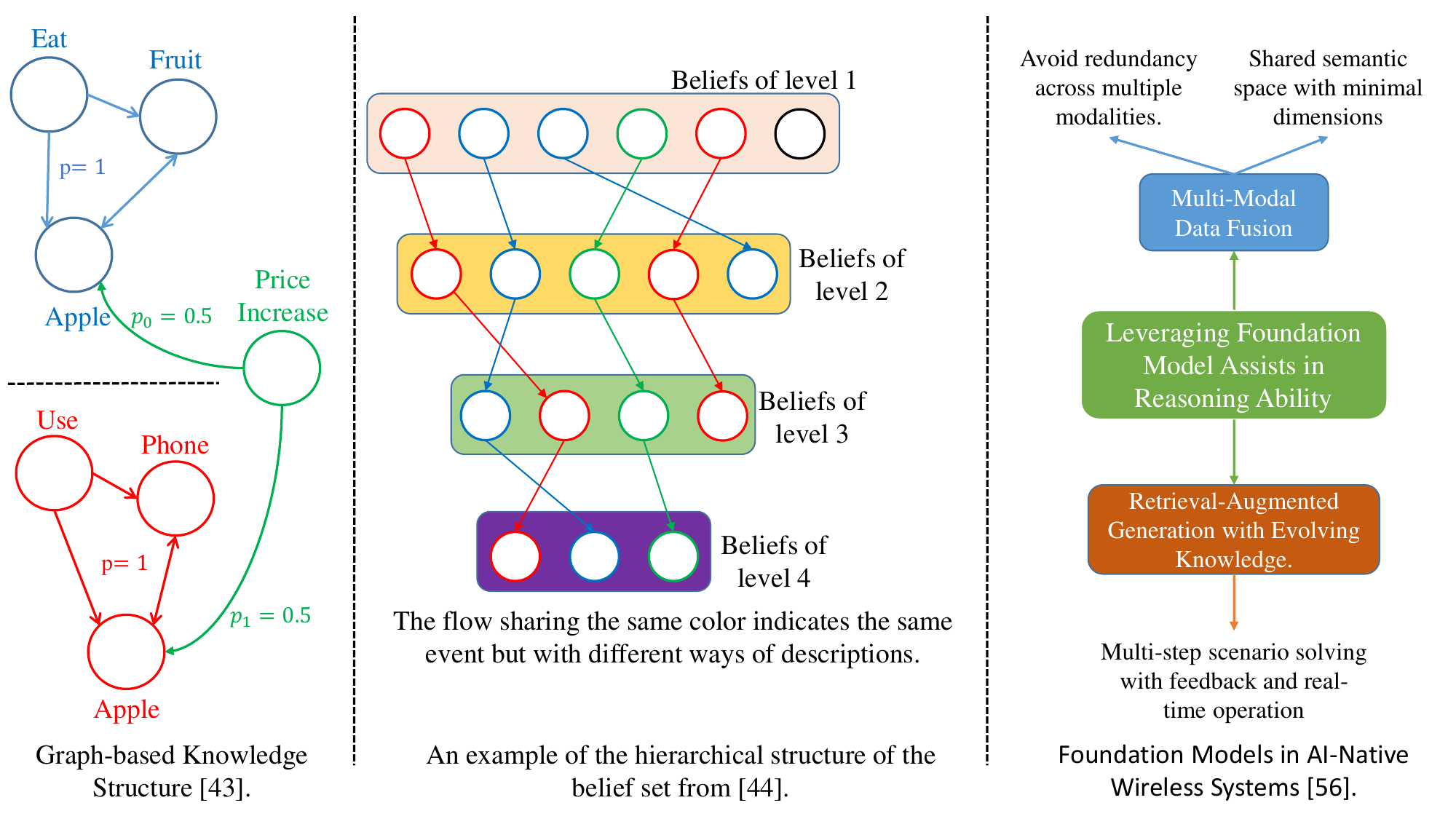}
\caption{The visualization of different knowledge representations.}
\label{TheoryofMindKnowledge}
\end{figure*}

Improving from the foundation of this work, Liang et al. \cite{Liang} proposed a different approach for reasoning-based semantic communication, where it adopts the graph-based knowledge structure to represent the semantic entity, relationship, and reasoning rules instead of a set of predefined symbols, which limit the capacity of the system. To be specific, the proposed graph contains three main components: entities, relations, and reasoning rules. By considering the graph-based structure to represent the semantic meaning of a message, they offer a more versatile approach to capturing hidden relationships among objects, a more accurate interpretation of semantic meaning. For instance, by delivering a message that has two words, ``apple" and  ``human", without any additional information, it is unclear what is the connection between them, and difficult to interpret the semantic meaning of the message. Now, consider that the receiver acquires a graph-based message, which also has exactly those two entities but a relation ``eat" between them; it would be perfectly clear that the human eats an apple (fruit) and not ``apple" the company due to the common sense as shown in the left side of Fig.~\ref{TheoryofMindKnowledge}. In addition to the proposed graph-based structure for the transmitted message, they further embed them to low-dimension representation for the improvement in communication efficiency, an innovative inference approach to tackle the missing entities/relations problem.  Finally, a life-long model update is proposed to help the communication agent learn from the previous experience (message) together with the newly received information, thus improving the reasoning rules.

Until now, the reasoning ability of communication agents has been limited in clarifying the actual semantic meaning of an embedded word by considering its context alongside other words. Seeing much more potential in the reasoning ability, the authors in \cite{KurisummoottilReasoningNeuro} proposed a framework to build an \textit{emergent language} to transmit fewer bits while containing the same semantic meaning compared with the conventional system. Instead of measuring the relations among words sent, they go directly to the root and identify the causal structure behind the data generation, find the best way to present the most meaningful way to describe the data by leveraging the Neuro-Synbolic AI framework and the generative flow networks. The work in \cite{ThomasReason1} proposed an improved version for the framework by considering the signaling game for the transmitter and receiver, which consequently a more concrete emergent language, and \textit{Neuro-Symbolic} AI to combine current learning experience and the reasoning from the past. Furthermore, a chain of thought reasoning enables the generalized capabilities of the listener without fine-tuning or retraining.

Sharing the same idea of creating a common language between transmitter and receiver, authors in \cite{Farshbafan} proposed a different approach to construct a new language, where a bottom-up curriculum learning framework based on reinforcement learning for speakers to determine the perfect abstract description and listener take the action that returns the highest reward. Their semantic communication system is goal-oriented, which means the model is designed to do certain tasks, and each task in their work is defined by a chain of multiple sequential events. The process can be described as follows: the speaker intends to guide the listener to do a task such as ``go to the door and open it". In the beginning, the listener and speaker have no common language; then the speaker has to break the task into multiple sequences of simple tasks such as step one step forward, turning right and step two steps, and rotating the door knob top open. Each event can be represented by multiple steps/beliefs as shown in the middle of Fig.~\ref{TheoryofMindKnowledge}. Due to the mismatch of a common language set, the listener may wrongly interpret the intended message of the speaker and step backward. Therefore, the speaker may encounter many difficulties in communicating with the listener, and it has to observe the listener's actions in the environment and adjust the message to steer it toward the right action. The task will be deemed complete once it ends after a predetermined period of time.

A different perspective is considered in work \cite{ThomasPragmatic}, where the author proposed the possibility of the mismatch in logic and reasoning between communication agents into consideration, which can severely affect the system's performance. To resolve the problem, they implement a two-level feedback mechanism to the transmitter network: conventional channel quality level and semantic feedback level. The first level is proposed to guarantee an efficient mapping process of semantic features to a finite constellation, while the latter one is used to directly estimate the mental state of the receiver through the semantic effectiveness metric. Different from the previous work, the transmitter can now understand the receiver's behavior more clearly by considering the effects of the environment. For instance, the receiver can behave differently to the same transmitted signal due to the environmental noise, thus causing confusion for the transmitter to understand and reason its listener. They leveraged the Theory of Mind and the Graph Neural Network (GNN) to facilitate the ability to extract causal relations and adapt its parameters quickly and effectively. In the simulation results, they show that their system only transmits the essential semantic representation and nearly achieves perfect semantic reliability.

\begin{table*}[t]
\centering
\renewcommand{\arraystretch}{1.00}
\caption{The summarization of contributions of various works related to the Theory of Mind-based semantic communication}
\label{sim_tab}
\begin{tabular}{|p{0.8cm}|p{2.2cm}|p{13.5cm}|} 
\hline
Works & Algorithm & Strengths \& Limitations \\ \hline
\cite{YongReasoning1} & 
GAML &
\begin{tabular}[c]{@{}l@{}}\textcolor{green}{\cmark} \: A novel generative adversarial imitation learning approach is proposed to guide the decoder to understand\\ the source user's reasoning mechanism.\\ 
\textcolor{green}{\cmark} \: Represent the hidden information through a graph-based structure.\\ 
\textcolor{red}{\xmark} \: Performance relies heavily on the quality and density of semantic graphs.\\ 
\textcolor{red}{\xmark} \: Implicit semantics are highly context-dependent, limiting generalization across diverse applications.\end{tabular} \\ \hline

\cite{YongReasoning2} & 
G-RML &
\begin{tabular}[c]{@{}l@{}}\textcolor{green}{\cmark} \: Developed from the \cite{YongReasoning1}, the authors capture both implicit and explicit semantics for the causal-reasoning\\ mechanism.\\ 
\textcolor{green}{\cmark} \: Maps the high-dimensional graphical into low-dimensional constellation space\\ 
\textcolor{green}{\cmark} \: The framework enables the destination users to map the observed explicit semantics to
the corresponding\\ implicit semantics with less transmitted data. \\ 
\textcolor{red}{\xmark} \: Single transmitter and receiver scenario.\\ 
\textcolor{red}{\xmark} \: Limited applications: finding the relations among entities.\\ 
\textcolor{red}{\xmark} \: Limit the number of reasoning paths: The performance decreases when a number of reasoning paths increases.\\ 
\end{tabular} \\ \hline

\cite{EmilioReasoning} & 
6G-GOALS &
\begin{tabular}[c]{@{}l@{}}\textcolor{green}{\cmark} \: Theoretical and algorithmic foundations of semantic and goal-oriented communication.\\ 
\textcolor{green}{\cmark} \: Integrate the semantic communication into an intelligent and adaptable Radio Access Network.\\ 
\textcolor{green}{\cmark} \: Improve the reasoning of communication agents with the Foundation Model.\\ 
\textcolor{red}{\xmark} \: Lack of numerical analysis. \\ 
\textcolor{red}{\xmark} \: The deployment of foundation model on IoT device is surreal.\\ 
\end{tabular} \\ \hline

\cite{FuhuiReasoning1} & 
Text2KG Alignment &
\begin{tabular}[c]{@{}l@{}}\textcolor{green}{\cmark} \: Cognitive semantic communication framework leverages a knowledge graph for semantic symbol abstraction\\ and error correction.\\ 
\textcolor{green}{\cmark} \: Demonstrate improved data compression rates and communication reliability compared to traditional system.\\ 
\textcolor{red}{\xmark} \: High resource requirements for fine-tuning large models (e.g., T5) and managing extensive knowledge graphs.\\ 
\textcolor{red}{\xmark} \: Single transmitter and receiver scenario.\\ 
\end{tabular} \\ \hline

\cite{FuhuiReasoning2} & 
\begin{tabular}[c]{@{}l@{}} Semantic Alignment\\ and Correction\\ Algorithms \\ \end{tabular} &
\begin{tabular}[c]{@{}l@{}}\textcolor{green}{\cmark} \: Developed from the \cite{FuhuiReasoning1}, the authors tackled the multiple-user problem in the cognitive semantic\\ communication system.\\ 
\textcolor{green}{\cmark} \: Leverage user context to determine the message ownership and avoid the wrong targeted sending.\\ 
\textcolor{red}{\xmark} \: It is more time-consuming than other techniques, twice as much as the JSCC due to the NLP module.\\ 
\end{tabular} \\ \hline

\cite{ShengzheReasoning} & 
RAG &
\begin{tabular}[c]{@{}l@{}}\textcolor{green}{\cmark} \: Fuse multi-modal sensing information to a common semantic space.\\ 
\textcolor{green}{\cmark} \: Connect the agent knowledge to real-world experience.\\
\textcolor{green}{\cmark} \: Facilitate the performance of logical and mathematical reasoning of the LLM. \\
\textcolor{red}{\xmark} \: Heavily depends on the performance of the LLM, which requires to be trained on the large amount of data.\\ 
\textcolor{red}{\xmark} \: Resource constraints to deploy the LLM on low-level device.\\ 
\end{tabular} \\ \hline

\cite{ChristinaReasoningDisentangling} & 
KG-MSF &
\begin{tabular}[c]{@{}l@{}}\textcolor{green}{\cmark} \: Fuse multi-modal data using shallow-level (direct) and deep-level (reasoning) semantic triplets. \\ 
\textcolor{green}{\cmark} \: Capture both direct and reasoning correlations among multi-modal data for tasks like visual question\\ answering (VQA). \\ 
\textcolor{red}{\xmark} \: Performance may vary significantly with channel conditions despite improved robustness. \\ 
\end{tabular} \\ \hline

\cite{BingyanReasoningKnowledge} & 
Transformer-based knowledge extractor &
\begin{tabular}[c]{@{}l@{}}\textcolor{green}{\cmark} \: Propose a knowledge graph-enhanced semantic communication framework that improves receiver-side\\ semantic decoding by integrating knowledge extraction.\\ 
\textcolor{green}{\cmark} \: Retrieve relevant factual triples from noisy received messages to enhance semantic decoding.\\ 
\textcolor{red}{\xmark} \: Low SNR conditions can increase false positives in triple extraction, affecting decoding reliability. \\ 
\end{tabular} \\ \hline

\cite{JinhoReasoningSemantic} & 
Q-learning &
\begin{tabular}[c]{@{}l@{}}\textcolor{green}{\cmark} \: Propose a game-theoretic model for semantic communication based on the Lewis signaling game to map\\ semantic messages to signals and responses.\\ 
\textcolor{green}{\cmark} \: Integrate correlated knowledge bases (KAs and KBs) at the sender and receiver to enhance semantic\\ agreement success rate.\\ 
\textcolor{red}{\xmark} \: When knowledge bases are insufficiently correlated or incomplete, semantic accuracy degrades significantly. \\ 
\end{tabular} \\ \hline

\cite{DylanReasoning} & 
Variational-AE with Wasserstein Loss &
\begin{tabular}[c]{@{}l@{}}\textcolor{green}{\cmark} \: Propose a novel conceptual SCCS that integrates reasoning for efficient and intelligent communication.\\ 
\textcolor{green}{\cmark} \: Can identify the true causal effect of one variable on another rather than the correlation.\\ 
\textcolor{red}{\xmark} \: Training VAE and performing causal inference requires significant computational resources. 
\end{tabular} \\ \hline

\end{tabular}
\end{table*}

Up to this point, all the discussed work above only examined the semantic communication system with one transmitter and one receiver, which makes the formation of new languages simple. When there are additional communication agents, the language formation among communication participants becomes more complex, and this will surely lead to the collapse of the semantic communication system due to the uninterpretable message at the receiver if they don't agree on the common language. The authors in \cite{MohamedMultiuser} are the first to raise the problem and referred it as the language mismatch. With the distinct languages, the receiver misinterprets the message from the receiver, which is referred to as semantic noise. The authors in the paper tried to tackle this problem by proposing a semantic channel equalizer, which measures the mismatch of knowledge representation and then propounds a framework based on optimal transport theory to transform the representation from a language to the targeted receiver language. In simple terms, their idea is translations among semantic languages. The remaining problem is how to translate correctly and effectively, which they achieve with a Bayes-optimal selection strategy. The authors in \cite{Nitisha} also proposed a multi-user semantic communication system but focused on a different problem instead of the mismatch language. Specifically, they focus on the tradeoff between the computing and communication resources among semantic communication users, where each user has the reasoning ability supported by the massive computing resources, in which they can fill in the missing information when receiving a message. In the proposed scenario, instead of transmitting the full message, only a part of it is transmitted, and the missing information is expected to be filled by the receiver using its reasoning ability. The shorter message transmitted indicates that there is less information the receiver has to start with, so more computing resources are dedicated to reasoning the remaining message. It is like solving a puzzle; the more hints we have, the easier the problem. Furthermore, the authors developed a noncooperative game model and designed an iterative algorithm to determine the best control actions and communication strategies that maximize the overall system's efficiency.

The mentioned research above mainly works on the physical layer designs while ignoring the essentials for designing the communication protocol in the Medium Access Control (MAC) layer. Acknowledging this, \cite{SeoProtocols} considered the need for developing protocols that can offer better flexibility and adaptability to the changing environment compared with the static control signaling message of the human-crafted classical protocol. Thus, they proposed a semantic protocol model constructed by improving neural network-based protocol models into an interpretable symbolic graph written in a logical language. Their proposed \textit{ProbLog-based Semantic Protocol Model} achieved not only communication but also memory efficiency while maintaining robust performance in non-stationary environments. Besides the above research, various works have been dedicated to studying the Theory of Mind-based semantic communication, which we briefly introduce in Table \ref{sim_tab}. There are existing works where the proposed scenarios overlap with different directions. For example, some works, such as \cite{FuhuiReasoning1, FuhuiReasoning2, ShengzheReasoning}, align with the AIGC direction. On the other hand, the works \cite{ChenlinReasoning,BingyanReasoningKnowledge} are the result of the combination of the Theory of Mind and deep joint source-channel coding directions.

\begin{figure*}[t]
\centering
\includegraphics[width=0.97\textwidth]{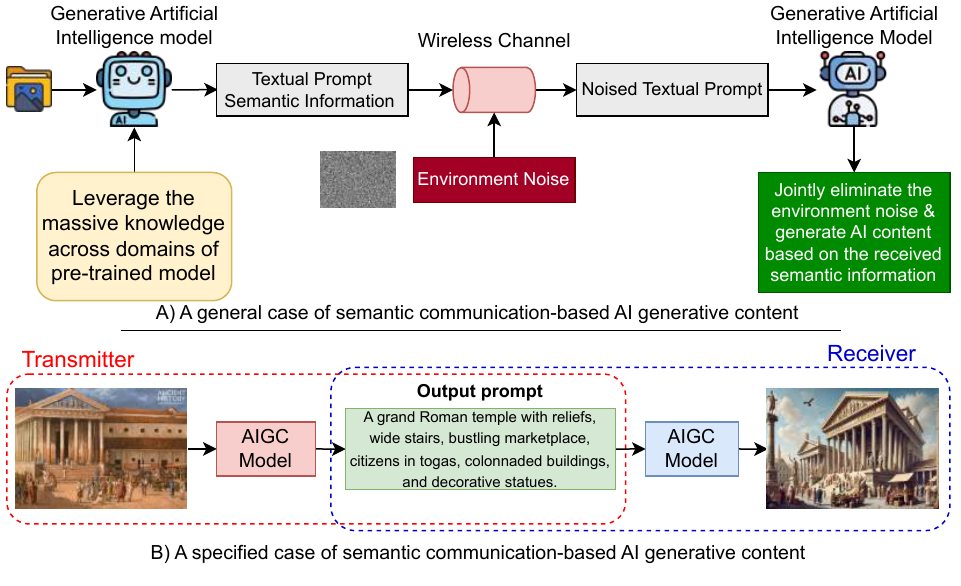}
\caption{The visualization of the architecture of Generative AI-based Semantic Communication Direction.}
\label{AIGC}
\end{figure*}

\subsection{Challenges of the Direction}
 In \cite{ChristinaLessdata}, the authors have provided the advancements of the methods compared to other directions, such as the flexibility or the proactive ability in decision making while escaping from the dependency of large availability of data. The direction is clearly interesting and getting more and more attention from the researchers within the field. However, there exist a couple of challenges preventing the direction from blooming. First of all, it is really difficult to provide casual reasoning like humans, for which the communication agent is required to form a large set of knowledge and beliefs. In addition, they also have to replace outdated beliefs/knowledge with new ones, not to mention the need to identify the wrong beliefs fed by adversarial agents. Furthermore, the infeasibility of achieving a large hierarchical set of causal and reasoning sequences is one main factor; for example, in \cite{Farshbafan}, authors considered three levels of beliefs, and they already have to employ the curriculum learning to reduce the complexity for the proposed solution. On the other hand, human beliefs are far more complex and do not follow the hierarchical structure as the assumption in the paper. Finally, causal reasoning is user-oriented, which might be fit for one particular user but completely wrong for other users. In order to obtain more attention and be widely developed, the Theory of Mind-based semantic communication researchers have to employ them in practical scenarios and demonstrate their effectiveness.
\vspace{-0.1in}
\subsection{Summary and Lesson Learned}

The section has discussed novelty research works under the Theory of Mind direction, providing examples to describe its concept and the evolving research direction. The development of the system to improve the reasoning ability of the communication agent has been provided throughout the literature review. The summary and key lessons learned from this section are listed as follows.
\begin{itemize}
    \item ToM-based Semantic communication is built up on the accumulated shared knowledge between the transmitter and the receiver. Therefore, it is essential to design the knowledge base for the communication agents, which can significantly improve the system's performance.
    \item The reasoning ability over the context and finding the causal relationship among entities is one of the most important properties of ToM-based Semantic Communication, which enables the transmitter to reduce the transmitted data and encourage the receiver to make the most semantic sentence out of it.
    \item Due to the limited capacity to train communication systems from scratch, current research leverages pre-trained large language models as communication agents to achieve powerful reasoning abilities across modalities, enabling broader task applications and improved performance.
\end{itemize}
\vspace{-0.1in}
\section{Direction II: Generative AI-Based Semantic Communication}
\label{DirectionII}
With the introduction of ChatGPT and other generative AI models, the communication agent is considered to be equipped with an AI module, which minimizes the transmitted signal length and leverages the AI module at the receiver site to generate the information. The system performance in this direction heavily depends on the capabilities of AI models, which is a drawback but is also a strength. The generated AI models era has already started, and enormous accomplishments have been achieved, such as ChatGPT, a generative artificial intelligence chatbot that has been proven useful across domains for humans or Gemini. Therefore, standing on the shoulders of giants will accelerate the development of semantic communication. As shown in Fig.~\ref{AIGC}, we demonstrate the overall workflow of the generative AI-based Semantic communication in both a general and specified case. The example of the specified case is implemented by using the GPT-4o version.
\vspace{-0.1in}
\subsection{Generative AI Models}
Since the introduction of the Large Language Model (LLM) to users at the end of 2022, it has created a significant impact in both academic and industrial circles due to its ability in logical thinking and also reasoning. It has successfully leveraged the massive obtained knowledge to answer most of the general questions and has demonstrated the capability to answer complex questions requiring advanced domain-specific understanding. As a result of massive attention, it has been developed to generate different modalities such as image, audio, and also video \cite{Keng-BoonSurvey} from text. One practical example, it has also been employed in creative endeavors, such as generating paintings or composing music, often producing results indistinguishable from those created by humans. This progress has opened up a new research area called generative AI, which goes beyond creating content from text. To be specific, unlimited research has studied the transformation from image modality into text description and also across different other modalities. With an enhanced knowledge base capable of capturing cross-modal relationships, the generative AI model has excelled in encoding ability compared to traditional AI methods; for instance, the generative AI model can encode one image in various ways, such as latent representations, image captioning, or even graph structure. On the other hand, the ability of traditional AI methods is limited to one form of representation. Another advantage is its ability to do logical thinking, where it encounters an unseen task but is still able to solve it efficiently. With a much broader knowledge about the abilities of generative models, readers are encouraged to refer to \cite{HanqunGenerative}. Most of the architecture of the generative AI models is built based on the combination advancement of diffusion model \cite{JonathanDiffusion} and Generative pre-trained Transformer technique, which step by step improves its ability in semantic encoding, content generating, and problem-solving in a certain field. In the next sub-section, we particularly focus on the deployment of generative AI in semantic communication and its benefits.
\vspace{-0.1in}
\begin{figure*}[t]
\centering
\includegraphics[width=1\textwidth]{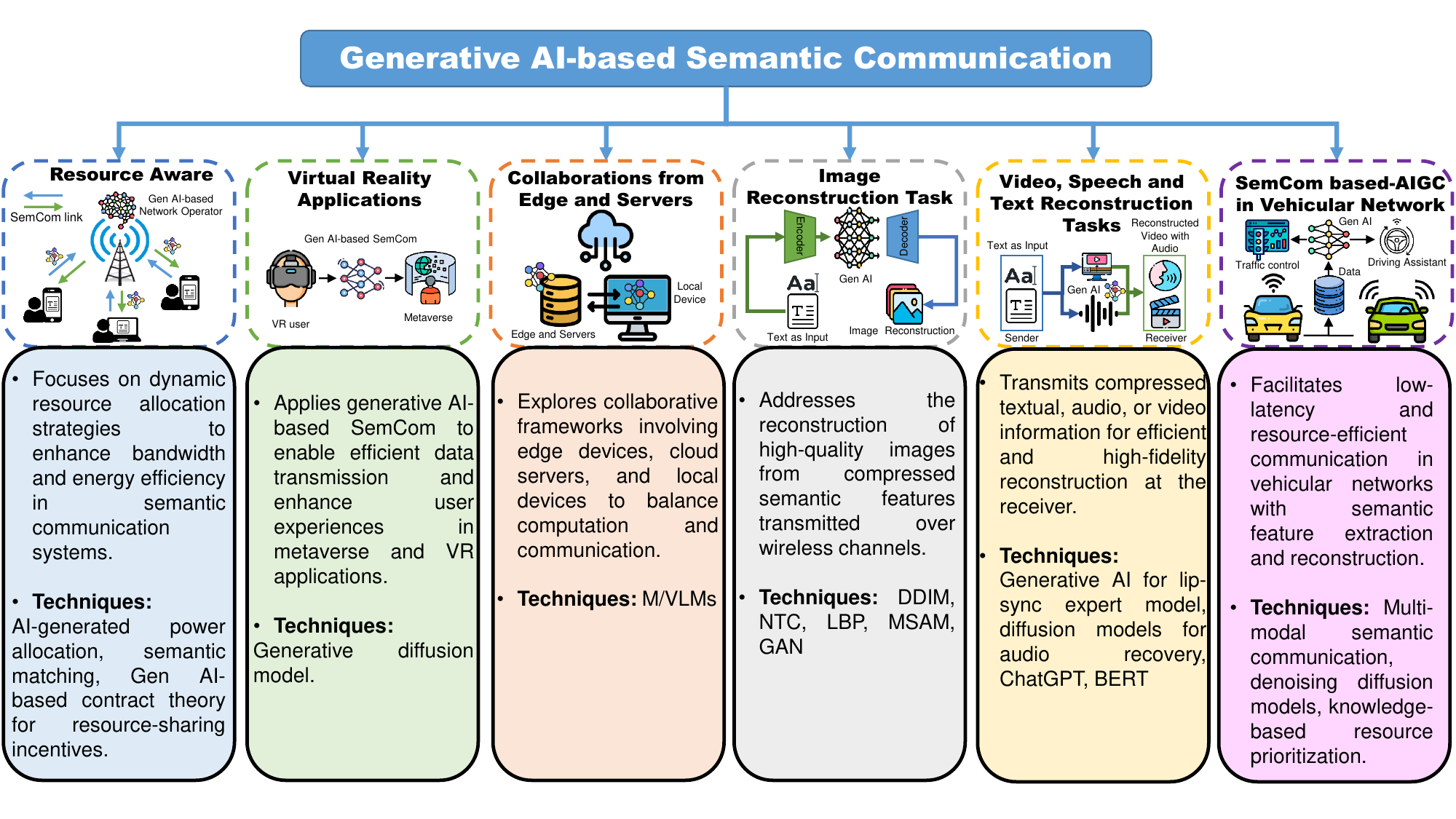}
\caption{The visualization of the architecture of generative AI-based semantic communication Direction.}
\label{AIGCService}
\end{figure*}
\subsection{The existing research of Generative AI-based Semantic Communication}
Recognizing the control and encoding ability of the generative AI model, researchers have widely addressed various challenges in semantic communication. These applications span a wide range of areas, including resource allocation, network management, and data encoding as shown in Fig.~\ref{AIGCService}, showcasing the versatility and impact of generative AI in this domain.
\subsubsection{Resource Aware}
Even though semantic communication has made great progress, wireless communication still faces challenges. Without a good strategy for allocating resources, it struggles to keep up with the rapid increase in the number of connected devices, which motivates the works in \cite{ChengsiAIGC,WangAIGC,BaoxiaAIGC,ChengAIGC}. In \cite{ChengsiAIGC}, the authors indicated that semantic communication and generative artificial intelligence technologies can complete each other. Specifically, semantic communication can facilitate the high demand for artificial intelligence-generated content (AIGC) services such as high data rates, throughput, and low latency under limited spectrum resources. On the other hand, the generative AI model can function as an elastic network operator, enabling a dynamic resource allocation strategy for the semantic communication system. For example, the AI-generated content application is inherently linked to GAI models, where it is easy to capture the immediate context so that it is suitable with the semantic communication framework. On the other hand, semantic communication can effectively address the massive data volume with different modalities under bandwidth and computing resources.

Authors in \cite{WangAIGC} considered the mixed reality application, where each individual user has to conduct the heavy and repetitive computational task to generate the artificial content for the view image individually in a normal case, and this leads to a massive amount of computing resources. Therefore, they proposed semantic communication among users, where the user transmits the generated content and semantic information extracted from its view image to nearby users, which can relax the computation constraints for the mounted head devices and enable efficient synchronization of free space. In addition, they knew that sharing the generated image view from one user to another user might not proceed without any incentive mechanism, so they designed a mechanism based on contract theory, where a generative AI is utilized to design optimal contracts. In order to avoid transmitting useless image views among users, first, the transmitting user distributed the interest points and descriptors to nearby users, and then they determined which images they wanted to receive by the semantic information matching process. After the desired image is specified, a contract is formulated to encourage sharing. 

Inspired by this work, \cite{BaoxiaAIGC} also proposed an AI-generated power allocation algorithm in a semantic communication system. Here, the authors considered object detection tasks and considered the semantic information within the image to decide which areas will be allocated more power to transmit. Specifically, a YOLO-based semantic communication is proposed to discard irrelevant and useless information after the image data acquisition, which tackles the huge volume of data transmission and exchanges between the physical world and the virtual in the digital twins application. On top of that, they enhance the performance of the detection module by designing a new layer aggregation network (ELAN-H) and SimAM attention. Lastly, they provided more power resources to transmit the regions with high semantic meaning while limiting the transmission of low information areas.

Unlike the works mentioned earlier, which proposed using generative AI models as operators to support semantic communication, the study in \cite{GenerativeAISem} showcased a practical example where the semantic communication system enhances the generative AI model in an AIGC task. In the proposed scenario, a mobile user offloads the content-generating task to a server for high-quality results. However, the high-quality image requires lots of bandwidth resources to download. This is where they proposed a semantic communication framework to address the problem, which prioritizes transmitting the regions with high semantic scores, eliminates the redundant information unrelated to the task, and meets the communication constraints.

Considering the same task, the authors in \cite{ChengAIGC} proposed a more comprehensive latency objective which includes the transmit latency, and the computing latency at both edge and local devices. By adjusting the diffusion steps of the generative AI model, they can control the generated image quality and the computing latency to effectively serve the user with different service requirements, such as latency over content quality or the reverse way. By doing this, the generative AI models can provide the network the dynamic availability it needs and make it a workload-adjustable transceiver, thereby allowing adjustment of computational resource utilization in edge and local. For further improvement, they formulated a resource-aware workload trade-off scheme by controlling resource availability, dynamic channel quality, and diverse AIGC service requirements and leveraged dueling double deep Q network (D3QN) to solve the problem.

\subsubsection{Virtual Reality Applications}
The use case of generative AI-based semantic communication has also grown, finding applications in Virtual Reality, where ensuring high-quality data transmission is the number one priority. A number of works \cite{YijingAIGC,Yijing2AIGC,ZheAIGC} have proposed a scenario to utilize the capacity of generative AI-based semantic communication in Metaverse. Specifically, the authors in \cite{YijingAIGC} proposed a unified framework that integrates semantic communication and the AIGC to facilitate the application requirement, which they referred to as integrated SemCom and AIGC (ISGC). In the paper, they illustrate the drawbacks in case the system is not well designed, such as inefficient use of resources and low-quality content. On the other hand, an efficient framework ISGC can facilitate visually appealing but also contextually relevant and meaningful, enhancing user experiences in the Metaverse while making sure the bandwidth resource is allocated to the user when it needs it. The proposed framework has three main modules: semantic module, inference module, and finally, rendering module. The semantic module is responsible for processing the data generated by the edge devices and conveying the semantic information before transmitting it to the metaverse service providers, which contain the other two modules. The received semantic information is first decoded by a semantic decoder, which normally results in low-quality or incomplete data (highly compressed ratio to save communication resources). This is where service providers come in and utilize the AIGC module to create superior digital content that enhances user experiences. The inference module is a pre-trained model and is able to deal with various modalities and distributions. Last but not least, the rendering module combines massive amounts of real or imagined information to create immersive and interactive virtual environments.

The work proposed in \cite{Yijing2AIGC} also considered the integration of the generative model in semantic communication for Metaverse application; however, instead of focusing on the resource allocation problem, they studied the problem of security against the attackers, whose purpose is to contaminate the Metaverse services by feeding malicious semantic data with similar semantic information but different desired content into the network. The article claims that the incorporation of blockchain technology into semantic communication can facilitate the decentralized, more effective, and safe transmission of semantic data among users in the undiscovered Metaverse. The blockchain-aided semantic communication can complement each other as follows: \textit{1)} semantic communication enables the participants to communicate, extract the semantic meaning from the data, and reduce the communication costs and storage memory, while \textit{2)} the blockchain technology allows them to construct trust among anonymous participants, thus they can freely share the extracted semantic features without the concern of information manipulation or false modification from the untrustworthy parties.

The most recent work that proposed generative AI-based semantic communication is \cite{ZheAIGC}, where the authors focused on an industrial factory scenario. They integrate the knowledge base at the receiver for the metaverse rendering, which contains useful information related to scenery, camera, and object bases. By utilizing these knowledge bases, the semantic communication system can get significant benefits; for example, by understanding the fixed and moving objects within the scenery, the system can dedicate computing and communicating resources to the moving objects while limiting resources for the fixed objects. Furthermore, the base camera knowledge and base object knowledge are used as condition input into the AI-generated content to construct a metaverse world, which facilitates a reliability construction. Finally, an optimal transport-enabled semantic denoiser was proposed to enhance the visual quality of the image metaverse scenery.


\subsubsection{Collaborations from Edge and Server}
Due to limited computational and communication resources, local devices occasionally fail to complete AIGC inference tasks, update the outdated generative AI model, or may be out of communication range with target devices. These challenges have driven researchers to leverage edge devices and cloud servers as alternative solutions in generative AI-based semantic communication \cite{MengmengAIGC,XiaLeAIGC,AIGCEdge0,AIGCEdge1,AIGCEdge2}.
An edge-device collaboration for generative semantic communication is proposed by authors in \cite{MengmengAIGC} to address a goal-oriented task, where both transmitter and edge devices are equipped with a pre-trained Multi-modal/Vision Language Models (M/VLMs). The considered task is to capture the main meaning of the image in textual modality, which can be easily achieved by the equipped generative AI model at the edge server or local devices. In several cases, the limited computation capacity of mobile devices cannot achieve high-quality performance, and thus, they propose the collaboration between the device and the edge server, where a larger M/VLM model is employed to generate a higher-quality prompt related to the communication intent. However, the collaboration provided by the edge server can incur extra latency due to the transmission of compressed source signals to the edge server; there are existing tradeoffs between latency and prompt quality. To balance the quality and the latency, the authors designed a joint optimization strategy to tackle prompt generation offloading and the allocation of communication and computation resources, with the goal of minimizing overall semantic communication latency while ensuring high semantic quality.
\begin{figure*}[t]
\centering
\includegraphics[width=0.85\textwidth]{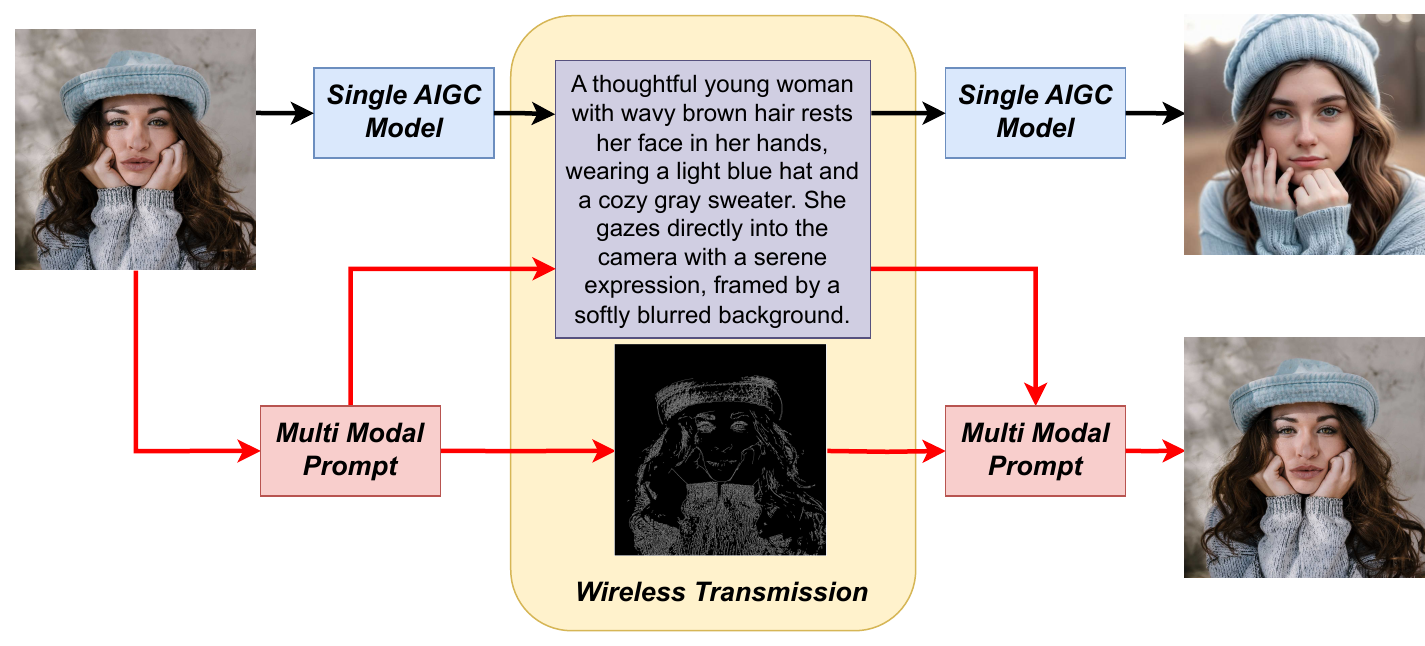}
\caption{The visualization of the problem in AIGC with a single modality (image-to-text) transmission.}
\label{ProblemAIGC}
\end{figure*}
Inspired by the work, \cite{XiaLeAIGC} proposes a broader collaboration scheme that includes a cloud server, edge servers, and finally, mobile devices. They illustrate the detail of the framework by sequentially implementing three stages: \textit{1) Initial Network Preparation Stage}, \textit{2) GAI-integrated SemCom Service Provisioning Stage}, and finally \textit{3) Model Synchronization and Update Stage}. An interesting point of the proposed framework is that they consider model synchronization and updating, which can enhance the collaboration performance among a large number of devices and guarantee up-to-date information. This fine-tuning model parameters process is executed at the cloud server, where user data are gathered.  Taking a more cautious approach, the authors of \cite{AIGCEdge0} propose fine-tuning the model on local devices and edge servers, thereby mitigating the risk of exposing private information during data collection.

The previous works simply assumed the local user devices are willing to participate in the model updating process, which is overly simple and less likely to happen. The authors in \cite{AIGCEdge1} proposed an incentive-based framework to encourage IoT devices to share their knowledge graphs. The proposed system introduces an IoT knowledge market that incorporates utility-driven incentives and fair pricing mechanisms, enabling effective and secure utilization of KG data.

\subsubsection{Image Reconstruction Task}
Previous works have demonstrated the outstanding capability of the AIGC model in terms of conveying the semantic meaning of the raw data and transforming it into different text modalities, which significantly reduce the size of the transmitted signal while guaranteeing oriented-task performance at the receiver. However, one drawback of the AIGC model is the reconstruction ability, where the receiver's goal is to reconstruct the original data. With only the textual problem being transmitted to the receiver, the generative AI content model will create an image according to the description but totally different from the original image, as shown in the upper part of Fig.~\ref{ProblemAIGC}. Therefore, various works \cite{HongyangAIGC2,ChunmeiAIGC,SenranAIGC} have been proposed to improve the reconstruction ability of the generative AI model. 

One of the earliest works, the authors in \cite{HongyangAIGC2} addressed this problem of the generative AI-based semantic communication by proposing a visual prompt (can be a contour map similar to the low branch in Fig.~\ref{ProblemAIGC}), which acts as a condition for the generative model that enables the accuracy of the reconstruction image. The design of the visual prompt needs to be informative for the receiver and also low in communication-consuming, which the authors acquire after the noise-adding phrase of the Denoising Diffusion Implicit Model \cite{JiamingDDIM}. Furthermore, they leverage the convert communication to protect the transmission of multi-modality prompts, whose target is to successfully transmit the image without detection by the warden. Therefore, an optimization problem is formulated to achieve this goal by controlling the transmission power, jamming power, and finally, the number of diffusion steps.

In contrast, the work in \cite{ChunmeiAIGC} and \cite{SenranAIGC} also tackles an image reconstruction task but under more relaxed constraints, where the reconstructed image only needs to preserve the structural features of the original. In \cite{ChunmeiAIGC}, the image semantic is extracted by the textual transform coding via prompt inversion \cite{YuxinPromt}, while the Holistically-nested Edge Detection (HED) with a non-linear transform code \cite{JohannesNTC} (NTC) model is used to detect and compress the edge map features. These two features will be transmitted and recovered at the receiver site. Then, they will be the input of the generative diffusion model. The text prompt and edge data map only provide the structure and the description of the image, which makes the output of the generative model share the semantic meaning with the original image and also the structure, but they are not identical. Lastly, they studied the semantic-aware resource allocation problem, whose objective is to minimize the total power consumption while ensuring the perceptual quality of the regenerated image.

On the other hand, \cite{SenranAIGC} offers a greater degree of control over a similar problem. Rather than relying solely on structural feature maps, the authors propose extracting diverse types of visual features from the diffusion process. Specifically, they decompose the image into different meaningful visual information such as color, textures, and edges, enabling more comprehensive and detailed image representation. The semantic color is derived by median filtering and downsampling operations, which results in the mosaic image and reflects the dominant color tone of their covered region. On the other hand, they obtain the texture feature maps by using the Local Binary Patterns (LBP) algorithm. These two semantic feature maps, along with the text description output from the BLIP model \cite{JunnanBLIP}, are encoded, transmitted, and decoded at the receiver. Then, the semantic feature maps go into ControlNet, which is an innovative model used to control the output of the generative AI model to have certain specified semantics. Their model achieves several advantages compared to the previous scenario, such as a higher compression ratio, stronger anti-noise ability, and magnificent ability to interpret and control the generated output image.

\subsubsection{Video, Speech and Text Reconstruction Tasks}
The Generative AI model is not limited to image reconstruction tasks and has also been applied to other modalities such as video, audio, and text \cite{PulkitAIGC,TianAIGCVideo,EleonoraAIGC,ZhengAIGCSpeech,ShuaishuaiAIGC}. Each modality has unique properties and is entirely distinct from the others, requiring separate handling and encoding approaches.

In terms of video transmission, \cite{PulkitAIGC} proposed a novel video compression pipeline where it takes advantage of the consistency of the setting environment. In the paper, they considered the application of webcam videos, which contain a human head talking. First, they transcribe the human speech into a text transcript, and then this text is transmitted and decoded at the receiver site. A special decoder is employed here, which takes the text as input and outputs the reconstructed video with audio. Specifically, the text is transformed into a speech signal by a state-of-the-art generative AI model, after that, based on the speech signal, they create the movement of the lip by a lip sync expert model \cite{PrajwalLIP}. The created lip sync is then applied to the driving video (head of a specified person) to create the talking head, which simulates the input video at the sender site. The idea of the paper is innovative and mining high-quality experience with significantly low transmitted data. Recognizing that the work lacks various aspects of human behavior, such as expressions, eye blinks, and head rotations, the authors of \cite{TianAIGCVideo} proposed frameworks for transmitting human head expressions, offering a more natural appearance in the video and enhancing the user experience. Furthermore, a specialized design was introduced to synchronize human expressions with speech, further improving the QoS.

Audio transmission problem is the main focus in the work \cite{EleonoraAIGC}, where it can encounter noise and also partially missing parts of the signal. The considered problem is challenging to solve and also troublesome to deal with the missing part of the continuous speech audio signal. Therefore, on the transmitter side, besides the audio encoder, a textual encoder (AI-generated based model) is proposed to capture the textual captions, embed them in dimension features, and finally transmit them to the receiver. A diffusion model-based framework is also equipped for the receiver; it takes both the text description and audio latent embedding as the input to recover the audio that resembles the original signal the most. The diffusion model can effectively eliminate the noise from the signal and inpainting the missing audio due to its training process. The textual description provides extra information and acts as a condition for the generative model, facilitating the full potential of the AI-generated model. While in \cite{ZhengAIGCSpeech}, the authors completely eliminate the speech signal and only transmit the text description to achieve better communication efficiency. With the proposed scenario, various modules have been adopted to synthesize the speech from the text at low communication and computation costs, such as Prior Encoder, Soundstream, and WavLM. 

In \cite{ShuaishuaiAIGC}, authors leverage pre-trained language models to quantify the semantic importance of frame in the transmitted signal. Two approaches implemented by the two models include a closed-source generative language model (ChatGPT) and an open-source Discriminative Language Model (BERT). They further define word/frame importance by using proposed semantic loss, which is calculated by the BERT model when that word/frame is missing from the sentence. To illustrate the applicability of the proposed framework, the paper provided three potential application scenarios and their properties independently. Last but not least, a semantic importance-aware priority-based communication strategy is designed to minimize semantic loss, where they allocate more power to transmit the important frame. In the simulation analysis, their proposed semantic loss and BERT achieved outstanding performance and outperformed the ChatGPT model and its randomness output by a large margin.

\subsubsection{Generative AI-based Semantic Communication in Vehicular Network}
One of the most compelling applications of generative AI-based semantic communication is in vehicular communication, which requires exceptionally high reliability and ultra-low latency. There has been a large amount of research that has devoted efforts to meet the high demands of vehicular network communication \cite{GuangyuanAIGC,RuichenAIGC,AviDebAIGC}. In \cite{GuangyuanAIGC}, the authors illustrate the advancement of generative AI by the ControlNet, which improved control and precision capabilities for various practical applications such as driving assistants and smart healthcare. However, the paper rarely talks about the various challenges of applying generative AI-based semantic communication in vehicular networks.

A detailed discussion about the real-time and fast decision-making requirements of vehicular networks is provided by the authors in \cite{RuichenAIGC}. In the proposed system, the authors introduced a semantic-aware framework for generative AI under vehicular networks, which leverages multimodal encoding capabilities to effectively reduce transmission length, lower communication latency, and improve traffic safety. In the paper, the information transmission related to car accidents is considered as a specific example. The typical process is as follows: the image is encoded by a traditional AI model before being transmitted to the target vehicle, the signal is decoded at the receiver's end, and the decision is made based on the image result. This process not only consumes significant communication bandwidth but also requires substantial computing resources at both the transmitter and receiver. To address these issues, the authors converted the image modality into text by leveraging the multimodal capabilities of the generative AI model for transmission. The receiving vehicle can then make decisions based on the received text alone while also re-creating the transmitted image from the text for additional information about the car accident. Besides, there are various possibilities for using generative AI in navigation, route optimization, traffic simulation, and prediction.

While the previous considered the V2V communication, the author in \cite{AviDebAIGC} considered the wireless transmission of the rode-side image from the surveillance camera to the server. Instead of capturing the image description as above, they utilized the Mobile Segment Anything Model (MSAM) and pix-to-pix GAN \cite{PixelGAN} to address the problem of high bandwidth consumption and latency in wireless communication systems. They acknowledged that the background remains largely unchanged, particularly in applications like Intelligent Transportation Systems (ITS). First, a lightweight Mobile Segment Anything Model (MSAM) is used at the transmitter to extract only the meaningful semantic information from images, significantly reducing the amount of data needing transmission. The reduction in data transmission eases the challenge of high resource consumption in edge devices, such as roadside units in ITS. Secondly, a Generative Adversarial Network (GAN) is employed at the receiver to reconstruct and denoise the received semantic data, ensuring high-quality image reconstruction even under noisy channel conditions.  
\begin{table*}[t]
\centering
\caption{the overview of contributions of research works related to generative AI-based semantic communication}
\label{AIGC_Table}
\renewcommand{\arraystretch}{1}
\begin{tabular}{|c|cccc|c|}

\hline
\multirow{3}{*}{Works} & \multicolumn{4}{c|}{The integration between semantic communication \& AIGC}                                                               & \multirow{3}{*}{Specialities} \\ \cline{2-5}
                       & \multicolumn{2}{c|}{The function of AIGC in SemCom}                                                & \multicolumn{2}{c|}{Modality \& Management}            &                               \\ \cline{2-5}
                       & \multicolumn{1}{c|}{As a Semantic Coder} & \multicolumn{1}{c|}{Colab W/ Semantic Coder} & \multicolumn{1}{c|}{Multi-Modal} & Network Management &                               \\ \hline

\cite{FeiboAIGCLarge}  & \multicolumn{1}{c|}{\textcolor{red}{\xmark}}                    & \multicolumn{1}{c|}{\textcolor{green}{\checkmark}}                  & \multicolumn{1}{c|}{\textcolor{green}{\checkmark}}            &       \multicolumn{1}{c|}{\textcolor{red}{\xmark}}                  &            \parbox[t]{5cm}{A multimodal alignment module is proposed along with a personalized LLM-based knowledge base}                    \\ \hline

\cite{FeiboAIGCLarge2}                   & \multicolumn{1}{c|}{\textcolor{red}{\xmark}}                    & \multicolumn{1}{c|}{\textcolor{green}{\checkmark}}                  & \multicolumn{1}{c|}{\textcolor{red}{\xmark}}            &    \multicolumn{1}{c|}{\textcolor{red}{\xmark}}                 &        \parbox[t]{5cm}{Split the image into different semantic segment by a KB and weight the semantic information by an attention module.}                              \\ \hline

\cite{FeiboAIGCLarge3}                   & \multicolumn{1}{c|}{\textcolor{red}{\xmark}}                    & \multicolumn{1}{c|}{\textcolor{green}{\checkmark}}                  & \multicolumn{1}{c|}{\textcolor{red}{\xmark}}            &    \multicolumn{1}{c|}{\textcolor{red}{\xmark}}                 &        \parbox[t]{5cm}{Focused on the 3D transmission scenario and proposed a 3D Semantic extractor based on generative AI model.}             \\ \hline

\cite{WantingAIGC}                   & \multicolumn{1}{c|}{\textcolor{red}{\xmark}}                    & \multicolumn{1}{c|}{\textcolor{green}{\checkmark}}                  & \multicolumn{1}{c|}{\textcolor{green}{\checkmark}}            &    \multicolumn{1}{c|}{\textcolor{green}{\checkmark}}                 &        \parbox[t]{5cm}{Tackle the challenging problem of multi-user scenario with the multi-modal LLM serves as a shared KB.}             \\ \hline

\cite{EleonoraAIGC2}                   & \multicolumn{1}{c|}{\textcolor{green}{\checkmark}}                    & \multicolumn{1}{c|}{\textcolor{red}{\xmark}}                  & \multicolumn{1}{c|}{\textcolor{red}{\xmark}}            &    \multicolumn{1}{c|}{\textcolor{red}{\xmark}}                 &        \parbox[t]{5cm}{A generative model is set up at the receiver to produce material that is semantically coherent with the message that was sent.}             \\ \hline

\cite{ZhenyiAIGC}                   & \multicolumn{1}{c|}{\textcolor{green}{\checkmark}}                    & \multicolumn{1}{c|}{\textcolor{red}{\xmark}}                  & \multicolumn{1}{c|}{\textcolor{red}{\xmark}}            &    \multicolumn{1}{c|}{\textcolor{red}{\xmark}}                 &        \parbox[t]{5cm}{It is the first study to use LLMs for physical layer coding and decoding without the need for extra fine-tuning or retraining.}             \\ \hline

\cite{JianhuaAIGC}                   & \multicolumn{1}{c|}{\textcolor{red}{\xmark}}                    & \multicolumn{1}{c|}{\textcolor{green}{\checkmark}}                  & \multicolumn{1}{c|}{\textcolor{red}{\xmark}}            &    \multicolumn{1}{c|}{\textcolor{red}{\xmark}}                 &        \parbox[t]{5cm}{A lightweight latent space transformation is proposed to resolve the generalization problem and obtain a real-time denoising system with consistency distillation.}             \\ \hline

\cite{WangAIGC}                   & \multicolumn{1}{c|}{\textcolor{red}{\xmark}}                    & \multicolumn{1}{c|}{\textcolor{green}{\checkmark}}                  & \multicolumn{1}{c|}{\textcolor{red}{\xmark}}            &    \multicolumn{1}{c|}{\textcolor{green}{\checkmark}}                 &        \parbox[t]{5cm}{Users transmitter the extracted semantic information from its viewport in virtual reality to the nearby user to determine the need for transmission, and then a contract AI-generated incentive mechanism is proposed.}             \\ \hline

\end{tabular}
\end{table*}
Up to this point, semantic communication has been deployed in various highly complex applications, while it is considered to apply in the Internet of Things service in \cite{ZhengAIGC}. In detail, their proposal focuses on the automotive market analysis, instead of transmitting the original image from the publicly accessible server (PAS), the user can acquire the textual representation from the PAS, which can significantly reduce the data transmitted and undisclosed the driving route of the PAS thus protecting their privacy on a certain degree. However, in the work, they considered the potential adversarial attacks that could disrupt semantic information extraction and then proposed a defense mechanism based on Generative Diffusion Models, which proactively eliminate the adversarial perturbations embedded within the image and thus obtain accurate textual information for semantic transmission. Finally, a resource allocation strategy to balance the energy use for transmission and defense mechanisms is also tackled by the generative model.
\vspace{-0.1in}
\subsection{Cross Direction Works}
With the powerful and enormous ability of the AIGC model, they have been adopted widely. Specifically, in \cite{FangzhouAIGC}, a novel generative artificial intelligence is proposed to enhance the reasoning ability of two communication participants - which is the main target of the Theory of Mind direction. The authors first determine the mutual information of the background knowledge and then enhance this mutual knowledge by its proposed adversarial training-assisted sample-generating policy, which creates sufficient samples for training the semantic communication transceivers. In the paper, they considered ChatGPT as the generative model to provide contextualized samples and, therefore, considered text modality for the semantic communication system. On the other hand, the authors in \cite{TongAIGC} highlight the denoising capability of the AIGC model, which is based on the Diffusion model. The Diffusion model operates by iteratively removing noise from noisy data over multiple steps to recover the original data, making it particularly well-suited for noise removal applications. In their model, instead of taking the reconstructed semantic features as the input to generate the close resembled image, it takes the original received signal, removes the noise from the channel environment along with the fading effect of the obstacles, and thus tries to obtain the transmitted signal at the receiver site. The successful denoising process can facilitate the joint source-channel decoder (Direction III) to improve its performance by a large margin. In the simulation results, they have shown that by utilizing Bayesian posterior probability with an appropriate sampling algorithm, their method successfully mitigates the channel noise.

Without a doubt, the adaptation of AI-generated content model for semantic communication is gaining more and more popularity in the research community, besides the mentioned works above, we provide some extra works along with a short description related to the special aspect of their works. We find that in \cite{HongyangAIGC3}, a well-structured and detailed survey about the role of the generative diffusion model in network optimization has been conducted by the authors, which is very interesting work and helpful in providing other viewing angles for AIGC model. Their application is not limited to the semantic communication direction but is feasible for various network scenarios such as incentive mechanism design, integrated sensing and communication, and the Internet of Everything networks. 
\vspace{-0.1in}
\subsection{Challenges of the Direction}

The impressive capabilities of generated AI models have demonstrated an unimagined performance for the next generation of communication and semantic communication. It is undeniable that there are advantages if we can employ the generative AI based semantic communication in the world. However, the deployment of a generative AI model for each communication user still has a long way to go, specifically due to the computing capacity of the AI generative model and also the energy-constrained on the device. Secondly, the content generated by AI models lacks interoperability, making it difficult for users to determine whether the transmission accurately conveys the intended meaning of the original message or merely represents a randomly generated result by the AI model. Lastly, generative AI is vulnerable to malicious attacks, which can mislead receivers with fabricated information, thereby increasing the risk of fraud and scams.
\vspace{-0.1in}
\subsection{Summary and Lesson Learned}

Throughout the section, we have demonstrated the enormous ability of generative AI to convey the semantic meaning of raw data, which enables the semantic communication paradigm. Additionally, the generative AI model has been proposed not only to encode the data but also to provide network management in semantic communication frameworks. The summary and key lessons are provided as follows:

\begin{itemize}
    \item Controllability: Even with the reduction in data transmission of semantic communication, the network remains difficult to catch up with the massive increase in the number of connected devices. The generative AI model is a promising solution to manage the network with its intelligence and can actively change the strategy according to the dynamic of the network.
    \item Creation: The generative AI-based semantic communication excels in generating high-quality content that adapts to user requirements while optimizing communication costs, which makes it a perfect fit for the application.
    \item Interpretation: With the ability to encode the raw data into a different modality depending on the requirement, the generative AI can massively drop the size of transmitting data but still ensure the high performance of tasks.
    \item Denoising and Reasoning Abilities: By simulating channel noise properties, it demonstrates exceptional performance in removing noise from received signals and offers explanations and reasoning for events, leveraging a deep knowledge of generative AI models.
\end{itemize}
\vspace{-0.1in}
\section{Direction III: Deep Joint Source-Channel Coding based Semantic Communication}
\label{DirectionIII}
In this section, we delve into the integration of semantic communication principles with deep joint source-channel coding (D-JSCC). This fusion aims to enhance communication efficiency by transmitting the essential semantic content of messages, rather than exact bit-level representations. We begin by discussing the traditional approach of separate source and channel coding, highlighting the limitations posed by the cliff effect. We then explore the concept of joint source-channel coding (JSCC), followed by the advancements brought about by deep learning techniques in D-JSCC. Finally, we review recent developments and applications that showcase the potential of semantic communication to transform communication system design.
\vspace{-0.1in}
\subsection{Separate Source Channel Coding}
Traditional communication systems follow the principals of the Shannon separation theorem, which advocates for separate source and channel coding stages \cite{Shannon}. Typically, source coding is performed at the application layer and aims to minimize the amount of data by eliminating redundant information from the original source. For instance, text compression methods generally focus on reducing redundancy in symbol representation using techniques like Huffman coding, arithmetic coding, or dictionary-based approaches such as Lempel-Ziv-Welch (LZW)\cite{salomon2007data}. For image data, traditional source coding methods typically employ three primary steps: transform, quantization, and entropy coding. Transform techniques such as Discrete Cosine Transform (DCT) and Discrete Wavelet Transform (DWT) form the basis of widely used standards like JPEG \cite{wallace1992jpeg}, JPEG2000 \cite{taubman2001jpeg2000}, and BPG \cite{bellard2014bpg}. 

For audio data, traditional source coding methods aim to exploit both temporal and spectral redundancies in the audio signal. Typically, these methods involve three major steps. First, the time-domain signal is transformed into the frequency domain using techniques such as the Fast Fourier Transform (FFT) or Modified Discrete Cosine Transform (MDCT) \cite{johnston1988transform}. Next, psychoacoustic models are applied to identify and discard components that are imperceptible to the human ear, thereby reducing data size without affecting perceived audio quality. Finally, the remaining data is compressed using entropy coding methods like Huffman coding. Standards such as MP3 \cite{pan1995tutorial} and Advanced Audio Coding (AAC) \cite{bradenburg1999aac} exemplify these principles and are widely used in digital audio applications.

In the case of video data, traditional compression techniques extend the principles of image coding by incorporating temporal redundancy reduction. This is achieved through inter-frame and intra-frame compression. Intra-frame compression applies image compression techniques such as DCT or DWT within individual video frames, while inter-frame compression reduces redundancy between successive frames using motion estimation and compensation techniques. The compressed data is further refined using entropy coding, such as Context-Adaptive Binary Arithmetic Coding (CABAC). These techniques are the foundation of widely used video coding standards like H.264/AVC \cite{wiegand2003h264}, H.265/HEVC \cite{sullivan2012h265}, and AV1 \cite{bankoski2013av1}, which achieve high compression efficiency while preserving video quality for streaming and storage applications. 
\begin{figure*}[t]
\centering
\includegraphics[width=0.90\textwidth]{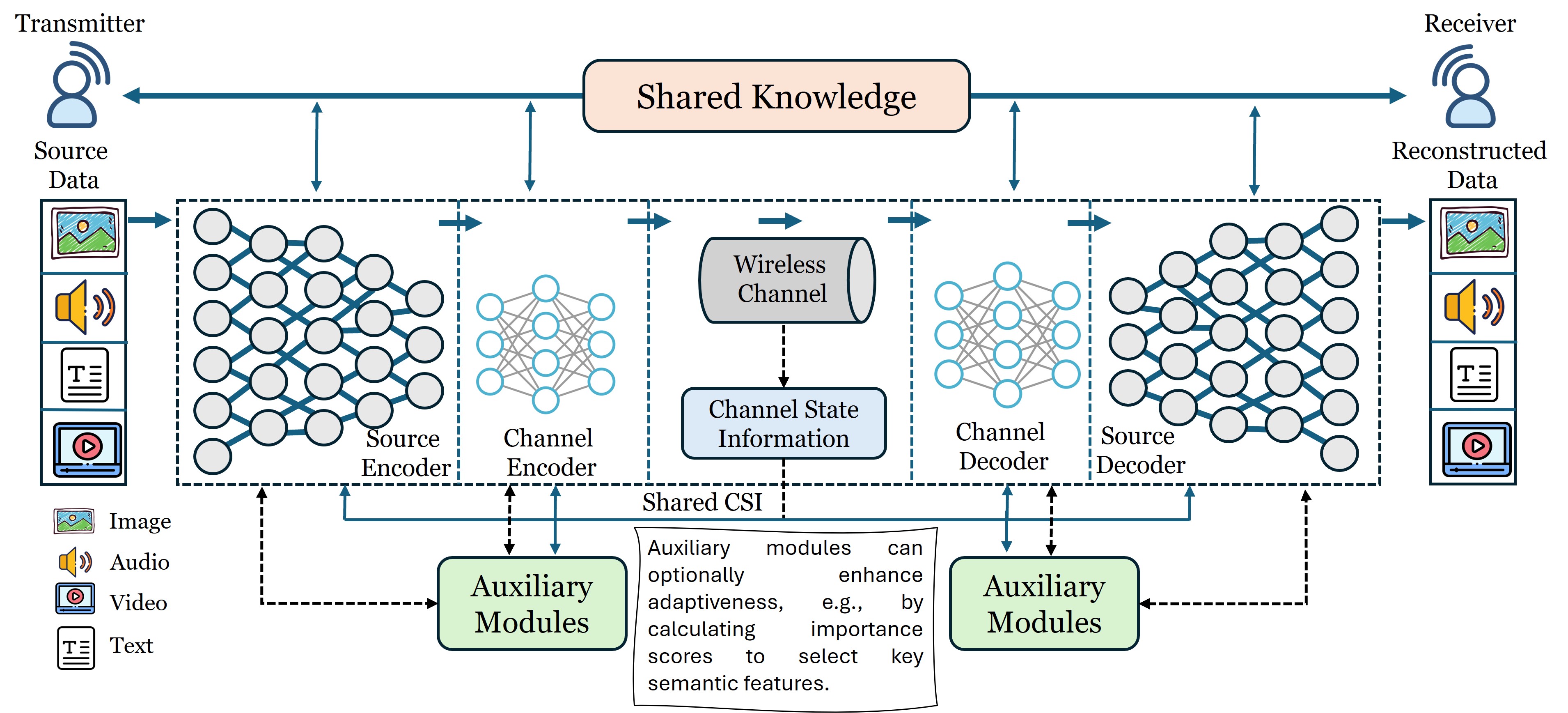}
\caption{General framework of Deep Joint Source-Channel Coding (D-JSCC) for efficient data transmission.}
\label{DJSCC}
\end{figure*}
\subsection{Joint Source Channel Coding}
While separate source and channel coding have been highly effective, the strict separation advocated by Shannon's theorem is optimal only under ideal conditions, such as infinite block lengths and error-free communication \cite{Shannon}. In practical scenarios, particularly under resource-constrained or noisy environments, this separation leads to inefficiencies and susceptibility to the cliff effect, where minor increases in channel noise can cause a sudden collapse in communication quality \cite{cover1991elements}.

Joint source-channel coding (JSCC) emerges as an alternative paradigm that integrates the source and channel coding processes into a unified framework. Instead of separately optimizing for compression and error correction, JSCC jointly encodes the source data for robust communication. This approach inherently accounts for the characteristics of both the source and the channel, enabling better performance under real-world conditions. One significant advantage of JSCC is its ability to provide graceful degradation in communication quality as channel conditions worsen, avoiding the abrupt cliff effect observed in separate coding systems \cite{skoglund2003jscc}. This gradual degradation is particularly beneficial in wireless communication scenarios, where channel conditions can fluctuate dynamically. By jointly optimizing the source and channel representations, JSCC systems achieve higher efficiency in bandwidth- and power-constrained environments. This resource efficiency is critical for applications involving Internet of Things (IoT) devices and low-power wireless networks \cite{goldsmith2005wireless}.

Furthermore, JSCC designs offer adaptability to specific applications, channel conditions, and quality-of-service (QoS) requirements. For instance, image and video transmission systems often employ JSCC to enhance real-time streaming over unreliable networks \cite{modestino1984image}. By integrating source coding redundancy directly into the transmission process, JSCC enables enhanced error resilience, allowing the system to maintain communication quality under challenging conditions.

Despite these advantages, traditional JSCC methods face limitations in scalability and adaptability to the growing complexity and diversity of modern communication systems. Conventional JSCC techniques often struggle to handle high-dimensional multimedia data or rapidly changing channel conditions in heterogeneous environments \cite{gallager1988information}. To overcome these limitations, advancements in machine learning and neural networks have paved the way for deep learning-based JSCC systems. These systems leverage data-driven techniques to optimize source-channel coding jointly, achieving remarkable robustness and efficiency even in complex and dynamic communication scenarios.

\subsection{Deep Joint Source-Channel Coding}
Deep Joint Source-Channel Coding (D-JSCC) leverages neural networks to jointly optimize source and channel coding, offering significant performance improvements over traditional separate coding schemes, especially in resource-constrained and noisy environments. Fig.~\ref{DJSCC} shows a general framework for D-JSCC, illustrating its unified approach to source compression and channel encoding.

One of the pioneering works in this area is the application of deep learning for wireless image transmission. In \cite{Bourtsoulatze2019DeepJSCC}, Bourtsoulatze \textit{et al.} introduced a deep neural network architecture that jointly performs source compression and channel encoding for image transmission over noisy channels. The proposed scheme, termed D-JSCC, employs a convolutional autoencoder that maps input images directly to channel symbols suitable for transmission over a wireless channel, without the need for explicit source or channel codes.

Since the introduction of D-JSCC, this concept has been extended to various domains within the semantic communication paradigm. These include applications in image transmission, audio and speech transmission, vedio streaming, text communication, and multimodal data transmission. Each of these extensions utilizes the ability of D-JSCC to jointly compress and encode source data while capturing essential semantic information, which enhances communication efficiency and robustness across different types of media. The following subsections provide an in-depth exploration of each of these application areas hightlighting how D-JSCC has been adapted to meet the unique challenges and requirements of each domain. 

\subsubsection{Image Transmission}
The concept of D-JSCC has significantly advanced the field of image transmission over wireless channels. In contrast to traditional separate source and channel coding schemes that often suffer from a cliff effect where image quality sharply deteriorates once the channel conditions drop below a certain threshold. D-JSCC provides a graceful degradation of image quality as channel conditions worsen. This robustness is achieved through end-to-end optimization, where the system learns how to distribute redundancy within the encoded representation during training. As a result, D-JSCC-based methods have demonstrated remarkable performance gains, especially under severe channel impairments, consistently outperforming conventional digital communication schemes. Building upon the foundational principles of D-JSCC, researchers have explored deploying joint source-channel coding in increasingly realistic and challenging wireless environments. 

\textit{Channel Modeling Works:} A key aspect of this endeavor is channel modeling, which captures the impact of noise, fading, and interference during signal propagation. By appropriately modeling or adapting to these channel effects, end-to-end deep learning architectures can learn robust encoding and decoding strategies that outperform traditional techniques. Some obstacles need to be paved for adapting D-JSCC to more practical communication setups \cite{9500996,FayçalAitImage,MingyuImage,YehaoImageDeep2,YehaoImageDeep}. 
In \cite{9500996}, they proposed a D-JSCC scheme for wireless image transmission over orthogonal frequency division multiplexing (OFDM) channels, which are prevalent in modern wireless systems. Their approach addresses the challenges posed by frequency-selective fading channels and demonstrates that D-JSCC can adapt to varying channel conditions without explicit channel state information. Similarly, \cite{FayçalAitImage} addresses the challenge of training end-to-end communication systems using neural networks (NNs) without requiring a differentiable channel model.
Traditional methods \cite{9500996, Bourtsoulatze2019DeepJSCC} rely on such models for training, which are often unavailable or inaccurate in real-world scenarios. To overcome this limitation, the authors propose a novel alternating training algorithm. This algorithm iteratively trains the receiver using the true gradient of the loss function and the transmitter using an approximation of the gradient, eliminating the need for channel model knowledge. The proposed method allows for simultaneous optimization of both transmitter and receiver, even with unknown or non-differentiable channel components. 

On the other hand, Mingyu \textit{et al.} \cite{MingyuImage} further enhances D-JSCC for OFDM by incorporating explicit channel estimation and equalization to handle multipath fading. Their study proposes a novel OFDM-guided D-JSCC scheme to overcome the limitations of multipath fading, where signals arriving at the receiver through multiple paths with varying delays and attenuations introduce inter-symbol interference, significantly degrading image quality.
The OFDM mitigates the effects of multipath fading by dividing the signal into orthogonal subcarriers, while the deep learning model, implemented using convolutional neural networks (CNNs), directly maps source images to OFDM symbols for transmission. At the receiver, the system reconstructs the images, effectively handling both compression and error correction within a unified framework. The authors incorporate explicit channel estimation and equalization in the decoder to enhance robustness against diverse channel conditions.

The challenge of unknown CSI without relying on traditional pilot signals was not considered in the previous works, which motivates Ye \textit{et al.} \cite{YehaoImageDeep2} to introduce a Conditional GAN-based end-to-end wireless communication system to address the problem. In this approach, a Conditional GAN is employed to model the wireless channel by learning the conditional distribution of the channel output based on input signals and any received pilot data. This surrogate channel model allows the transmitter and receiver DNNs to be trained jointly in an end-to-end manner, effectively eliminating the need for explicit CSI estimation. The integration of CNNs enhances the system’s ability to handle larger block sizes and mitigate the curse of dimensionality. In \cite{YehaoImageDeep} the authors further enhance the system by introducing a stochastic convolutional layer-based end-to-end wireless communication system. This improved approach models the wireless channel as an untrainable stochastic convolutional layer within the deep neural network architecture, streamlining the system by removing the complexity associated with GANs. By embedding the channel effects directly into the network as a fixed stochastic layer, the transmitter and receiver DNNs can implicitly learn and adapt to channel characteristics more efficiently. This method not only simplifies the architecture but also enhances adaptability and robustness in moderately to highly variable channel environments.

\begin{figure*}[t]
\centering
\includegraphics[width=0.80\textwidth]{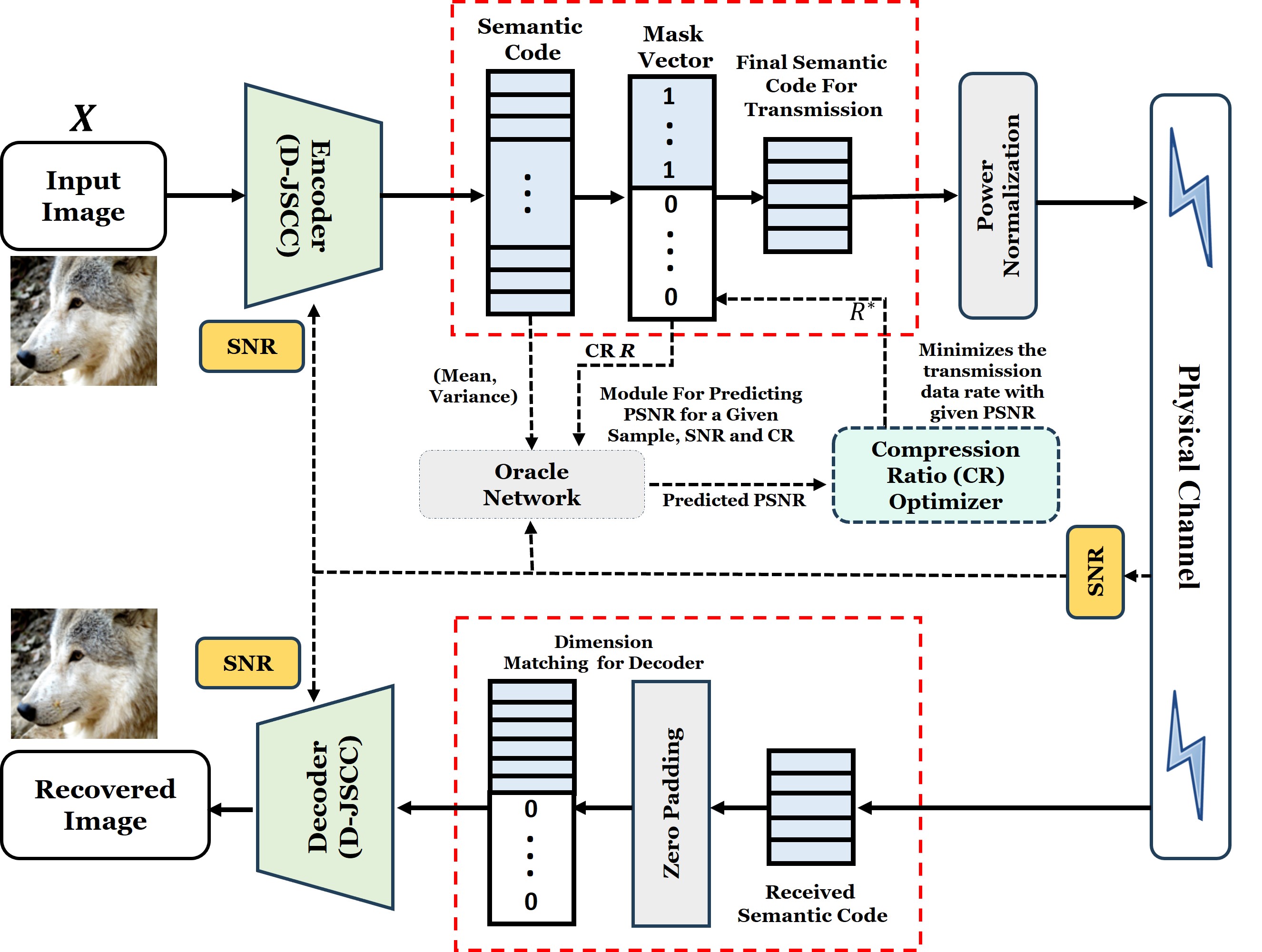}
\caption{The example framework of Deep Joint Source-Channel Coding (D-JSCC) for efficient data transmission.}
\label{ExJSCC}
\end{figure*}
\textit{Related Works on Semantic Noise:} Beyond physical-layer noise and fading, semantic noise presents additional challenges for semantic communication systems. Semantic noise indicates the mismatches between intended and received semantic symbols, which lead to task failures. The authors of \cite{QiyuImageRobust} initially proposed a robust end-to-end semantic communication framework to handle semantic noise by applying a masked autoencoder and adversarial training, which helped the model better handle noisy samples and recover from semantic errors. To further enhance robustness, adversarial training with weight perturbation is employed, which exposes the model to noisy samples during training, improving its ability to handle such samples. The encoded features are represented using a shared discrete codebook between the transmitter and receiver, ensuring efficient and reliable communication. Building on this foundation, the extended version \cite{QiyuImage} of the paper dives deeper into the complexities of semantic noise. It broadens the definition to include both sample-dependent and sample-independent noise, addressing disturbances from both the transmitter and receiver sides. It introduces a Feature Importance Module (FIM) that smartly identifies and suppresses irrelevant or noise-related features. Moreover, the authors refine their masking strategy within the VQ-VAE to better handle areas affected by semantic noise, ensuring that only the most critical information is sent. Finally, they make the system more practical by designing a discrete codebook that works seamlessly with existing digital communication systems.

Another work \cite{SongjieImageRobust} introduces robustness through the Robust Information Bottleneck (RIB) framework, which formulates a trade-off between informativeness and robustness in task-oriented communication systems. The motivation stems from the vulnerability of task-relevant information to distortion caused by channel variations, especially in learning-based JSCC systems. 

The authors derive a tractable variational upper bound for the RIB objective function and overcome the intractability of direct mutual information calculations.
This allows for effective optimization of the informativeness-robustness trade-off.
The proposed Discrete Task-oriented JSCC (DT-JSCC) method leverages this framework by encoding informative messages for inference tasks into discrete representations compatible with digital modulation schemes. This approach provides several advantages over existing solutions such as low latency and high inference performance, compatible with digital modulation.

\textit{Related Works on Compression Rate:} To achieve high-efficiency semantic compression in modern communication systems, researchers have proposed several innovative frameworks that address the challenges of transmitting high-dimensional data efficiently while preserving semantic relevance \cite{JinchengImage,JiaweiImage,DavidImageBandwidth-agile,MingyuImageDeep,ZhangWenyuImagePredictive}. Firstly, in \cite{JinchengImage}, the authors tackle the limitations of traditional D-JSCC methods in handling high-dimensional data and adapting to source distributions, particularly within the context of semantic communications. They introduce Nonlinear Transform Source-Channel Coding (NTSCC), a framework designed to optimize compression and transmission in two stages. First, NTSCC employs a nonlinear analysis transform to map high-dimensional source data into a lower-dimensional latent space. Second, the latent representation undergoes adaptive rate transmission via a deep JSCC encoder, which allocates different coding rates to parts of the latent representation based on an estimated entropy model. This dynamic allocation optimizes bandwidth usage and enhances transmission efficiency. Additionally, the framework incorporates a hyperprior-aided codec refinement mechanism that improves decoding performance at the receiver, further enhancing rate-distortion efficiency.

Building on the need for efficient feature compression in resource-constrained environments, \cite{JiaweiImage} presents BottleNet++ to address the challenge of optimizing feature compression in systems where neural networks are split between mobile devices with limited resources and powerful edge servers. Traditional approaches treat model splitting, feature compression, and communication as separate problems, leading to inefficiencies. BottleNet++ overcomes these limitations by proposing an end-to-end trainable architecture that utilizes joint source-channel coding with lightweight convolutional neural networks as encoders and decoders. This design leverages the fault-tolerant nature of deep neural networks to achieve high compression ratios without significant accuracy loss. The method integrates feature compression with communication theory, achieving bandwidth reductions of up to 64 times and compression ratios of up to 256 bits. In addition, its efficient compression enables earlier network splitting, significantly reducing computational demands on the device. It represents a breakthrough in edge-assisted inference, offering a solution for efficient and adaptive device-edge co-inference.

In modern communication systems, adaptive and progressive transmission has become increasingly important for addressing the dynamic nature of wireless channels and diverse application requirements. 

In the context of semantic communication, adaptive and progressive transmission takes on an even more critical role. 
Progressive transmission allows the system to refine semantic content incrementally, ensuring that the receiver gains meaningful information at each stage, even under degraded channel conditions. Several works illustrate how these principles can be applied to enhance semantic communication systems.

In \cite{DavidImageBandwidth-agile}, \textit{Kurka et al.} introduce DeepJSCC-\textit{l} for adaptive-bandwidth image transmission over wireless channels. This approach enables images to be transmitted in progressive layers, allowing receivers to incrementally improve reconstruction quality as more layers are received. The study addresses two scenarios: successive refinement, where layers are sent sequentially to refine the image, and multiple descriptions, where layers are independent and can be received in any order. By employing convolutional autoencoders, DeepJSCC-\textit{l} effectively learns to transmit images progressively with minimal loss in end-to-end performance compared to single transmission. It also demonstrates comparable performance to SOTA digital progressive transmission schemes, particularly in low signal-to-noise ratio and limited bandwidth conditions, while offering the advantage of graceful degradation with channel SNR.

Approaching the image refinement problem from a different perspective, in \cite{MingyuImageDeep}, the authors solved the issue of needing multiple trained models for multi-rate JSCC by developing a single deep neural network model capable of dynamically adjusting its transmission rate based on both channel conditions (SNR) and image content.

The adaptive rate control significantly improves bandwidth efficiency, particularly in high SNR scenarios or when transmitting images with less information, without sacrificing significant image quality compared to models trained for specific fixed rates. The use of a policy network and the Gumbel-Softmax trick enables end-to-end training and differentiable rate control.

Along with the compression ratio problem, the authors in \cite{ZhangWenyuImagePredictive} resolved another headache problem in semantic communication simultaneously, which is the SNR adaptation. 

To address these issues, the authors propose a Predictive and Adaptive Deep Coding (PADC) framework. PADC comprises three modules: a variable code length enabled D-JSCC model (DeepJSCC-V), an Oracle Network (OraNet) for PSNR prediction, and a CR optimizer. DeepJSCC-V allows for flexible code length adjustment, solving the CR-adaptation problem. OraNet predicts PSNR based on image content, channel SNR, and CR, addressing the transmission quality prediction problem. The CR optimizer uses OraNet's predictions to minimize the CR while meeting a target PSNR, thus providing adaptive and quality-guaranteed rate control. This combined approach improves bandwidth efficiency significantly compared to previous models.

Another major challenge is the incompatibility of existing D-JSCC methods with the limited set of channel inputs allowed by practical hardware. Previous D-JSCC approaches assumed the ability to transmit any complex value, which is unrealistic for commercially available hardware that typically uses predefined constellations like those specified in standards such as IEEE 802.11ad. 

To overcome this, the authors \cite{Tze-YangImageDeepJSCC} propose DeepJSCC-Q, a novel approach that incorporates a finite channel input alphabet. The adopted approach involves a quantization layer that maps the encoder's output to a discrete set of constellation points. They introduce two quantization strategies: one using a fixed constellation and another where the constellation itself is a learned parameter. This allows for optimization of both the channel input geometry and distribution. Furthermore, they utilize Kullback-Leibler (KL) divergence as a regularization term to encourage exploration of the constellation points during training, leading to improved performance.

\textit{Literature on task-oriented semantic communication:} By extracting and transmitting only the essential semantic content required for task execution, task-oriented semantic communication approaches optimize communication resources while ensuring high performance in task-specific applications. Task-oriented semantic communication has shown immense potential for applications like UAV-based sensing, AI-driven IoT networks, and image classification\cite{JiawenImage,KangXuImage,HaijunImageDRL,ChenImageSemantic}. In the work \cite{JiawenImage}, authors proposed a novel framework, Personalized Saliency Fused Semantic Communication (PERSF-SEMCOM), to address the inefficiencies and lack of personalization in existing UAV-based image-sensing communication systems. PERSF-SEMCOM \cite{JiawenImage} extracts semantic information in the form of fixed-size (subject-relation-object) triplets using a triplet detector. These compact triplets capture the essential semantic details of the image and are transmitted to users for matching against their specific queries.

To ensure that critical semantic content is preserved and tailored to each user's interests, PERSF-SEMCOM employs a personalized attention-based mechanism. This mechanism assigns different weights to triplets based on user preferences, prioritizing the transmission of information most relevant to individual users and increase likelihood of successfully transmitting high-priority information. 

In multi-user scenarios, the authors tackle the challenge of resource allocation by introducing a game-theoretic resource allocation scheme. This scheme models the impact of dynamic wireless fading channels on semantic transmission and optimizes power allocation to maximize resource utilization and maintain fairness among users.

Similarly, in \cite{KangXuImage}, the authors propose a task-oriented aerial image transmission paradigm for UAV-to-MEC (Mobile Edge Computing) server communication. This framework is designed to address the limitations of traditional image reconstruction-focused methods by transmitting only semantically relevant image blocks necessary for downstream scene classification tasks. Using deep reinforcement learning (DRL), the framework dynamically selects the most informative image blocks based on content and channel conditions, achieving an optimal balance between transmission latency and classification accuracy. This task-driven approach significantly improves communication efficiency, making it particularly suited for resource-constrained UAV applications.

Previous works focused on the design for the task-oriented semantic communication scenario, while \cite{HaijunImageDRL} the authors considered the resource allocation scheme tailored for Task-Oriented Semantic Communication Networks (TOSCN), specifically applied to image classification tasks. The authors formulate a joint optimization problem that considers factors such as semantic compression ratio, transmit power, and bandwidth allocation. The objective is to maximize the long-term transmission efficiency of tasks. To solve this complex optimization problem, they employ DRL, specifically the Deep Deterministic Policy Gradient (DDPG) method. The DDPG agent is trained to dynamically allocate resources based on the semantic value of the data and the current network conditions. This dynamic allocation achieves an optimal balance between data transmission and task accuracy, ensuring efficient use of communication resources while maintaining high performance in task execution.

The authors of \cite{ChenImageSemantic} introduce a novel semantic communication system tailored for AI-driven IoT applications, addressing inefficiencies in traditional methods that neglect semantic meaning. This framework emphasizes the extraction, transmission, and recovery of essential semantic information, ensuring that only task-relevant data is communicated. By achieving substantial bit rate reductions compared to methods like JPEG and JPEG2000 (10\% and 2\%, respectively), it optimizes the bandwidth necessary for transmitting data critical to task execution. The system incorporates semantic slice-models (SeSM), enabling flexible model adaptations to meet the specific requirements of tasks based on performance needs, channel conditions, and transmission goals. These advancements reduce communication resource demands while improving system efficiency, making it particularly well-suited for resource-constrained environments where efficient task execution is a priority. 

\begin{table*}[t]
\centering
\caption{Advantages and Disadvantages of Deep JSCC for Wireless Image Transmission}
\label{tab:deep_jscc_summary}
\begin{tabular}{|p{1.1cm}|p{1cm}|p{14.5cm}|}
\hline
\centering \textbf{Reference}                            & \centering\textbf{Method} &  \textbf{Advantages and Disadvantages}                                                                                                \\ \hline

\centering \cite{Bourtsoulatze2019DeepJSCC}  & \centering  D-JSCC                                         & \textcolor{green}{\cmark} \: Direct mapping of image pixels to complex channel symbols, bypassing the inefficiencies of separate source and channel coding. \newline
\textcolor{green}{\cmark} \: Graceful degradation under varying channel conditions, avoiding the  ``cliff effect" observed in traditional systems.\newline
\textcolor{green}{\cmark} \: Simplified system design by integrating compression, error correction, and modulation into a single NN framework. \newline
\textcolor{red}{\xmark} \: Performance generalizability to unseen or drastically different channel conditions remains uncertain, with limited exploration of alternative architectures and hyperparameters.
\\ \hline

\centering \cite{9500996}  & \centering  D-JSCC with OFDM                                       & \textcolor{green}{\cmark} \: Direct mapping of source images to complex baseband OFDM symbols, simplifying the system architecture by eliminating separate coding steps.\newline
\textcolor{green}{\cmark} \: Superior performance in low SNR and low-rate scenarios, outperforming  SOTA separate coding schemes in constrained environments.\newline
\textcolor{green}{\cmark} \: Robustness against non-linear signal clipping in OFDM systems, achieved by incorporating clipping as a differentiable layer in the training process. \newline
\textcolor{green}{\cmark} \: Incorporation of domain knowledge through differentiable OFDM layers accelerates training convergence and improves performance over purely data-driven methods.\newline
\textcolor{red}{\xmark} \: Generalizability to unforeseen channel conditions still remains limited, with reliance on specific channel models during training.\newline
\textcolor{red}{\xmark} \: Risk of overfitting, particularly when training with limited data, necessitating careful regularization and architecture selection.
\\ \hline

\centering \cite{FayçalAitImage}  & \centering  Model-Free Training                                       & \textcolor{green}{\cmark} \: Eliminates the need for a differentiable channel model, making it robust to real-world channel uncertainties and complexities.\newline
\textcolor{green}{\cmark} \: Achieves performance comparable to model-aware methods, as shown in simulations on AWGN and Rayleigh block-fading (RBF) channels.\newline
\textcolor{green}{\cmark} \: Alternating training algorithm enables adaptive learning, allowing the system to adjust to actual channel behavior during training.\newline
\textcolor{red}{\xmark} \: Slower convergence compared to model-aware methods, especially in the initial stages of training, due to gradient approximations.
\\ \hline

\centering \cite{MingyuImage} & \centering  OFDM-Guided Deep JSCC                                       & \textcolor{green}{\cmark} \: Provides significant performance improvement with 2.5–4 dB SNR gain over traditional schemes for equivalent image quality in low SNR scenarios.\newline
\textcolor{green}{\cmark} \: Incorporates OFDM blocks to mitigate multipath fading and inter-symbol interference, enhancing robustness and training performance.\newline
\textcolor{green}{\cmark} \: Uses explicit MMSE-based channel estimation and equalization with residual connections to improve image reconstruction quality and computational efficiency.\newline
\textcolor{green}{\cmark} \: Demonstrates robustness to parameter mismatches, unseen channel conditions, and signal clipping for PAPR reduction.\newline
\textcolor{green}{\cmark} \: Integrates adversarial training to enhance perceptual image quality, especially for low SNR scenarios with improved high-frequency details.\newline
\textcolor{red}{\xmark} \: Assumes time synchronization between transmitter and receiver, which may not hold in real-world scenarios, neglecting timing offsets and carrier frequency offset (CFO).\newline
\textcolor{red}{\xmark} \: Architectural complexity with OFDM blocks and additional modules for CSI feedback and PAPR reduction increases training time and resources.
\\ \hline

\centering \cite{QiyuImage} & \centering Masked VQ-VAE &
\textcolor{green}{\cmark} \: Enhanced robustness against semantic noise through the masked VQ-VAE architecture, adversarial training with weight perturbation, and the FIM.\newline
\textcolor{green}{\cmark} \: Reduced transmission overhead by sending only codebook indices, significantly cutting down on transmitted symbols compared to methods like JPEG+LDPC.\newline
\textcolor{green}{\cmark} \: Adaptability to multiple downstream tasks, including classification, image retrieval, and reconstruction, with minimal fine-tuning of the decoder.\newline
\textcolor{green}{\cmark} \: Demonstrates robustness against sample-dependent and sample-independent semantic noise, improving classification accuracy across SNR levels.\newline
\textcolor{red}{\xmark} \: Computational complexity due to the FIM, adversarial training, and iterative optimization increases resource demands, limiting applicability in resource-constrained environments.\newline
\textcolor{red}{\xmark} \: The reconstructed image appearance can differ from the transmitted image due to reconstruction being based on fixed concepts from the codebook. This dependency on the codebook may lead to deviations from the original image, especially in cases of unseen or complex patterns.
\\ \hline
\end{tabular}
\end{table*}

\setcounter{table}{3}  
\begin{table*}[t]
\centering
\caption{(Cont.) Advantages and Disadvantages of Deep JSCC for Wireless Image Transmission}
\label{AAA}
\begin{tabular}{|p{1.1cm}|p{1.2cm}|p{14.5cm}|}
\hline
\centering \textbf{Reference}                            & \centering\textbf{Method} &  \textbf{Advantages and Disadvantages}                                                                                                \\ \hline

\centering \cite{DavidImageBandwidth-agile} & \centering  DeepJSCC-\textit{l} & 
\textcolor{green}{\cmark} \: Bandwidth agility and adaptability via progressive layered transmission, allowing image reconstruction at different resolutions depending on available bandwidth.\newline
\textcolor{green}{\cmark} \: Robustness and graceful degradation under noisy channel conditions, ensuring gradual rather than abrupt quality loss compared to traditional digital transmission schemes.\newline
\textcolor{green}{\cmark} \: Successive refinement and multiple description capabilities, enabling robust reconstruction from any subset of received layers, useful in unreliable or fragmented channel conditions.\newline
\textcolor{red}{\xmark} \: Large model size and memory requirements, particularly for architectures with multiple decoders, which may limit deployment in resource-constrained devices.\newline
\textcolor{red}{\xmark} \: Multi-objective trade-offs between successive refinement and multiple description problems, requiring further research on optimal weighting strategies for balancing image quality across layers.
\\ \hline

\centering \cite{Tze-YangImageDeepJSCC}  & \centering  DeepJSCC-Q &
\textcolor{green}{\cmark} \: Practical applicability due to constellation constraints, ensuring compatibility with real-world hardware and standardized modulation schemes (e.g., OQPSK, QAM).\newline
\textcolor{green}{\cmark} \: Trainable constellations allow learning optimized modulation symbols, adapting to specific channel conditions and improving transmission efficiency over standard modulation formats.\newline
\textcolor{red}{\xmark} \: Limited understanding of learned constellations and their optimization, requiring further research into why learned constellations outperform standard QAM constellations.
\\ \hline

\centering \cite{ZhangWenyuImagePredictive} & \centering PADC &
\textcolor{green}{\cmark} \: Adaptive and flexible code rate optimization using DeepJSCC-V, allowing real-time adjustment based on SNR and image content.\newline
\textcolor{green}{\cmark} \: Data-level and instance-level optimization strategies for bandwidth-efficient transmission while maintaining target PSNR.\newline
\textcolor{green}{\cmark} \: Simple and efficient implementation using code mask operations for variable code length transmission without requiring additional training techniques.\newline
\textcolor{red}{\xmark} \: Suboptimal performance in high SNR regimes, where traditional separation-based schemes may still provide better results.\newline
\textcolor{red}{\xmark} \: Computational complexity in training DeepJSCC-V, requiring more training epochs compared to ADJSCC.
\\ \hline

\end{tabular}
\end{table*}

\textit{Literature Reviews on Performance Improvement:} Enhancing perceptual quality and leveraging advanced architectures are important for improving the effectiveness of semantic communication systems, especially in scenarios demanding high-quality image reconstruction and efficient resource utilization. 
In \cite{EcenazImage}, the authors propose two novel JSCC schemes, InverseJSCC and GenerativeJSCC, aimed at improving the perceptual quality of reconstructed images during wireless transmission, particularly under low bandwidth and low SNR conditions. Existing D-JSCC methods primarily minimize distortion metrics like MSE, often neglecting perceptual quality as experienced by humans. To address this, the authors utilize the perceptual capabilities of Deep Generative Models (DGMs). InverseJSCC employs a pre-trained StyleGAN to denoise and enhance the output of a traditional D-JSCC model, improving perceptual quality without affecting distortion performance. GenerativeJSCC, on the other hand, is a supervised end-to-end approach that integrates a StyleGAN-based decoder with a jointly trained encoder, achieving superior results in both distortion and perceptual metrics like LPIPS.

Building on the use of advanced architectures, \cite{YangkeImageWITT} introduces WITT, a novel wireless image transmission scheme that leverages Swin Transformers to enhance the efficiency and performance of semantic image transmission, particularly for high-resolution images. Swin Transformers effectively extracts long-range information through the hierarchical representation structure. Additionally, a spatial modulation module dynamically adapts the transmission process to varying channel conditions, optimizing performance across different scenarios. It achieves significant improvements in wireless image transmission efficiency, underscoring the role of advanced models in driving next-generation semantic communication systems.

In \cite{LokumarambageImageWireless}, the authors aimed to develop a semantic communication system for efficient image transmission over bandwidth-limited wireless channels while maintaining image quality. Their primary solution was a novel approach that transmits a semantic segmentation map of the image instead of the full-resolution image itself. This semantic map represents the object classes and their locations within the image and is used by a pre-trained GAN at the receiver to reconstruct the original image. The authors in \cite{DanlanImage} proposed an innovative approach that first divides image into different feature maps using a semantic encoder, which represents various objects such as buildings, cars, etc. Then, these feature maps are independently quantized using adaptive semantic coding based on reinforcement learning before being transmitted to the receiver. The received feature maps are input into a GAN-based semantic decoder to effectively fuse all local and global features into a single reconstructed image.

The problem of multi-user semantic communication in wireless networks is increasingly important due to the expanding number of connected devices and the rising demand for efficient communication technologies. The authors of \cite{WeizhiImageNon} noticed the missing works on multi-user scenarios, which motivated them to develop a novel solution, NOMASC (Non-Orthogonal Multiple Access Enhanced Multi-User Semantic Communication), designed to improve spectral efficiency and transmission rates in multi-user settings using NOMA techniques. NOMASC addresses the challenge by incorporating an asymmetric quantizer and a neural network-based model for modulation and demodulation. The asymmetric quantizer discretizes the continuous semantic features, reducing the hardware demands. The neural network handles intelligent multi-user detection (MUD) and adapts to diverse data types and modalities. This integrated approach allows for more efficient and robust transmission, particularly under low-to-medium SNR conditions. 

The aforementioned works have considered multi-user semantic communication scenarios but do not consider the difference in computing capacity, which is a significant drawback. In the study \cite{LocSemantic0}, authors have been the first to consider different computing capacities of multi-users, which is reflected in their equipped DL models. The raised scenario is a challenging problem for the optimization of semantic communication. In the work, they proposed the conditional encoder to solve the problem, by acknowledging the received user, the joint source-channel encoder can effectively output the semantic features for the particular user. In addition, the work also considered the wireless channel and network conditions in the transmission process to provide a dynamic semantic communication system that can change the transmission length according to the channel condition or network traffic. Authors in \cite{LocSemantic1} proposed a different approach to solve the problem, where it considered a collaboration among users in the training process. Moreover, the joint source-channel encoder is only trained one time by a trustworthy decoder to prevent the catastrophic forgetting properties of deep neural networks when being trained to serve a mass amount of users. In Table \ref{tab:deep_jscc_summary}, we provide a comprehensive summary of major advancements in D-JSCC for image transmission.

\subsubsection{Text Communication}
Effective and reliable text transmission is at the heart of countless modern applications, from real-time messaging and voice assistants to mission-critical communications in healthcare and emergency services. Ensuring the accurate and efficient delivery of text data has always been an important area of communication systems.

Semantic communication systems offer a transformative solution by prioritizing the transmission of meaningful content. In the context of text communication, these systems must overcome unique challenges such as variable sentence structures, limited bandwidth, and multi-user environments.

The authors of \cite{HuiqiangText} introduce DeepSC, a novel semantic communication system designed for text transmission. DeepSC shifts the focus from bit-level to semantic-level accuracy by leveraging the Transformer architecture to extract and encode meaningful information from text. This approach enables joint semantic-channel coding, which optimizes both the extraction of semantic content and its reliable transmission over noisy channels, thereby maximizing system capacity and reducing semantic errors. A significant innovation of DeepSC is the incorporation of cross-entropy and mutual information loss functions in the receiver's optimization process, ensuring accurate meaning recovery alongside high data transmission rates. Additionally, the authors introduce sentence similarity as a new performance metric to better evaluate semantic accuracy compared to traditional metrics like the BLEU score.

\textit{Channel Modeling Works:} Similarly with image transmission, it is a primary challenge is the accurate transmission of data over fading wireless channels, which can corrupt signals and hinder model training. To overcome this, the authors of \cite{HuiqiangText2} introduce a CSI-aided training process that leverages CSI to mitigate the effects of fading during model training, thereby enhancing the robustness and accuracy of data transmission. Additionally, a refined least squares (LS) estimator with reduced pilot overhead is developed to further improve CSI estimation. Another significant challenge is the large size and computational demands of deep learning models, which makes them unsuitable for deployment on devices with limited processing power and memory, such as IoT devices, which can quickly drain their batteries and limit their operational lifespan. The authors address this by implementing model compression techniques, specifically weight pruning, and quantization, to reduce both the model size and computational complexity. Furthermore, they design a finite-bits constellation to lower the hardware costs in data transmission.

\textit{Network Management for Text Semantic Communication:} The study in \cite{LeiTextresource} addresses the problem in resource allocation for text semantic communications by introducing a novel metric, semantic spectral efficiency (S-SE), which measures communication efficiency from a semantic perspective. This metric considers both channel assignment and the number of transmitted semantic symbols to optimize resource allocation. The researchers formulated an optimization problem aimed at maximizing the overall S-SE and employed a combination of techniques, including decoupling the problem into smaller, independent parts and utilizing methods like exhaustive search and the Hungarian algorithm, to obtain an optimal solution. Their simulations validated the effectiveness of the proposed resource allocation model, demonstrating the superiority of semantic communications in terms of S-SE compared to traditional wireless communication, on average 19 bits fewer.

On the other hand, \cite{XidongText} addresses the challenge of designing efficient multiple access (MA) schemes for future networks that must integrate semantic and conventional bit-based communication. To tackle this, the authors introduce a novel Semi-NOMA scheme that balances spectrum efficiency and interference management by splitting the bit stream into shared and orthogonal sub-bands, outperforming traditional OMA and NOMA schemes. The study uses a regression-based approach with a generalized logistic function to approximate semantic similarity, enabling the analytical characterization of system performance. Metrics like the Semantic-versus-Bit (SvB) Rate Region and Power Region are employed to evaluate these schemes, with results showing that Semi-NOMA consistently achieves superior performance across various conditions, especially when the semantic-interested user (S-user) has better channel conditions than the bit-interested user (B-user).
The authors analyze the impact of key parameters like semantic similarity, target rates, and semantic encoding schemes, demonstrating the robustness and adaptability of Semi-NOMA.

The above works only considered fixed codeword lengths, which are inflexible and inefficient for varying sentence lengths. This motivates the authors in \cite{PeiwenText} to consider a more efficient framework for text transmission. The authors solved this problem by proposing a novel architecture, SC-RS-HARQ, which combines semantic coding (SC) with Reed-Solomon (RS) channel coding and Hybrid Automatic Repeat Request (HARQ). This approach allows for variable-length coding, dynamically adjusting code lengths based on sentence length and channel conditions. Furthermore, they developed an end-to-end architecture called SCHARQ to further enhance performance, significantly reducing bit consumption and sentence error rate. A key innovation is the introduction of a new error detection method, Sim32, which allows the receiver to accept semantically similar sentences even if they contain errors, thus conserving transmission resources while maintaining semantic integrity.

\subsubsection{Audio and Speech Transmission}
The importance of audio and speech data transmission has grown exponentially with the advent of advanced technologies such as online gaming, virtual reality (VR), and video conferencing. The volume of audio and speech data transmitted globally has increased dramatically. As these technologies evolve, the demand for efficient, robust, and scalable audio transmission systems continues to grow.

Several works have laid the foundation for D-JSCC by showcasing its ability to outperform traditional systems under low SNR scenarios for speech modality. The framework DeepSC-S in  \cite{ZhenziSpeech0, ZhenziSpeech1} tackles some of the most persistent shortcomings of traditional speech transmission—specifically, the unnecessary emphasis on bit accuracy and the system’s inability to thrive in dynamic environments. DeepSC-S integrates semantic and channel coding through a deep learning-based architecture, enriched with a squeeze-and-excitation (SE) network for attention. By focusing on the essential features of the audio and ignoring irrelevant details, it avoids overburdening the channel. Notably, it achieves consistently robust performance across varying SNR conditions without the need for retraining, as validated through metrics like Signal-to-Distortion Ratio (SDR) and Perceptual Evaluation of Speech Quality (PESQ).

Another framework called SemAudio is proposed in \cite{WeiSpeech}, which identifies three key challenges in real-time audio transmission: extracting essential semantic information from streaming audio signals, achieving reliable recovery performance, and jointly designing streaming semantic and channel coding to address channel distortion and attenuation. To address the first challenge, SemAudio, a deep learning-based system, leverages the Transformer-XL architecture to capture long-distance dependencies in audio data and uses a chunk-based masking attention strategy for real-time processing. The second challenge of reliable recovery is tackled by jointly designing the semantic and channel encoder/decoder, minimizing errors in both the time and frequency domains. Furthermore, the joint design of the semantic and channel coding components also addresses the third challenge of mitigating channel distortion and attenuation. The combination of techniques allows SemAudio to outperform traditional communication systems in terms of recovery accuracy and latency.

\begin{table*}[t]
\centering
\caption{Summary of Major Works in Audio and Speech Semantic Communication}
\label{tab:audio_speech_summary}
\begin{tabular}{|p{1.1cm}|p{1.2cm}|p{14.5cm}|}
\hline
\centering \textbf{Reference}                            & \centering\textbf{Method} &  \textbf{Advantages and Disadvantages}                                                                                                \\ \hline

\centering \cite{ZhenziSpeech0} & \centering  DeepSC-S & 
\textcolor{green}{\cmark} \: Focuses on semantic-level information recovery rather than bitstream reconstruction, enabling efficient bandwidth usage and task-specific transmission.\newline
\textcolor{green}{\cmark} \: Integrates SE-ResNet attention mechanism to enhance critical speech feature recovery, ensuring robustness under low SNR conditions.\newline
\textcolor{green}{\cmark} \: Employs joint source-channel coding for end-to-end optimization, simplifying the communication pipeline and improving efficiency.\newline
\textcolor{red}{\xmark} \: Requires high computational resources for the attention mechanism, limiting deployment on low-power devices. \newline
\textcolor{red}{\xmark} \: Performance heavily depends on large, high-quality training datasets, making real-world generalization challenging.
\\ \hline

\centering \cite{WeiSpeech} & \centering  SemAudio & 
\textcolor{green}{\cmark} \: Designed for real-time streaming, addressing latency issues through a chunk-based mask attention strategy within Transformer-XL architecture.\newline
\textcolor{green}{\cmark} \: Capable of processing streaming signals with significantly lower latency compared to traditional methods.\newline
\textcolor{green}{\cmark} \: Adaptable mask design enables flexibility for various practical scenarios.\newline
\textcolor{green}{\cmark} \: Integrates a combined time and frequency domain loss function to leverage phonetic information and improve noise suppression.\newline
\textcolor{red}{\xmark} \: Assumes perfect CSI, which is rarely achievable, and robustness under imperfect CSI conditions remains unexplored.\newline
\textcolor{red}{\xmark} \: Experiments limited to the LibriSpeech corpus, raising concerns about generalizability to other datasets and real-world audio types.
\\ \hline

\centering \cite{TongSpeech} & \centering  FL Based D-JSCC & 
\textcolor{green}{\cmark} \: Leverages wav2vec-based autoencoder architecture for efficient semantic information extraction, reducing bandwidth consumption and improving spectrum efficiency.\newline
\textcolor{green}{\cmark} \: Enhances privacy using FL, enabling collaborative training without sharing local data.\newline
\textcolor{green}{\cmark} \: Exhibits adaptability and scalability through FL, supporting growing numbers of devices and diverse network conditions without major architectural changes.\newline
\textcolor{red}{\xmark} \: High computational complexity due to the wav2vec autoencoder and FL training dynamics, posing challenges for resource-constrained edge devices.\newline
\textcolor{red}{\xmark} \: Communication overhead during FL training can be significant, particularly in large-scale deployments with limited network bandwidth.
\\ \hline

\centering \cite{ChenSpeech} & \centering  Perceptual D-JSCC & 
\textcolor{green}{\cmark} \: Achieves a 60\% reduction in symbols required for transmission, enhancing transmission efficiency.\newline
\textcolor{green}{\cmark} \: Optimizes human auditory perception using a multiresolution joint loss function, improving perceived speech quality.\newline
\textcolor{green}{\cmark} \: Reduces model complexity and processing time compared to traditional methods, making it practical for resource-constrained applications.\newline
\textcolor{red}{\xmark} \: Subjective assessments (e.g., MUSHRA tests) have variability due to a small participant pool.\newline
\textcolor{red}{\xmark} \: The multiresolution loss function introduces additional computational complexity and is highly sensitive to the tuning of multiple hyperparameters, impacting efficiency and robustness.
\\ \hline

\centering \cite{ZhouSpeech} & \centering DeepSC-TS & 
\textcolor{green}{\cmark} \: Achieves significant data transmission efficiency, transmitting smaller data sizes compared to DeepSC-S without requiring extensive pre-processing.\newline
\textcolor{green}{\cmark} \: Utilizes Swin Transformer for enhanced feature extraction, leveraging W-MSA and SW-MSA mechanisms for better semantic integration and adaptability.\newline
\textcolor{green}{\cmark} \: Exhibits superior robustness in low SNR environments, outperforming traditional systems in AWGN, frequency deviation, and Rayleigh fading channels.\newline
\textcolor{red}{\xmark} \: Limited adaptability in extremely low-bandwidth environments and varying speech durations.\newline
\textcolor{red}{\xmark} \: Performance degrades significantly in extremely low SNR conditions, particularly in Rayleigh fading channels.
\\ \hline

\centering \cite{XiaoSpeech} & \centering DSST & 
\textcolor{green}{\cmark} \: Achieves up to 75\% bandwidth savings compared to traditional methods like AMR-WB + 5G LDPC and Opus + 5G LDPC, while maintaining or improving speech quality.\newline
\textcolor{green}{\cmark} \: Demonstrates robustness across various channel conditions (e.g., AWGN, COST2100 fading) and performs well even in low SNR environments.\newline
\textcolor{green}{\cmark} \: Employs end-to-end optimization with flexible rate-distortion trade-offs, allowing adaptation to different bandwidth constraints and quality requirements.\newline
\textcolor{red}{\xmark} \: High computational complexity poses challenges for real-time implementation on resource-limited devices.\newline
\textcolor{red}{\xmark} \: Sensitive to hyperparameter tuning (e.g., Lagrange multiplier, entropy model settings), requiring extensive experimentation for optimization.
\\ \hline

\end{tabular}
\end{table*}

Instead of considering a reconstruction task like previous works, a semantic-oriented speech transmission is developed for a speech recognition task. In the paper, they implemented several key modules: a soft alignment module using an attention mechanism and LSTM layers to identify and prioritize semantically important parts of the speech spectrum; a redundancy removal module to discard irrelevant data; and a semantic corrector leveraging a pre-trained language model to refine transcription accuracy. Furthermore, they employed subwords as tokens instead of full words to enhance semantic preservation and handle unseen words. A two-stage training scheme accelerated the model's training process, and for speech-to-speech transmission, an additional module extracted compact, supplementary information (duration, pitch, power) to improve reconstruction quality.

An interesting scenario is proposed in \cite{TongSpeech}, where they adopted a distributed training framework for audio semantic communication using the federated learning technique, which addressed the challenge of transmitting large audio data with limited bandwidth. Then, the authors proposed a novel system employing a wav2vec-based autoencoder, leveraging CNN to extract and transmit only the semantic information of audio signals. The approach significantly reduces data transmission volume by not gathering the local data at one server.  Additionally, the distributed training framework prevents the privacy thread. This method not only protects user privacy but also allows the model to learn from diverse audio characteristics.

Previous works mainly considered the audio reconstruction quality as the metric to evaluate the system performance and ignored the perceptuality. 
\cite{ChenSpeech} proposed a perceptually motivated, low-complexity approach to enhance speech semantic communication. This approach significantly improves both the transmission efficiency and the quality of received speech by incorporating a low-complexity fully convolutional semantic encoder to extract semantic information efficiently, reducing the amount of transmitted data by 60\%. Especially, a multi-resolution joint loss function optimizes the reconstruction of speech to better align with human auditory perception, enhancing the listening experience. The proposed method paves the way for more efficient and user-friendly communication technologies, particularly in applications like video conferencing where accurate and high-quality speech transmission is crucial. The reduction in model complexity and improved transmission efficiency are key advancements that enhance the feasibility and scalability of semantic communication systems for real-world applications.

The authors in \cite{ZhouSpeech} recognized the high computing complexity of conventional transformer technique, and adopted the Swin Transformer architecture to lower the computing complexity of the DL model. Unlike previous methods, DeepSC-TS, the proposed system, simplifies the semantic encoder design using 2D CNN layers while maintaining the original data transmission load. This approach results in significant reductions in computational and storage complexity, improved perceptual evaluation scores (PESQ) indicating enhanced audio quality, and superior noise resilience, particularly in low SNR conditions, which enable an efficient and robust solution for future communication systems. 

Dynamic adaptation to the physical environment is essential for audio transmission systems to ensure efficient and robust communication across varying channel conditions. Two notable works, \cite{XiaoSpeech} and \cite{EleonoraAIGC}, proposed innovative solutions for this challenge. Deep Speech Semantic Transmission (DSST), a high-efficiency semantic coded transmission method for end-to-end speech transmission over wireless channels is introduced in \cite{XiaoSpeech}.
This system utilizes a nonlinear transform to map speech sources into a semantic latent space, thereby improving the efficiency of speech data representation. A key innovation is the implementation of an entropy model on this latent space, enabling the allocation of coding rates based on the significance of different semantic features, which optimizes bandwidth usage. Furthermore, DSST incorporates a channel signal-to-noise ratio (SNR) adaptation mechanism, allowing a single model to adapt effectively to diverse channel conditions. This holistic approach, combining nonlinear transformation, entropy-guided rate allocation, and SNR adaptation, results in significant bandwidth savings while maintaining high audio quality compared to existing engineered and neural speech transmission methods. 


Previous works only consider channel noise and ignore the incomplete received signal problem, \cite{EleonoraAIGC} propose a novel generative audio semantic communication framework to address this scenario. This framework reformulates the communication problem as an inverse problem, prioritizing the transmission of lower-dimensional representations of the audio signal and its associated semantics instead of the full bitstream. A conditional diffusion model is employed at the receiver to reconstruct the audio content, effectively handling noise and missing data while maintaining semantic consistency. This approach leverages latent representations and semantic information to enhance audio restoration quality under severely degraded channel conditions, outperforming prior state-of-the-art methods. A key innovation is the automatic computation of optimal denoising parameters, which adapts to various channel conditions without requiring prior knowledge, making the system robust and adaptable to real-world communication scenarios. In Table \ref{tab:audio_speech_summary}, we provide a detailed summary of major advancements in audio and speech semantic communication.

\subsubsection{Video Streaming} In the subsection, we provided a comprehensive literature review on semantic communication for video transmission. Authors in \cite{Tze-YangVideo} have been one of the first works to consider video transmission using the joint source-channel coding process in an end-to-end manner. In the paper, they formulated a bandwidth allocation problem among video frames, which considered the difference between the current frame with the previous frame. If the two frames are almost identical to each other, it would be a waste to allocate too many communication resources to it. Oppositely, if the two frames are completely different, the next frame needs to be allocated more resources than normal. To optimize the bandwidth, they have adopted the reinforcement learning solution and achieved outstanding performances compared with the traditional communication scenario.

Developing on \cite{Tze-YangVideo} idea, the authors in \cite{SixianVideo} have developed a high-efficiency deep JSCC for video transmission. Firstly, they utilized the nonlinear transform and the conditional coding architecture to create an adaptive entropy model, which later is used to control channel bandwidth for each region within the frame. This rate-adaptive feature transmission is achieved by adopting the variable-length coding for each embedding vector. Similarly, the information from the previous frame also participated in the encoding process of the current frame as the context information. The improvement between this work and the \cite{Tze-YangVideo} is the consideration of bandwidth allocation within the same frame.

By adopting the advancement of DL, authors in \cite{ZhangVideo} propose a novel architecture for the semantic encoder/decoder. The contributions of the paper are three-fold as follows: \textit{1) Bi-optical flow extractor}, \textit{2) noise attention module}, and \textit{3) feature choice and fusion modules}. They were the first to leverage the bi-optical flow extractor to extract the fine-grained semantic inter-frame information, while noise attention can mitigate channel noise degradation by taking the SNR value into the encoding/decoding process. Finally, the feature choice module drops the features containing low semantic information, while the fusion module concentrates on the critical semantic features for better reconstruction performance.

Different from the aforementioned works, the work in \cite{NiuVideo} considered a multi-dimension noise that consists of channel noise and semantic noise. While the channel noise degrades the performance of the reconstruction video, the semantic visual noise causes semantic misunderstanding. A channel-spatial attention mechanism is considered in the architecture design of the joint-source channel encoder/decoder to protect the signal from channel noise. On the other hand, an additional module named semantic representation and refinement is proposed to particularly deal with the visual semantic noise; the background knowledge contains information from the previous frame, which also participates in the semantic refinement process. The receiver can leverage the knowledge background to correct the semantic information and synthesize the video frame. A perceptual quality metric is proposed to guarantee the visual quality of the reconstruction video. 

In \cite{BaoVideo}, the authors considered the transmission of a group of frames at a time, which is an intriguing idea. By transmitting a group of frames, they extracted a common feature map along with the individual feature maps from those frames. These extracted feature maps have a much lower dimension size than the original frame, which improves the communication efficiency of the proposed semantic system. To further reduce the communication cost, they first transformed these groups of frames into latent space and formed latent representation. With all the feature maps ready, they will be fed into an entropy-based semantic importance, which outputs an importance value associated with each feature map. The maps associated with high values will be encoded by longer code. Last but not least, the proposed system aims to balance reconstruction video quality and transmission length, motivating the design of a loss function that incorporates both aspects. 

Authors in \cite{QiyuanVideo} introduced the Object-Attribute-Relation (OAR) mechanism to achieve an extremely low bit-rate coding for video transmission. In the work, instead of transmitting low-latent representations of each video frame like other works, they only transmit the first frame of each GoP (group of pictures) and the OAR graphs, which achieve a CBR value of 1/300. The OAR graph can characterize the motion and state changes of the objects in the video. At the receiver, they have a generative model to generate new frames using reference frames to provide appearance information of objects. Primary image prediction is performed based on the motion information of OARs through optical flow estimation. In the simulation result, their model outperformed H.265 coding at lower bit-rates and synergized with the JSCC to deliver robust and efficient video transmission.

Sharing the same idea with \cite{QiyuanVideo}, the author approached a different approach, their first proposal was frame interpolation, which only needs to transmit a subset of frames from the receiver, and the receiver will leverage the key frames to interpolate the missing frames. The essential key-frames are extracted by the proposed \textit{Adaptive Key-frame Extractor}, which can be retransmitted multiple times in case of low SNR. The position of these key frames in the video sequence is also recorded and later used to interpolate other middle frames. The key frames are then encoded by the Semantic Vector Quantization Encoder, which outputs the index of the nearest vector in the latent embedding space $E$. By transmitting only the index, the receiver utilizes the shared latent embedding space to retrieve and reconstruct multi-channel features efficiently. This approach can maintain the integrity of the recovered features. In the simulation section, the authors have effectively demonstrated outstanding performance compared with the H.265 standard, especially in multi-path channels. 

In the \cite{GuoVideo}, the author studied a video question-answering task, where the receiver can answer a question related to the video without the need to reconstruct that video. Therefore, the author transmitted the semantic features that were related to the question and can improve the accuracy of the receiver only. Both the encoder and decoder at the transmitter and receiver sites are redesigned and trained so that the receiver can achieve an accurate answer and not waste the communication and computation resources for video reconstruction. These task-related features are represented in graph-based spatial correlation, which lowers the transmission signal length. 
\begin{table*}[t]
\centering
\caption{The overview of contributions of research works related to multi-modal DeepJSCC-based semantic communication.}
\begin{tabular}{|c|c|p{12.6cm}|} 
\hline
\textbf{Reference} & \textbf{Method} & \textbf{Advantages and Disadvantages} \\ \hline
\cite{JingMultimodalSemantic} & 
MSC System &
\begin{tabular}[c]{@{}l@{}} 
\textcolor{green}{\cmark} \: Significant reduction in data transmission volume by leveraging shared and private (S\&P) semantic \\ \:\: representation models, reducing redundancy and achieving efficiency at various compression ratios.\\
\textcolor{green}{\cmark} \: Adaptable to multiple downstream tasks (e.g., classification, retrieval, and reconstruction) with minimal \\ \:\: fine-tuning, broadening applicability. \\
\textcolor{red}{\xmark} \: Sensitive to noise at extremely low bandwidth ratios (k/n $<$ 0.2), where high compression factors result \\ \:\: in reduced detail storage and increased vulnerability to information loss.\\
\textcolor{red}{\xmark} \: Limited modality support, with unclear adaptability to additional modalities like audio, video, or sensor \\ \:\: data, requiring further exploration of semantic representation models for new modalities.\\
\end{tabular} \\ \hline
\cite{TianYunMultimodal} & 
SyncSC &
\begin{tabular}[c]{@{}l@{}} 
\textcolor{green}{\cmark} \: Achieves synchronous multimodal transmission of video and speech by employing timestamps and a \\ \:\:visual-guided speech synthesis module, ensuring temporal and semantic alignment.\\
\textcolor{green}{\cmark} \: Enhanced robustness to packet loss using PacSC for video semantics and TextPC for text, leveraging\\ \:\:  semantic redundancy and pre-trained models to reconstruct lost data.\\
\textcolor{green}{\cmark} \: PacSC, based on masked autoencoders, reconstructs lost video packets more effectively than traditional\\ \:\:  Reed-Solomon codes, especially under high packet loss rates.\\
\textcolor{green}{\cmark} \: Visual-guided speech synthesis integrates facial expression coefficients and text to generate temporally\\ \:\:  and semantically aligned speech, improving naturalness and synchronization in audio-visual outputs.\\
\textcolor{red}{\xmark} \: High computational complexity in modules like video generation and visual-guided speech synthesis limits\\ \:\:real-time performance, requiring further optimization to reduce latency and resource usage.\\
\end{tabular} \\ \hline
\cite{LiYuandiMultimodal} & 
MMTrustSC &
\begin{tabular}[c]{@{}l@{}} 
\textcolor{green}{\cmark} \: Enhanced security and reliability through a two-level coding scheme combining Reed-Solomon codes with\\ \:\: conventional encoders, offering resilience against noise  and data corruption.\\
\textcolor{green}{\cmark} \: Hybrid encryption mechanism combining ECC for secure key exchange and AES for efficient data \\ \:\: encryption, ensuring confidentiality and integrity in hostile network environments.\\
\textcolor{green}{\cmark} \: Effective inter-modal complementarity management using cross-modal attention mechanisms, such as\\ \:\: AGVA, to align audio and visual data for superior performance in noisy environments.\\
\textcolor{green}{\cmark} \: Adaptability to multi-user settings with architecture designed to handle multiple transmitting users and a \\ \:\: multi-antenna receiver, enhancing scalability for real-world applications.\\
\textcolor{red}{\xmark} \: Limited dataset scope, with evaluation restricted to the AVE subset of Audioset, raising questions about \\ \:\: generalizability to other datasets and applications.\\
\end{tabular} \\ \hline
\cite{WangPenghongMultimodal} & 
Distributed SC &
\begin{tabular}[c]{@{}l@{}} 
\textcolor{green}{\cmark} \: Improved performance in AVP tasks, with superior F-scores, Type@AV, and Event@AV metrics compared\\ \:\:to SOTA methods, even under low SNR conditions.\\
\textcolor{green}{\cmark} \: Effective integration of auxiliary feedback, where low-dimensional audio semantics guide compression of \\ \:\:high-dimensional visual data, minimizing redundant transmissions. \\
\textcolor{red}{\xmark} \: Potential performance loss in extremely noisy conditions due to aggressive compression of audio features,  \\ \:\: particularly at low SNR, resulting in reduced accuracy for certain metrics.\\
\textcolor{red}{\xmark} \: Assumption of high SNR (20 dB) for the feedback channel, which might not hold in real-world scenarios,\\ \:\: potentially impacting performance under lower feedback channel SNR conditions. \\
\end{tabular} \\ \hline
\cite{GuoJieMultimodal} & 
DTCN &
\begin{tabular}[c]{@{}l@{}} 
\textcolor{green}{\cmark} \: Employs Deep Joint Source-Channel Relay Coding (Deep-JSCRC) to complement missing information, \\ \:\:mitigate semantic noise, and enhance communication reliability, particularly in noisy environments.\\
\textcolor{green}{\cmark} \: Collaborative device-server optimization dynamically redistributes workloads using edge intelligence\\ \:\: and FL-based parameter sharing, improving resource utilization and privacy protection.\\
\textcolor{green}{\cmark} \: Adaptable to multitasking scenarios, with experimental results showing improved accuracy compared to \\ \:\:baseline methods on the UPMC Food-101 dataset.\\
\textcolor{red}{\xmark} \: Heavy reliance on edge devices and servers, with performance directly tied to their computational\\ \:\: capabilities and connectivity. Limited or unreliable edge infrastructure could impact system effectiveness.\\
\textcolor{red}{\xmark} \: Potential for bottlenecks at semantic relays or edge servers during high traffic or computational demand,\\ \:\:  depending on workload distribution and FL efficiency.\\
\end{tabular} \\ \hline
\cite{ZhangGuangyiMultimodal} & 
U-DeepSC &
\begin{tabular}[c]{@{}l@{}} 
\textcolor{green}{\cmark} \: Reduced model complexity and size by employing a unified framework for multiple tasks, eliminating \\ \:\: the need for separate models, significantly reducing parameter count and computational overhead.\\
\textcolor{green}{\cmark} \: Improved transmission efficiency through a vector-wise dynamic scheme and lightweight Feature \\ \:\: Selection Module that optimizes feature transmission based on task requirements and channel conditions.\\
\textcolor{green}{\cmark} \: Enhanced adaptability and flexibility by integrating task embedding vectors, allowing easy adaptation\\ \:\: to new tasks without requiring extensive retraining.\\
\textcolor{red}{\xmark} \: Two-phase training complexity increases computational cost, with separate encoder-decoder optimization \\ \:\: and fine-tuning phases that add to training time.\\
\textcolor{red}{\xmark} \: Potential performance trade-offs, as a single unified model may not achieve the same level of task-specific\\ \:\: optimization as dedicated models for each task.\\
\end{tabular} \\ \hline

\end{tabular}
\label{Multimodaltable}
\end{table*}
\subsubsection{Multi-modal Data Transmission} The need for multi-modal data has raised significantly in the last couple of years, which drives the requirement for developing a system that is capable of transmitting multi-modal data across users. Acknowledging this trend, numerous investigations have proactively studied and proposed semantic communications for multi-modal data transmission for both task-oriented and data reconstruction approaches \cite{ZhuZengleMultimodal,MinqiMultimodal,RenChaoMultimodal,FuWeiqiMultimodal}.

The scenario proposed in \cite{HuiqiangMultimodalVQA} is one of the earliest studies for multi-modal transmission in semantic communication, which involves one image and one text transmitter, and a multi-modal receiver. In the paper, they considered the transmitted text to be sampled from a set of questions related to the image transmitted by the other user. At the user, each modality is decoded by a separate channel decoder and passed to a memory, attention, and composition NN to capture the correlation among modalities and generate the correct answer. 

Building on the success of the multi-user multi-modal semantic communication of the previous work, the authors in \cite{HuiqiangMultimodal} extend the works to facilitate two more intelligent tasks: image retrieval and machine translation. Moreover, a transformer-based semantic encoder/decoder is leveraged to better capture the main information within the data; then they proposed an information fusion module to fuse the keywords and the corresponding image region to achieve high-accuracy answers for the VQA task. While the machine translation and image retrieval tasks only involve a single modality thus, there is no need to implement the data fusion process.

In \cite{LiAngMultimodal}, authors focused on the answers to two questions: \textit{1) How do we get the true meanings?} and \textit{2) how do we precisely interpret semantics?} to facilitate the high demand for multi-modal services. Different from the two mentioned works above, the authors considered a cross-modal encoder, which can uncover potential relationships and shared semantic information across modalities. A cross-modal knowledge graph is proposed to further improve the cross-modal encoding and decoding process by providing the same or similar knowledge among participants. This knowledge base guarantees quick and thorough information retrieval by storing comprehensible descriptions of the intra- and inter-modal entities and relations. 

Sharing the same intuition related to inherent correlation among the multi-modal data, authors of \cite{WangWenjunMultimodal} presented cross-modal alignment between two modalities at both encoding and decoding process instead of a single module combining the multi-modal at the end \cite{HuiqiangMultimodal}. To be specific, the Shapley value is utilized as the loss function for the alignment. Last but not least, they create a cross-modal amendment network that dynamically modifies the auxiliary semantic information weights across various modalities. Its goal is to improve the overall quality of semantic transmission by fixing semantic mismatches that occur during communication. More diversity in a number of modalities is considered in \cite{ChenMingkaiMultimodal}, where it solves the problem related to three modalities: image, text, and audio, in which each modality is encoded by a specialized semantic extractor. They assumed that identical immediate feature vectors existed between modalities; thus, they took advantage of the Frobenius norm to identify these collapse features and eliminate them from the transmission process to improve communication efficiency. In \cite{ChaoweiMultimodal}, an integration between multi-modal semantic communication with the MEC system has been proposed to accelerate the improve the quality of service and user experience in interactive AR. To be specific, the multi-modal semantic has the amazing ability to eliminate channel noise, which can improve the image/audio quality, while the MEC is responsible for content caching due to its massive memory. By offloading the computing and caching the popular content at the edge server, the users can obtain a good experience. 

The authors in \cite{HeYangshuoMultimodal} saw the incompatibility problem between end-to-end analog semantic communication with modern digital communication, even with the quantization technique, often resulting in sub-optimality. The second problem is the need to retrain or redesign the network to execute a different task. In order to tackle the first problem, they designed a framework that incorporates the multi-modal semantic coder with the conventional channel encoder, which leverages the compatibility of existing channel coding technologies. Due to the independence between the source encoder and channel encoder, they can avoid having to retrain from scratch with a pre-trained multi-modal source encoder. Additionally, a rate-adaptive coding mechanism is designed to provide unequal error protection for different semantic information. 

In Table~\ref{Multimodaltable}, we provide a summarization of current existing works on multi-modal transmission and their contributions. Most of them considered two modalities, while only a limited number proposed solutions for the three modalities. Within the table, several innovative ideas and novel scenarios have been proposed. It includes cross-modal learning techniques, where the model improves the learning capacity in one modality by exploiting the understanding from other modalities, to execute real-time decision making as in a disaster scenario \cite{GaoYunMultimodal}.
\vspace{-0.15in}
\subsection{Challenges within the Directions}
With the advancement of DL models and techniques, deep JSCC-based semantic communication has outperformed traditional wireless communication by a large margin in both performance metrics and compression ratios. However, the direction is nowhere near the perfect solution due to one major drawback, which is the difference in model capacity equipped with multi-transmitter and multi-receiver. So far, most of the work only proposed one transmitter with one receiver, a limited number of works considered one transmitter with multiple receivers \cite{NguyenOneTransmitterMultiple}, multiple transmitters with one receiver \cite{ZhiyiOneReciver}, and multiple transmitters with multiple receivers (if there is they consider same model is shared to all transmitter and receiver). Assuming that all transmitters and receivers are equipped with identical models and possess the same computational capacity represents an overly idealized and unrealistic scenario. In a general scenario, all transmitters and receivers possess different models for semantic communication, which makes it extremely difficult to optimize the model parameters. DL models exhibit a catastrophic forgetting property, causing performance degradation on the first task when the network is trained to perform a second task.
\vspace{-0.15in}

\subsection{Summary and Lesson Learned}

Throughout the section, we have reviewed various works across different tasks and modalities to showcase the outstanding ability of Deep-JSCC direction in improving communication efficiency and the robustness against the channel noise from the physical environment.

\begin{itemize}
    \item Encoding Ability: The encoding ability in this direction is closely linked to advancements in deep learning (DL) techniques. With the rapid growth of DL models, the development of DeepJSCC has progressed at an unprecedented pace, evolving from low-energy-consuming designs to high-performance architectures, addresses various requirements for next communication generation.
    \item Robustness: The integration of wireless conditions and the semantics of data into the encoding process can avoid the cliff effect under low SNR scenarios and also improve the performance when the channel condition improves.
    \item Modality: Each modality possesses unique properties and requires tailored encoding techniques to effectively capture the semantic meaning within the data. By considering the distinctive characteristics of each modality and DL model, transmission data can be significantly reduced while achieving better performance compared to conventional wireless communication systems.
\end{itemize}
\vspace{-0.1in}
\IEEEpubidadjcol
\section{Other Promising Semantic Communication Works and Potential Solutions}
\label{OtherDirections}
\subsection{Other Promising Semantic Communication Works}
Besides the works that fall under the three aforementioned above, there exists some communication work-related semantic information that does not belong to those categories but is worth discussing due to the benefits they bring to the communication field \cite{Yang_semantic,Huang_semantic,Yang_semantic_Sensing,Avi_Semantic,Puligheddu_SEM-O-RAN,Sun_S-RAN,Zhang_Channel_Semantics,Loc_SemanticFederated,Lotfi_Semantic,Zhao_Semantic}. In the scenario described in \cite{Yang_semantic}, the authors equipped the system with a camera to capture images of the environment. They subsequently utilized the semantic information extracted from these images to form the mmWave beam and predict blockages. This semantic information-aided mmWave beamforming can help the system achieve ultra-reliable and low-latency communications. The real-world application for their proposal framework is autonomous driving.

In \cite{AviDebAIGC2}, the authors tackled the inefficiencies in communication within Connected Autonomous Vehicular Networks (CAVNs). Traditional methods transmit large amounts of raw data, like traffic sign images, consuming excessive bandwidth and slowing decision-making. The authors address this by focusing on the semantic meaning of the data instead of the raw information. Their solution extracts key semantic features from traffic signs using a Convolutional Autoencoder (CAE), compressing the data into smaller, meaningful representations or ``concepts". These concepts are transmitted to a Macro Base Station (MBS), which uses a Proximal Policy Optimization (PPO) algorithm to interpret them and guide autonomous vehicles efficiently.

Authors of \cite{Huang_semantic} designed an RIS-aided semantic communication to actively and passively navigate the signal from edge device to edge server when a direct connection is blocked by obstacles. In the paper, a unified design framework is the learning of encoder/decoder with the design objective of the physical layer, which is built based on the \textit{Infomax principle}. A combination between semantic communication and integrated sensing technology has been proposed in \cite{Yang_semantic_Sensing}. To be specific, the base station transmits the sensing and communication signals at the same time, and the sensing signals are considered artificial noise and improve the security of the communication signal. Semantic communication is proposed as a secure technique, where only a receiver that shares a knowledge base with the BS can decode the signal.  

The authors of \cite{Avi_Semantic} address the challenge of achieving robust beamforming in millimeter-wave (mmWave) communications for 6G networks, specifically focusing on maintaining consistent Quality of Service (QoS) under diverse and dynamic environmental conditions. The authors recognize that while deep learning models using RGB camera images show promise in reducing beam training overhead, their performance degrades significantly due to sensitivity to lighting and environmental changes, leading to QoS fluctuations and unreliable network performance. To overcome this limitation, they propose a novel beamforming technique that integrates semantic localization (using K-means clustering and YOLOv8) and optimal beam selection (using a lightweight hybrid transformer and CNN architecture). This hybrid approach aims to leverage multimodal data (RGB images and GPS data) to accurately predict optimal beams even under challenging conditions, thereby ensuring consistent QoS and maximizing data rates for mobile users. A new metric, Accuracy-Complexity Efficiency (ACE), is introduced to evaluate the trade-off between model accuracy and computational efficiency. 

Authors in \cite{Puligheddu_SEM-O-RAN} were the first to propose a \textit{semantic} and \textit{flexible} slicing framework for 5G and beyond Open Radio Access Networks. In their proposed scenario, they considered the objective of executing edge-assisted deep learning (DL) tasks and acknowledged that different tasks can tolerate varying levels of image compression. This flexibility in the slicing algorithm allows for the allocation of different amounts of resources to meet the performance requirements of each task.
\begin{table}[t]
\centering
\caption{The challenges of each semantic communication direction.}
\label{DirectionChallenges}

\begin{tabular}{|p{2.5cm}|p{5.5cm}|}
\hline
\textbf{Directions}                           & \textbf{Challenges Need to be Solved}                                                                                                                                                                     \\ \hline
Theory of Mind-based Semantic Communication           & Large amount of interactions required to form semantic language, mismatch languages when scaling the number of communication parties.                    \\ \hline
Generative AI-based Semantic Communication       & The limitations of communication device hardware for deploying semantic-based AI models, the lack of message interoperability, and the vulnerability to model attacks \\ \hline
Joint source-channel coding-based Semantic Communication & The scalability of the system when resource differences exist in communication devices, and a unified framework for cross-architecture deep learning models.    \\ \hline
Quantum-based Semantic Communication      & The scarcity of available quantum computers and the complexity of quantum systems.                                                                               \\ \hline
\end{tabular}
\end{table}
\vspace{-0.3in}
\subsection{Future Direction}
The advantages of semantic communication are enormous compared with traditional wireless communication, which motivates the attention shift from academic research institutes. Each semantic communication direction remains challenges and have room for development.

\subsubsection{Theory of Mind-based Semantic Communication}
The need to build an effective knowledge base mechanism for the communication agent is crucial, as the system must update its knowledge base to incorporate new beliefs and facts while eliminating outdated information. Secondly, the challenge of designing an appropriate representation for the knowledge base can directly affect the reasoning ability of the network, and it should be low-dimensional to optimize storage. Additionally, it is important to address the challenges related to the knowledge base and reasoning ability when the networks scale up with an increasing number of communication agents. Finally, the need to interpret the multi-modality data is increasing and becoming inevitable, while the current studies of the directions is limited to textual modality.

\subsubsection{Generative AI-based Semantic Communication}

It is still in the early stage of the direction, which indicates various aspects that need to be improved. One of the aspects is to deploy an energy-efficient generative AI model for the transmitter and receiver; the communication devices are constrained by resources compared to the cloud server. On the other side, the generative AI model consumes a large amount of energy and requires high computing capacity and large hardware resources to operate. Secondly, it is significant that we have control over the generative AI model in semantic communication and can explain its output to ensure reliable, trustworthy, and ethical communication scenarios. Without control and explainability, it is impossible to compile the system with regulations and standardization of real-world implementation systems.

\subsubsection{Deep JSCC-based Semantic Communication}

Even joint source-channel coding techniques were proposed very early, but the performance back then was not impressive. However, with the advancement in DL and its integration into the direction, DJSCC-based semantic communication has achieved outstanding performance compared to traditional one, which has encouraged researchers to shift their attention to it and foster its rapid progress. Despite these achievements, the development direction still has many aspects to develop. Especially the problem of networks containing users with different DL models. Along with that is the ability to perform many different tasks to strengthen the flexibility of the system, simulating the changing needs of users.

\subsubsection{Quantum-based Semantic Communication}

Quantum communication and computing have emerged as a promising technology due to several advantages such as unconditional security, tamper detection, information capacity, and long-distance communication. These benefits are achieved due to the unique properties of the quantum: \textit{a) entanglement, b) superposition, c) no-cloning, and finally d) parallelism}. First, the entanglement refers to the interconnected correlation among quantum qubits, where the state of one qubit has a direct influence on the state of the others regarding the distance. Entanglement is a crucial technique for sharing state information through quantum teleportation between two communication parties. Secondly, superposition allows a qubit to exist as a probabilistic combination of \textit{zero} and \textit{one}, in contrast to classical systems where a bit can only have a definite value of either 0 or 1. The superposition of qubits is the key principle that enables the co-existence of multiple states simultaneously and thus facilitates the system's ability to compute many possibilities in parallel. Seeing these tremendous potentials, researchers have integrated quantum into semantic communication \cite{Khalidquantum,Nunavathquantum,Chehimiquantum,Tariqquantum}.

The authors in \cite{Khalidquantum} have laid the foundation for developing a quantum semantic communication system (QSC), which comprises five modules: quantum embedding, quantum semantic encoding, quantum anonymous teleportation, quantum semantic decoding, and quantum semantic detection. The quantum embedding stage can encode the classical data in various ways, including amplitude, angle, basis, and Hamiltonian. Then, the quantum semantic encoding leverages the quantum machine learning \cite{Biamontequantum} to extract the data semantics, which are later transported using anonymous entanglement and broadcast. Similarly, authors in \cite{Nunavathquantum} took advantage of the entanglement property in qubits states for transmitting the knowledge graph. Specifically, the KG is presented as vector space, encoded by Quantum Amplitude Embedding to derive the quantum states, transmitted and decoded at the receiver. 

Meanwhile, \cite{Chehimiquantum} improved the resource efficiency of the quantum network by clustering the data samples into different clusters before feeding them into the quantum-encoding circuit. A SOTA prototype for quantum anonymous semantic broadcast is designed, where a lightweight AI model is deployed to semantic retrieval processing, and the quantum anonymous protocol protects the identity of communication parties (an extra quantum layer of untraceable privacy). Despite the hype of these works, it is unimaginable that we can implement an actual quantum semantic communication at this stage due to the limitations in various aspects of quantum. However, with the attention of many researchers from all fields working together, quantum problems will be solved, leading to a new era for quantum semantic communication. In Table~\ref{DirectionChallenges}, we present a summary to briefly highlight the current challenges in each direction.

\vspace{-0.1in}
\section{Conclusion and Discussion}
\label{Conclusion}
Semantic communication is destined to become an essential pillar of the next generation of communication, addressing various challenges faced by current 5G systems without compromising any aspects. By looking into the meaning of the data and transmitting only the important ones, we can significantly reduce the transmission data while leveraging channel conditions into the transmission process. With the enormous advantages offered by semantic communication, researchers have been racing to allocate resources and develop new techniques to leverage its full potential. This rapid progression, however, has led to a fragmentation of approaches, creating a divide into three main directions: each is built on a different theory, such as the Theory of Mind, AI-generated content, and joint source-channel coding. Unfortunately, there is no single standardized protocol or unified framework that has been agreed upon among researchers or legitimate communication companies. Therefore, it is difficult to determine which direction is leading in the field. In the survey, we first introduce the foundation idea of each direction, explain its general target, and provide the details of existing works within the directions. At the end of each direction, we provide the challenges that must be addressed before deployment into real-world scenarios. These challenges vary widely across directions, ranging from the long learning process for the causal reasoning ability of the Theory of Mind, the requirement for a powerful model for the AIGC model within the second direction, and finally, the scalability for multiple transmitters and receivers for the joint source-channel coding. Through the detailed description of existing works and the remaining challenges, we want to deliver to the reader the overall picture of semantic communication and what is left to improve. While it is difficult to predict which the future of semantic communication will fall into which direction, or all three directions will be merged into one, the one thing we can ensure is semantic communication is going to be the core technology in next-generation communication.

\newpage












\vfill


\vspace{-0.1in}


\begin{thebibliography}{1}
\bibliographystyle{IEEEtran}

\bibitem{Shannon}C. E. Shannon, ‘‘A mathematical theory of communication,’’ \emph{Bell Syst. Tech. J.}, vol. 27, no. 3, pp. 379–423, Jul./Oct. 1948.
\bibitem{Weaver}
W. Weaver, “Recent contributions to the mathematical theory of communication,” \emph{ETC: a review of general semantics,} pp. 261–281, 1953.
\bibitem{Huffman}
David A Huffman. ``A method for the construction of minimum-redundancy codes," \emph{Proceedings of the IRE}, vol. 40, no. 9, pp. 1098–1101, Sep. 1952.
\bibitem{Golomb}
S. Golomb, “Run-length encodings (corresp.),” \emph{IEEE Trans. Inf. Theory,} vol. 12, no. 3, pp. 399–401, Jul. 1966.
\bibitem{Rissanen}
 J. Rissanen and G. G. Langdon, “Arithmetic coding,” \emph{IBM J. Res. Develop.,} vol. 23, no. 2, pp. 149–162, 1979.
\bibitem{DeVore}
R. A. DeVore, B. Jawerth, and B. J. Lucier, “Image compression through wavelet transform coding,” \emph{IEEE Trans. Inf. Theory,} vol. 38, no. 2, pp. 719–746, Mar. 1992.
\bibitem{Hamming}
R. W. Hamming, ‘‘Error detecting and error correcting codes,’’ \emph{Bell Syst. Tech. J.,} vol. 29, no. 2, pp. 147–160, Apr. 1950.
\bibitem{Reed}
Irving S. Reed and Gustave Solomon. ‘‘Polynomial codes over certain finite fields.’’ \emph{Journal of the Society for Industrial and Applied Mathematics,} vol. 8, no. 2, pp. 300–304, Jun. 1960.
\bibitem{Gallager}
R. G. Gallager, “Low-density parity-check codes,” \emph{IRE Trans. Inf. Theory,} vol. 8, no. 1, pp. 21–28, Jan. 1962.
\bibitem{Elias}
Peter Elias, “Universal codeword sets and representations of the integers,” \emph{IEEE Trans. Inf. Theory,} vol. 21, no. 2, pp. 194–203, Mar. 1975.
\bibitem{Berrou}C. Berrou, A. Glavieux, and P. Thitimajshima, “Near Shannon limit error-correcting coding and decoding: Turbo-codes. 1,” in \emph{Proc. IEEE Int. Conf. Commun.,} pp. 1064–1070, Geneva, Switzerland, May 1993.

\bibitem{Arikan}E. Arikan, “Channel polarization: A method for constructing capacity achieving codes for symmetric binary-input memoryless channels,” \emph{IEEE Trans. Inf. Theory,} vol. 55, no. 7, pp. 3051–3073, Jul. 2009.
\bibitem{SemanticSurvey0}
S. Guo et al., ``A Survey on Semantic Communication Networks: Architecture, Security, and Privacy," \emph{IEEE Commun. Surveys Tuts.,} Early Access, Dec. 2024. DOI: 10.1109/COMST.2024.3516819.

\bibitem{ChristinaLessdata}
C. Chaccour, W. Saad, M. Debbah, Z. Han and H. V. Poor, ``Less Data, More Knowledge: Building Next Generation Semantic Communication Networks," \emph{IEEE Commun. Surveys Tuts.,} Early Access 2024, DOI: 10.1109/COMST.2024.3412852. 

\bibitem{SemanticSurvey1}
Zhang, P., Liu, Y., Song, Y. and Zhang, J., 2024. ``Advances and challenges in semantic communications: A systematic review," \emph{National Science Open,} vol. 3, no. 4, Nov. 2023.

\bibitem{SemanticSurvey2}
Meng, R., Gao, S., Fan, D., Gao, H., Wang, Y., Xu, X., Wang, B., Lv, S., Zhang, Z., Sun, M. and Han, S., 2025. ``A Survey of Secure Semantic Communications." \emph{arXiv:2501.00842}, Jan. 2025.
\bibitem{SemanticSurvey3}
P. Zhang et al., ``Intellicise Wireless Networks From Semantic Communications: A Survey, Research Issues, and Challenges," \emph{IEEE Commun. Surveys Tuts.,} Early Access, Aug. 2024. DOI 10.1109/COMST.2024.3443193.
\bibitem{SemanticSurvey4}
W. Yang et al., ``Semantic Communications for Future Internet: Fundamentals, Applications, and Challenges," \emph{IEEE Commun. Surveys Tuts.,} vol. 25, no. 1, pp. 213-250, First-quarter 2023.
\bibitem{SemanticSurvey5}
T. M. Getu, G. Kaddoum and M. Bennis, ``A Survey on Goal-Oriented Semantic Communication: Techniques, Challenges, and Future Directions," \emph{IEEE Access,} vol. 12, pp. 51223-51274, Mar. 2024.
\bibitem{SemanticSurvey6}
T. M. Getu, G. Kaddoum and M. Bennis, ``Making Sense of Meaning: A Survey on Metrics for Semantic and Goal-Oriented Communication," \emph{IEEE Access,} vol. 11, pp. 45456-45492, May 2023.

\bibitem{ChengsiAIGC}
C. Liang et al., ``Generative AI-driven Semantic Communication Networks: Architecture, Technologies and Applications," \emph{IEEE Trans. Cogn. Commun. Netw.,} Early Access, Jul. 2024, DOI: 10.1109/TCCN.2024.3435524. 

\bibitem{WangSurvey}
Y. Wang, H. Han, Y. Feng, J. Zheng and B. Zhang, ``Semantic Communication Empowered 6G Networks: Techniques, Applications, and Challenges," \emph{IEEE Access,} Early Access, Jan. 2025.  DOI:10.1109/ACCESS.2025.3532797.

\bibitem{MengSurvey}
S. Meng, S. Wu, J. Zhang, J. Cheng, H. Zhou and Q. Zhang, ``Semantics-Empowered Space-Air-Ground-Sea Integrated Network: New Paradigm, Frameworks, and Challenges," \emph{IEEE Commun. Surveys Tuts.,} Early Access, Jun. 2024.  DOI: 10.1109/COMST.2024.3416309.
\bibitem{DWonSurvey}
D. Won et al., ``Resource Management, Security, and Privacy Issues in Semantic Communications: A Survey," \emph{IEEE Commun. Surveys Tuts.,} Early Access, Oct. 2024. DOI: 10.1109/COMST.2024.3471685.
\bibitem{ZLuSurvey}
Z. Lu et al., ``Semantics-Empowered Communications: A Tutorial-Cum-Survey," \emph{IEEE Commun. Surveys Tuts.,} vol. 26, no. 1, pp. 41-79, First-quarter 2024.

\bibitem{Frith}
C. Frith and U. Frith, “Theory of Mind,” \emph{Current Biol.,} vol. 15, no. 17, pp. R644–R645, Sep. 2005.

\bibitem{ThomasReason1}
C. Kurisummoottil Thomas and W. Saad, ``Neuro-Symbolic Causal Reasoning Meets Signaling Game for Emergent Semantic Communications," \emph{IEEE Trans. Wireless Commu.,} vol. 23, no. 5, pp. 4546-4563, May 2024. 
\bibitem{GenerativeAISem}
Y. Liu et al., ``Cross-Modal Generative Semantic Communications for Mobile AIGC: Joint Semantic Encoding and Prompt Engineering," \emph{IEEE Trans. Mobile Comput.} vol. 23, no. 12, pp. 14871-14888, Dec. 2024.
\bibitem{GenerativeAISem1}
Xia, L., Sun, Y., Liang, C., Zhang, L., Imran, M.A. and Niyato, D., ``Generative AI for semantic communication: Architecture, challenges, and outlook," \emph{arXiv:2308.15483,} Aug. 2023.

\bibitem{Bourtsoulatze2019DeepJSCC}
E. Bourtsoulatze, D. Burth Kurka and D. Gündüz, ``Deep Joint Source-Channel Coding for Wireless Image Transmission," \emph{IEEE Trans. Cogn. Commun. Netw.,} vol. 5, no. 3, pp. 567-579, Sep. 2019.
\bibitem{GenerativeAI_Ref}
R. Zhang et al., ``Generative AI for Space-Air-Ground Integrated Networks," \emph{IEEE Wireless Commun.} vol. 31, no. 6, pp. 10-20, Dec. 2024, DOI: 10.1109/MWC.016.2300547. 
\bibitem{GenerativeAI_Ref2}
M. Moor, O. Banerjee, Z. S. H. Abad, H. M. Krumholz, J. Leskovec, E. J. Topol, and P. Rajpurkar, “Foundation models for generalist medical artificial intelligence,” \emph{Nature,} vol. 616, no. 7956, pp. 259–265, Apr. 2023.
\bibitem{GenerativeAI_Ref3}
 P. Jiang, C.-K. Wen, X. Yi, X. Li, S. Jin, and J. Zhang, “Semantic communications using foundation models: Design approaches and open issues,” \emph{IEEE Wireless Commun.} vol. 31, no. 3, pp. 76–84, Jun. 2024.
\bibitem{GenerativeAI_Ref4}
H. Nam, et al., ``Sequential Semantic Generative Communication for Progressive Text-to-Image Generation," in \emph{Proc. IEEE Annu. Int. Conf. Sens. Commun. Netw.,} Madrid, Spain, Sep. 2023.


\bibitem{LeCun}
Y. LeCun and Y. Bengio, ‘‘Convolutional networks for images, speech, and time series,’’ \emph{The Handbook of Brain Theory and Neural Networks,} vol. 3361, Cambridge, MA, USA: MIT Press, 1995.

\bibitem{Kaiming}
K. He, X. Zhang, S. Ren, and J. Sun, “Deep residual learning for image recognition,” in \emph{Proc. IEEE Conf. Comput. Vis. Pattern Recognit.,} pp. 770–778, NV, Jun. 2016.

\bibitem{Vaswani}
A. Vaswani et al., “Attention is all you need,” \emph{in Proc. 31st Int. Conf. Neural Inf. Process. Syst.,} pp. 6000–6010, CA, USA, Dec. 2017.
\bibitem{Dosovitskiy}
A. Dosovitskiy et al., “An image is worth 16x16 words: Transformers for image recognition at scale,” \emph{in Proc. Int. Conf. Learn. Representations,} Virtual Conference, May 2021.

\bibitem{SemanticDJSCCLength}
B. Zhang, Z. Qin and G. Y. Li, ``Semantic Communications With Variable-Length Coding for Extended Reality," \emph{IEEE J. Sel. Topics Signal Process.,} vol. 17, no. 5, pp. 1038-1051, Sep. 2023,
\bibitem{Kahneman}
Kahneman, Daniel. ``Thinking, fast and slow." \emph{Farrar, Straus and Giroux,} Oct. 2011.
\bibitem{Hyowoon}
H. Seo et al., “Semantics-native communication via contextual reasoning,” \emph{IEEE Trans. Cognit. Commun. Netw.,} vol. 9, no. 3, pp. 604--617, Jun. 2023.
\bibitem{Ogden}
C. K. Ogden and I. A. Richards, “The Meaning of Meaning: A Study of the Influence of Thought and of the Science of Symbolism.” \emph{Harcourt, Brace \& World, Inc.} NY, USA, Jan. 1923.
\bibitem{Liang}
Liang, Jingming, et al., ``Life-long learning for reasoning-based semantic communication." in \emph{Proc. IEEE Int. Conf. Commun. Workshops (ICC Workshops),} pp. 271-276., Seoul, South Korea, May  2022.
\bibitem{Farshbafan}
M. K. Farshbafan, W. Saad, and M. Debbah, “Curriculum learning for goal-oriented semantic communications with a common language,” \emph{IEEE Trans. Commun.,} vol. 71, no. 3, pp. 1430–1446, Mar. 2023.
\bibitem{KurisummoottilReasoningNeuro}
C. K. Thomas and W. Saad, “Neuro-symbolic artificial intelligence (AI) for intent based semantic communication,” in \emph{Proc. IEEE Global Commun. Conf.,} Rio de Janeiro, Brazil, Dec. 2022.
\bibitem{ThomasPragmatic}
C. K. Thomas, E. C. Strinati and W. Saad, ``Reasoning with the Theory of Mind for Pragmatic Semantic Communication," in \emph{Proc. IEEE 21st Consum. Commun. Netw. Conf. (CCNC),} NV, USA, Jan. 2024.


\bibitem{Thomasdigital}
C. K. Thomas, W. Saad, and Y. Xiao, “Causal semantic communication for digital twins: A generalizable imitation learning approach,” \emph{IEEE J. Sel. Areas Inf. Theory,} vol. 4, pp. 698–717, Nov. 2023.
\bibitem{MohamedMultiuser}
Sana, Mohamed, and Emilio Calvanese Strinati. ``Semantic channel equalizer: Modelling language mismatch in multi-user semantic communications." in \emph{Proc. IEEE Global Commun. Conf.,} pp. 2221-2226, Kuala Lumpur, Malaysia, Dec. 2023.
\bibitem{Nitisha}
Singh, Nitisha, et al., ``On the Computing and Communication Tradeoff in Reasoning-Based Multi-User Semantic Communications." \emph{arXiv:2406.15199,} Jun. 2024.
\bibitem{SeoProtocols}
S. Seo, J. Park, S.-W. Ko, J. Choi, M. Bennis, and S.-L. Kim, “Toward semantic communication protocols: A probabilistic logic perspective,” \emph{IEEE J. Sel. Areas Commun.,} vol. 41, no. 8, pp. 2670–2686, Aug. 2023.


\bibitem{YongReasoning1}
Y. Xiao, et al., “Reasoning on the air: An implicit semantic communication architecture,” in \emph{Proc. IEEE Int. Conf. Commun. Workshops (ICC Workshops),} pp. 289-294, Seoul, South Korea, May 2022.

\bibitem{YongReasoning2}
Y. Xiao et al., “Reasoning over the air: A reasoning-based implicit semantic-aware communication framework,” \emph{IEEE Trans. Wireless Commun.,} vol. 23, no. 4, pp. 3839–3855, Apr. 2024.

\bibitem{EmilioReasoning}
Strinati, Emilio Calvanese, Paolo Di Lorenzo, Vincenzo Sciancalepore, Adnan Aijaz, Marios Kountouris, Deniz Gündüz, Petar Popovski et al. ``Goal-oriented and semantic communication in 6G AI-native networks: The 6G-GOALS approach." \emph{arXiv:2402.07573,} Feb. 2024.

\bibitem{FuhuiReasoning1}
Zhou, Fuhui, et al., ``Cognitive semantic communication systems driven by knowledge graph." in  \emph{Proc. IEEE Int. Conf. Commun. (ICC),} pp. 4860-4865, Seoul, South Korea, May 2022.

\bibitem{FuhuiReasoning2}
F. Zhou, et al., “Cognitive semantic communication systems driven by knowledge graph: Principle, implementation, and performance evaluation,” \emph{IEEE Trans. Commun., vol.} 72, no. 1, pp. 193–208, Jan. 2024.

\bibitem{ShengzheReasoning}
Xu, Shengzhe, et al., ``Large multi-modal models (LMMs) as universal foundation models for AI-native wireless systems." \emph{Preprint arXiv:2402.01748,} Feb. 2024.

\bibitem{ChenlinReasoning}
C. Xing, J. Lv, T. Luo and Z. Zhang, ``Representation and Fusion Based on Knowledge Graph in Multi-Modal Semantic Communication," \emph{IEEE Wireless Commun. Lett.,} vol. 13, no. 5, pp. 1344-1348, May 2024. 


\bibitem{ChristinaReasoningDisentangling}
C. Chaccour and W. Saad, “Disentangling learnable and memorizable data via contrastive learning for semantic communications,” in \emph{Proc. IEEE 56th Asilomar Conf. Signals, Syst., Comput.,} Asilomar, CA, Oct. 2022.

\bibitem{BingyanReasoningKnowledge}
B. Wang, R. Li, J. Zhu, Z. Zhao, and H. Zhang, “Knowledge enhanced semantic communication receiver,” \emph{IEEE Commun. Lett.,} vol. 27, no. 7, pp. 1794–1798, Jul. 2023.



\bibitem{JinhoReasoningSemantic}
J. Choi and J. Park, “Semantic communication as a signaling game with correlated knowledge bases,” in \emph{Proc. IEEE 96th Veh. Technol. Conf.,}  London, UK, Sep. 2022.


\bibitem{DylanReasoning}
Wheeler, Dylan, and Balasubramaniam Natarajan. ``Conceptual Learning and Causal Reasoning for Semantic Communication." \emph{Authorea Preprints,} Nov. 2024.









\bibitem{Keng-BoonSurvey}
K.-B. Ooi et al., “The potential of generative artificial intelligence across disciplines: Perspectives and future directions,” \emph{J. Comput. Inf. Syst.,} pp. 1–32, Oct. 2023.

\bibitem{HanqunGenerative}
H. Cao et al., “A Survey on Generative Diffusion Models," \emph{IEEE Trans. Knowl. Data Eng.,} vol. 36, no. 7, pp. 2814-2830, Jul. 2024.


\bibitem{JonathanDiffusion}
 J. Ho, A. Jain, and P. Abbeel, “Denoising diffusion probabilistic models,” in \emph{Proc. Int. Conf. Neural Inf. Process. Syst.,} pp. 6840–6851, Virtual Conference, Dec. 2020.




\bibitem{WangAIGC}
H. Du, J. Wang, et al., ``AI-Generated Incentive Mechanism and Full-Duplex Semantic Communications for Information Sharing," \emph{IEEE J. Sel. Areas Commun.,} vol. 41, no. 9, pp. 2981-2997, Sep. 2023.

\bibitem{BaoxiaAIGC}
B. Du et al., ``YOLO-Based Semantic Communication With Generative AI-Aided Resource Allocation for Digital Twins Construction," \emph{IEEE Internet Things J.,} vol. 11, no. 5, pp. 7664-7678, Mar. 2024. 


\bibitem{ChengAIGC}
R. Cheng, et al., ``A Wireless AI-Generated Content (AIGC) Provisioning Framework Empowered by Semantic Communication," \emph{IEEE Trans. Mobile Comput.,} Early Access, Nov. 2024, DOI: 10.1109/TMC.2024.3493375. 

\bibitem{YijingAIGC}
Y. Lin et al., ``A Unified Framework for Integrating Semantic Communication and AI-Generated Content in Metaverse," \emph{IEEE Network,} vol. 38, no. 4, pp. 174-181, Jul. 2024.



\bibitem{Yijing2AIGC}

Y. Lin et al., ``Blockchain-Aided Secure Semantic Communication for AI-Generated Content in Metaverse," \emph{IEEE Open J. Comput. Soc.,} vol. 4, pp. 72-83, Mar. 2023. 

\bibitem{ZheAIGC}
Wang, Zhe, Nan Li, Yansha Deng, and A. Hamid Aghvami. ``Goal-oriented Semantic Communications for Metaverse Construction via Generative AI and Optimal Transport."  \emph{arXiv:2411.16187,} Nov. 2024.

\bibitem{ZhengAIGC}
J. Zheng, B. Du, H. Du, J. Kang, D. Niyato and H. Zhang, ``Energy-Efficient Resource Allocation in Generative AI-Aided Secure Semantic Mobile Networks," \emph{IEEE Trans. Mobile Comput.} vol. 23, no. 12, pp. 11422-11435, Dec. 2024. 



\bibitem{FangzhouAIGC}
F. Zhao, et al., ``Enhancing Reasoning Ability in Semantic Communication Through Generative AI-Assisted Knowledge Construction," \emph{IEEE Commun. Lett.,} vol. 28, no. 4, pp. 832-836, Apr. 2024.



\bibitem{YuxinPromt}
Wen, Yuxin, et al., ``Hard prompts made easy: Gradient-based discrete optimization for prompt tuning and discovery." in \emph{Proc. Advances in Neural Information Processing Systems 36,} BC, Canada, Dec. 2024.


\bibitem{JohannesNTC}
J. Ballé et al., “Nonlinear transform coding,” \emph{IEEE J. Sel. Topics Signal Process.,} vol. 15, no. 2, pp. 339–353, Oct. 2020.




\bibitem{MengmengAIGC}
Ren, Mengmeng, et al., ``Generative Semantic Communication via Textual Prompts: Latency Performance Tradeoffs." \emph{arXiv:2409.09715,} Sep. 2024.
\bibitem{XiaLeAIGC}
Xia, Le, Yao Sun, Chengsi Liang, Lei Zhang, Muhammad Ali Imran, and Dusit Niyato. ``Generative AI for semantic communication: Architecture, challenges, and outlook."  \emph{arXiv:2308.15483,} Aug. 2023.
\bibitem{AIGCEdge0}
H. Wang, H. Li, M. Sheng and J. Li, ``Collaborative Fine-Tuning of Mobile AIGC Models with Wireless Channel Conditions," \emph{IEEE Wireless Commun.} vol. 31, no. 4, pp. 32-38, Aug. 2024.
\bibitem{AIGCEdge1}
G. Hao, Q. Pan and J. Wu, ``Incentive Distributed Knowledge Graph Market for Generative Artificial Intelligence in IoT," \emph{IEEE Internet Things J.,} Early Access, Dec. 2024. DOI: 10.1109/JIOT.2024.3522191.
\bibitem{AIGCEdge2}
Tang, Shunpu, et al., ``Retrieval-augmented Generation for GenAI-enabled Semantic Communications." \emph{arXiv:2412.19494,} Dec. 2024.


\bibitem{HongyangAIGC2}
Du, Hongyang, et al., ``Generative Al-aided Joint Training-free Secure Semantic Communications via Multi-modal Prompts." in \emph{Proc. IEEE Int. Conf. Acoust., Speech, Signal Process.,}  Seoul, South Korea, May 2024.


\bibitem{ChunmeiAIGC}
Xu, Chunmei, Mahdi Boloursaz Mashhadi, Yi Ma, and Rahim Tafazolli. ``Semantic-aware power allocation for generative semantic communications with foundation models." \emph{arXiv:2407.03050,} Jul. 2024.

\bibitem{SenranAIGC}
Fan, Senran, et al., ``Semantic Feature Decomposition based Semantic Communication System of Images with Large-scale Visual Generation Models."  \emph{arXiv:2410.20126,} Oct. 2024.




\bibitem{JiamingDDIM}
J. Song, C. Meng, and S. Ermon, “Denoising diffusion implicit models,” in \emph{Proc. Int. Conf. Learn. Representations,} Virtual Conference, May 2021.



\bibitem{JunnanBLIP}
J. Li, et al., “BLIP: Bootstrapping language-image pre-training for unified vision-language understanding and generation,” in \emph{Proc. 39th Int. Conf. Mach. Learn. (ICML),} Maryland, USA, Jul. 2022.


\bibitem{PulkitAIGC}
Tandon, Pulkit, et al., ``Txt2Vid: Ultra-low bitrate compression of talking-head videos via text." \emph{IEEE J. Sel. Areas Commun.,}, vol. 41, no. 1, pp. 107-118, Jan. 2023.
\bibitem{TianAIGCVideo}
Y. Tian, J. Ying, Z. Qin, Y. Jin and X. Tao, ``Synchronous Semantic Communications for Video and Speech," in \emph{Proc. IEEE Int. Conf. Commun. (ICC),} pp. 3396-3401, Denver, CO, USA, Jun. 2024.

\bibitem{EleonoraAIGC}
Grassucci, Eleonora, et al., ``Diffusion models for audio semantic communication." in \emph{Proc. IEEE Int. Conf. Acoust., Speech, Signal Process.,} pp. 13136-13140, Seoul, South Korea, May 2024.
\bibitem{ZhengAIGCSpeech}
Zheng, Jiahao, Jinke Ren, Peng Xu, Zhihao Yuan, Jie Xu, Fangxin Wang, Gui Gui, and Shuguang Cui. ``Generative semantic communication for text-to-speech synthesis." arXiv preprint arXiv:2410.03459 (2024).

\bibitem{ShuaishuaiAIGC}
S. Guo, Y. Wang, S. Li and N. Saeed, ``Semantic Importance-Aware Communications Using Pre-Trained Language Models," \emph{IEEE Commun. Lett.,} vol. 27, no. 9, pp. 2328-2332, Sep. 2023. 





\bibitem{PrajwalLIP}
K. R. Prajwal, et al., “A lip sync expert is all you need for speech to lip generation in the wild,” in \emph{Proc. ACM Int. Conf. Multimedia,} Seattle, WA, pp. 484–492, Oct. 2020.



\bibitem{GuangyuanAIGC}
Liu, Guangyuan, Hongyang Du, Dusit Niyato, Jiawen Kang, Zehui Xiong, Dong In Kim, and Xuemin Shen. ``Semantic communications for artificial intelligence generated content (AIGC) toward effective content creation." \emph{IEEE Network,} vol. 38, no. 5, pp. 295-303, Sep. 2024.

\bibitem{RuichenAIGC}
Zhang, Ruichen, Ke Xiong, Hongyang Du, Dusit Niyato, Jiawen Kang, Xuemin Shen, and H. Vincent Poor. ``Generative AI-enabled vehicular networks: Fundamentals, framework, and case study." \emph{IEEE Network,} vol. 38, no. 4, pp. 259-267, Jul. 2024.

\bibitem{AviDebAIGC}
Raha, Avi Deb, Md Shirajum Munir, Apurba Adhikary, Yu Qiao, and Choong Seon Hong. ``Generative AI-driven semantic communication framework for NextG wireless network."  \emph{arXiv:2310.09021,} Oct. 2023.
\bibitem{PixelGAN}
 P. Isola et al., “Image-to-image translation with conditional adversarial networks,” in \emph{Proc. IEEE/CVF Conf. Comput. Vis. Pattern Recognit.,} pp. 1125–1134, Honolulu, HI, USA, Jul. 2017.
\bibitem{AviDebAIGC2}
Raha, Avi Deb, et al., ``An artificial intelligent-driven semantic communication framework for connected autonomous vehicular network." in \emph{Proc. Int. Conf. Inf. Netw.,} pp. 352-357, Bangkok, Thailand, Jan. 2023.


\bibitem{TongAIGC}
Wu, Tong, et al., ``CDDM: Channel denoising diffusion models for wireless semantic communications." \emph{IEEE Trans. Wireless Commu.,} vol. 23, no. 9, pp. 11168-11183, Sep. 2024. 



\bibitem{HongyangAIGC3}
H. Du et al., ``Enhancing Deep Reinforcement Learning: A Tutorial on Generative Diffusion Models in Network Optimization." \emph{IEEE Commun. Surveys Tuts.,} vol. 26, no. 4, pp. 2611-2646, Fourth-quarter 2024. 

\bibitem{FeiboAIGCLarge}
Jiang, F., Dong, L., Peng, Y., Wang, K., Yang, K., Pan, C. and You, X., ``Large AI model empowered multimodal semantic communications." \emph{IEEE Commun. Mag.,}  vol. 63, no. 1, pp. 76-82, Jan. 2025.
\bibitem{FeiboAIGCLarge2}
F. Jiang et al., ``Large AI Model-Based Semantic Communications," \emph{IEEE Wireless Commun.} vol. 31, no. 3, pp. 68-75, Jun. 2024. 
\bibitem{FeiboAIGCLarge3}
Jiang, Feibo, Yubo Peng, Li Dong, Kezhi Wang, Kun Yang, Cunhua Pan, and Xiaohu You. ``Large Generative Model Assisted 3D Semantic Communication." \emph{arXiv:2403.05783,} Mar. 2024.

\bibitem{WantingAIGC}
Yang, Wanting, et al., ``Rethinking generative semantic communication for multi-user systems with multi-modal LLM." \emph{Preprint arXiv:2408.08765,} Aug. 2024.
\bibitem{EleonoraAIGC2}
Grassucci, Eleonora, el al., ``Generative AI meets semantic communication: Evolution and revolution of communication tasks." \emph{Preprint arXiv:2401.06803,} Jan. 2024.
\bibitem{ZhenyiAIGC}
Wang, Zhenyi, et al., ``Large Language Model Enabled Semantic Communication Systems." \emph{Preprint arXiv:2407.14112,} Jul. 2024.
\bibitem{JianhuaAIGC}
Pei, Jianhua, et al., ``Latent Diffusion Model-Enabled Real-Time Semantic Communication Considering Semantic Ambiguities and Channel Noises." \emph{Preprint arXiv:2406.06644,} Jun. 2024.

\bibitem{ChenMingkaiMultimodalAIGC}
Chen, Mingkai, et al., ``Cross-Modal Graph Semantic Communication Assisted by Generative AI in the Metaverse for 6G." \emph{Research,}
vol. 7, pp. 0342, Mar. 2024.



\bibitem{salomon2007data}
D. Salomon, ‘‘Data compression: The complete reference (by D. Salomon; 2007) [book review],’’ \emph{IEEE Signal Process. Mag.,} vol. 25, no. 2, pp. 147–149, Mar. 2008.
\bibitem{wallace1992jpeg}
G. K. Wallace, ``The JPEG still picture compression standard," \emph{IEEE Trans. Consum. Electron.,} vol. 38, no. 1, pp. xviii-xxxiv, Feb. 1992.
\bibitem{taubman2001jpeg2000}
D.S. Taubman and M.W. Marcellin, ``JPEG2000: Image Compression Fundamentals, Standards and Practice," \emph{Springer}, Nov. 2001.
\bibitem{bellard2014bpg}
F. Bellard. “Better portable graphics.” 2014. Accessed: Mar. 13, 2020.
[Online]. Available: \url{https://bellard.org/bpg/}

\bibitem{johnston1988transform}
J. D. Johnston, ``Transform coding of audio signals using perceptual noise criteria,'' \emph{IEEE J. Sel. Areas Commun.,} vol. 6, no. 2, pp. 314--323, Feb. 1988.

\bibitem{pan1995tutorial}
D. Pan, ``A tutorial on MPEG/audio compression,'' \emph{IEEE Multimedia,} vol.2, no.2, pp. 60--74, Aug. 1995.

\bibitem{bradenburg1999aac}
K. Brandenburg and M. Bosi, ``Overview of MPEG audio: Current and future standards for low-bit-rate audio coding,'' \emph{J. Audio Eng. Soc.,} vol. 45, no. 12, pp. 4--21, Feb. 1997.


\bibitem{wiegand2003h264}
T. Wiegand, G. J. Sullivan, G. Bjontegaard, and A. Luthra, ``Overview of the H.264/AVC video coding standard,'' \emph{IEEE Trans. Circuits Syst. Video Technol.,} vol. 13, no. 7, pp. 560--576, Jul. 2003.

\bibitem{sullivan2012h265}
G. J. Sullivan, J. Ohm, W. Han, and T. Wiegand, ``Overview of the high efficiency video coding (HEVC) standard,'' \emph{IEEE Trans. Circuits Syst. Video Technol.,} vol. 22, no. 12, pp. 1649--1668, Dec. 2012.

\bibitem{bankoski2013av1}
J. Bankoski, P. Wilkins, and Y. Xu, ``Technical overview of VP8, an open source video codec for the web,'' in \emph{Proc. IEEE Int. Conf. Multimedia Expo (ICME),} pp. 1--6, Barcelona, Spain, Jul. 2013.


\bibitem{cover1991elements}
T. M. Cover and J. A. Thomas, ``Elements of Information Theory." \emph{Wiley-Interscience}, Hoboken, NJ, USA: Wiley, Jul. 2006.
  

\bibitem{skoglund2003jscc}  
M. Skoglund, N. Phamdo and F. Alajaji, ``Hybrid Digital–Analog Source–Channel Coding for Bandwidth Compression/Expansion," \emph{IEEE Trans. Inf. Theory,} vol. 52, no. 8, pp. 3757-3763, Aug. 2006.

\bibitem{goldsmith2005wireless}
A. Goldsmith, ``Wireless Communications," \emph{Cambridge Univ. Press,} Cambridge, U.K, Aug. 2005.

\bibitem{modestino1984image}
J. Modestino and D. Daut, “Combined source-channel coding of images,” \emph{IEEE Trans. Commun.,} vol. 27, no. 11, pp. 1644–1659, Nov. 1979.


\bibitem{gallager1988information}
R. G. Gallager, Information Theory and Reliable Communication. \emph{Springer} Hoboken, NJ, USA: Wiley, Jul. 1970.




\bibitem{9500996}
M. Yang, C. Bian and H. -S. Kim, ``Deep Joint Source Channel Coding for Wireless Image Transmission with OFDM," in \emph{Proc. IEEE Int. Conf. Commun. (ICC),} pp. 1-6, Montreal, QC, Canada, Jun. 2021.


\bibitem{FayçalAitImage}
F. A. Aoudia and J. Hoydis, ``Model-Free Training of End-to-End Communication Systems," \emph{IEEE J. Sel. Areas Commun.,} vol. 37, no. 11, pp. 2503-2516, Nov. 2019.

\bibitem{MingyuImage}
M. Yang, C. Bian and H. -S. Kim, ``OFDM-Guided Deep Joint Source Channel Coding for Wireless Multipath Fading Channels," \emph{IEEE Trans. Cogn. Commun. Netw.,} vol. 8, no. 2, pp. 584-599, Jun. 2022.

\bibitem{YehaoImageDeep2}
H. Ye, et al., "``Deep Learning-Based End-to-End Wireless Communication Systems With Conditional GANs as Unknown Channels," \emph{IEEE Trans. Wireless Commu.,} vol. 19, no. 5, pp. 3133-3143, May 2020.


\bibitem{YehaoImageDeep}

H. Ye, G. Y. Li, and B.-H. Juang, “Deep learning based end-to-end wireless communication systems without pilots,” \emph{IEEE Trans. Cognit. Commun. Netw.,} vol. 7, no. 3, pp. 702–714, Sep. 2021.

\bibitem{QiyuImageRobust}
Q. Hu, G. Zhang, Z. Qin, Y. Cai, G. Yu and G. Y. Li, ``Robust Semantic Communications Against Semantic Noise," in \emph{Proc. IEEE 96th Veh. Technol. Conf., (VTC2022-Fall),} pp. 1-6, London, UK, Sep. 2022.

\bibitem{QiyuImage}
Q. Hu, G. Zhang, Z. Qin, Y. Cai, G. Yu and G. Y. Li, ``Robust Semantic Communications With Masked VQ-VAE Enabled Codebook," \emph{IEEE Trans. Wireless Commu.,} vol. 22, no. 12, pp. 8707-8722, Dec. 2023. 

\bibitem{SongjieImageRobust}
S. Xie, et al.,``Robust Information Bottleneck for Task-Oriented Communication With Digital Modulation," \emph{IEEE J. Sel. Areas Commun.,} vol. 41, no. 8, pp. 2577-2591, Aug. 2023.




\bibitem{JinchengImage}
J. Dai et al., ``Nonlinear Transform Source-Channel Coding for Semantic Communications," \emph{IEEE J. Sel. Areas Commun.,} vol. 40, no. 8, pp. 2300-2316, Aug. 2022.


\bibitem{JiaweiImage}
J. Shao and J. Zhang, ``BottleNet++: An End-to-End Approach for Feature Compression in Device-Edge Co-Inference Systems," in \emph{Proc. IEEE Int. Conf. Commun. Workshops,} pp. 1-6, Dublin, Ireland, Jun. 2020.


\bibitem{DavidImageBandwidth-agile}
D. B. Kurka and D. Gündüz, ``Bandwidth-Agile Image Transmission With Deep Joint Source-Channel Coding," \emph{IEEE Trans. Wireless Commu.,} vol. 20, no. 12, pp. 8081-8095, Dec. 2021.

\bibitem{MingyuImageDeep}
M. Yang and H. -S. Kim, ``Deep Joint Source-Channel Coding for Wireless Image Transmission with Adaptive Rate Control," in \emph{Proc. IEEE Int. Conf. Acoust., Speech, Signal Process.,} Singapore, May 2022. 

\bibitem{ZhangWenyuImagePredictive}
W. Zhang, et al., ``Predictive and Adaptive Deep Coding for Wireless Image Transmission in Semantic Communication," \emph{IEEE Trans. Wireless Commu.,} vol. 22, no. 8, pp. 5486-5501, Aug. 2023.

\bibitem{Tze-YangImageDeepJSCC}
T. -Y. Tung, D. B. Kurka, M. Jankowski and D. Gündüz, ``DeepJSCC-Q: Constellation Constrained Deep Joint Source-Channel Coding," \emph{ IEEE J. Sel. Areas Inf. Theory,} vol. 3, no. 4, pp. 720-731, Dec. 2022.




\bibitem{JiawenImage}
J. Kang et al., ``Personalized Saliency in Task-Oriented Semantic Communications: Image Transmission and Performance Analysis," \emph{IEEE J. Sel. Areas Commun.,} vol. 41, no. 1, pp. 186-201, Jan. 2023.

\bibitem{KangXuImage}
X. Kang, B. Song, J. Guo, Z. Qin and F. R. Yu, ``Task-Oriented Image Transmission for Scene Classification in Unmanned Aerial Systems," \emph{IEEE
Trans. Commun.,} vol. 70, no. 8, pp. 5181-5192, Aug. 2022.

\bibitem{HaijunImageDRL}
H. Zhang, H. Wang, Y. Li, K. Long and A. Nallanathan, ``DRL-Driven Dynamic Resource Allocation for Task-Oriented Semantic Communication," \emph{IEEE
Trans. Commun.,} vol. 71, no. 7, pp. 3992-4004, Jul. 2023.

\bibitem{ChenImageSemantic}
C. Dong, H. Liang, X. Xu, S. Han, B. Wang and P. Zhang, ``Semantic Communication System Based on Semantic Slice Models Propagation," \emph{IEEE J. Sel. Areas Commun.,} vol. 41, no. 1, pp. 202-213, Jan. 2023.


\bibitem{EcenazImage}
E. Erdemir, T. -Y. Tung, P. L. Dragotti and D. Gündüz, ``Generative Joint Source-Channel Coding for Semantic Image Transmission," \emph{IEEE J. Sel. Areas Commun.,} vol. 41, no. 8, pp. 2645-2657, Aug. 2023.


\bibitem{YangkeImageWITT}
K. Yang, et al., ``WITT: A Wireless Image Transmission Transformer for Semantic Communications," in \emph{Proc. IEEE Int. Conf. Acoust., Speech, Signal Process.,} pp. 1-5, Rhodes Island, Greece, Jun. 2023.

\bibitem{WeizhiImageNon}
W. Li, et al., ``Non-Orthogonal Multiple Access Enhanced Multi-User Semantic Communication," \emph{IEEE Trans. Cogn. Commun. Netw.,} vol. 9, no. 6, pp. 1438-1453, Dec. 2023.

\bibitem{LokumarambageImageWireless}
M. U. Lokumarambage, et al., ``Wireless End-to-End Image Transmission System Using Semantic Communications," \emph{IEEE Access,} vol. 11, pp. 37149-37163, Apr. 2023. 

\bibitem{DanlanImage}
D. Huang, F. Gao, X. Tao, Q. Du and J. Lu, ``Toward Semantic Communications: Deep Learning-Based Image Semantic Coding," \emph{IEEE J. Sel. Areas Commun.,} vol. 41, no. 1, pp. 55-71, Jan. 2023.

\bibitem{LocSemantic0}
L. X. Nguyen et al., ``Swin Transformer-Based Dynamic Semantic Communication for Multi-User With Different Computing Capacity," \emph{IEEE Trans. Veh. Technol.,} vol. 73, no. 6, pp. 8957-8972, Jun. 2024.

\bibitem{LocSemantic1}
L. X. Nguyen et al., ``Optimizing Multi-User Semantic Communication via Transfer Learning and Knowledge Distillation," \emph{IEEE Commun. Lett.,} Early Access, Nov. 2024, DOI: 10.1109/LCOMM.2024.3499956.


\bibitem{HuiqiangText}
H. Xie, Z. Qin, G. Y. Li and B. -H. Juang, ``Deep Learning Enabled Semantic Communication Systems," \emph{IEEE Trans. Signal Process.,} vol. 69, pp. 2663-2675, Apr. 2021.


\bibitem{HuiqiangText2}
H. Xie and Z. Qin, ``A Lite Distributed Semantic Communication System for Internet of Things," \emph{IEEE J. Sel. Areas Commun.,} vol. 39, no. 1, pp. 142-153, Jan. 2021.


\bibitem{LeiTextresource}
L. Yan, Z. Qin, R. Zhang, Y. Li and G. Y. Li, ``Resource Allocation for Text Semantic Communications," \emph{IEEE Wireless Commun. Lett.,} vol. 11, no. 7, pp. 1394-1398, Jul. 2022.


\bibitem{PeiwenText}
P. Jiang, C. -K. Wen, S. Jin and G. Y. Li, ``Deep Source-Channel Coding for Sentence Semantic Transmission With HARQ," \emph{IEEE Trans. Commun.,} vol. 70, no. 8, pp. 5225-5240, Aug. 2022.

\bibitem{XidongText}
X. Mu, Y. Liu, L. Guo and N. Al-Dhahir, ``Heterogeneous Semantic and Bit Communications: A Semi-NOMA Scheme," \emph{IEEE J. Sel. Areas Commun.,} vol. 41, no. 1, pp. 155-169, Jan. 2023.

\bibitem{ZhenziSpeech0}
Z. Weng, Z. Qin and G. Y. Li, ``Semantic Communications for Speech Signals," in \emph{Proc. IEEE Int. Conf. Commun. (ICC),} pp. 1-6, Montreal, QC, Canada, Jun. 2021.

\bibitem{ZhenziSpeech1}
Z. Weng and Z. Qin, ``Semantic Communication Systems for Speech Transmission," \emph{IEEE J. Sel. Areas Commun.,} vol. 39, no. 8, pp. 2434-2444, Aug. 2021. 

\bibitem{WeiSpeech}
H. Wei, et al., ``SemAudio: Semantic-Aware Streaming Communications for Real-Time Audio Transmission," in \emph{Proc. IEEE Global Commun. Conf.,} pp. 3965-3970, Rio de Janeiro, Brazil, Dec. 2022.

\bibitem{TianxiaoSpeech}
T. Han, Q. Yang, Z. Shi, S. He and Z. Zhang, ``Semantic-Preserved Communication System for Highly Efficient Speech Transmission," \emph{IEEE J. Sel. Areas Commun.,} vol. 41, no. 1, pp. 245-259, Jan. 2023. 



\bibitem{TongSpeech}
 H. Tong et al., “Federated learning for audio semantic communication,” \emph{Front. Commun. Netw.,} vol. 2, Art. no. 734402., Sep. 2021.



\bibitem{ChenSpeech}
X. Chen, J. Wang, L. Xu, J. Huang and Z. Fei, ``A Perceptually Motivated Approach for Low-Complexity Speech Semantic Communication," \emph{Internet Things J.,} vol. 11, no. 12, pp. 22054-22065, Jun. 2024.


\bibitem{ZhouSpeech}
Z. Zhou, S. Zheng, J. Chen, Z. Zhao and X. Yang, ``Speech Semantic Communication Based on Swin Transformer," \emph{IEEE Trans. Cogn. Commun. Netw.,} vol. 10, no. 3, pp. 756-768, Jun. 2024.




\bibitem{XiaoSpeech}
Z. Xiao, S. Yao, J. Dai, S. Wang, K. Niu and P. Zhang, ``Wireless Deep Speech Semantic Transmission," in \emph{Proc. IEEE Int. Conf. Acoust., Speech, Signal Process.,} pp. 1-5, Rhodes Island, Greece, Jun. 2023.




\bibitem{Tze-YangVideo}
T. -Y. Tung and D. Gündüz, ``DeepWiVe: Deep-Learning-Aided Wireless Video Transmission," \emph{IEEE J. Sel. Areas Commun.,} vol. 40, no. 9, pp. 2570-2583, Sep. 2022.

\bibitem{SixianVideo}
S. Wang et al., ``Wireless Deep Video Semantic Transmission," \emph{IEEE J. Sel. Areas Commun.,} vol. 41, no. 1, pp. 214-229, Jan. 2023.

\bibitem{ZhangVideo}
Z. Zhang, Q. Yang, et al., ``Deep Learning Enabled Semantic Communication Systems for Video Transmission," in \emph{Proc. IEEE 98th Veh. Technol. Conf., (VTC2023-Fall),} pp. 1-5, Hong Kong, Oct. 2023.


\bibitem{NiuVideo}
H. Niu, et al., ``Deep Learning Enabled Video Semantic Transmission Against Multi-Dimensional Noise," in \emph{Proc. IEEE Globecom Workshops (GC Wkshps),} pp. 1267-1272, Kuala Lumpur, Malaysia, Dec. 2023.

\bibitem{BaoVideo}
Z. Bao, et al., ``MDVSC—Efficient Wireless Model Division Video Semantic Communication," \emph{IEEE Internet Things J.,} Early Access, Sep. 2024, DOI: 10.1109/JIOT.2024.3464614.

\bibitem{QiyuanVideo}
Du, Qiyuan, Yiping Duan, Qianqian Yang, Xiaoming Tao, and Mérouane Debbah. ``Object-Attribute-Relation Representation based Video Semantic Communication." \emph{arXiv:2406.10469,} Jun. 2024.
\bibitem{GuoVideo}
Guo, Jiangyuan, Wei Chen, Yuxuan Sun, Jialong Xu, and Bo Ai. ``VideoQA-SC: Adaptive Semantic Communication for Video Question Answering." \emph{arXiv:2406.18538,} Jun. 2024.







\bibitem{ZhuZengleMultimodal}
Zhu, Zengle, Rongqing Zhang, Xiang Cheng, and Liuqing Yang. ``Multi-Modal Fusion-Based Multi-Task Semantic Communication System." \emph{arXiv:2407.00964,} Jul. 2024.

\bibitem{MinqiMultimodal}
M. Zhu, ``Multi-Modal Semantic Communication System with Multiplicative Combination," in \emph{Proc. International Conference on Electronics and Information Technology (EIT),} Chengdu, China, Sep. 2024.

\bibitem{RenChaoMultimodal}
C. Ren et al., ``Multimodal Virtual Semantic Communication for Tiny-Machine-Learning-Based UAV Task Execution," \emph{IEEE Internet Things J.,} vol. 11, no. 19, pp. 30864-30874, Oct. 2024. 


\bibitem{FuWeiqiMultimodal}
W. Fu et al., ``Multimodal Generative Semantic Communication Based on Latent Diffusion Model," in \emph{Proc. IEEE 34th Int. Workshop Mach. Learn. Signal Process. (MLSP),} pp. 1-6, London, UK, Sep. 2024.





\bibitem{HuiqiangMultimodalVQA}
H. Xie, Z. Qin and G. Y. Li, ``Task-Oriented Multi-User Semantic Communications for VQA," \emph{IEEE Wireless Commun. Lett.,} vol. 11, no. 3, pp. 553-557, Mar. 2022. 

\bibitem{LiAngMultimodal}
A. Li et al., ``Cross-Modal Semantic Communications," \emph{IEEE Wireless Commun.} vol. 29, no. 6, pp. 144-151, Dec. 2022.

\bibitem{HuiqiangMultimodal}
H. Xie, Z. Qin, X. Tao and K. B. Letaief, ``Task-Oriented Multi-User Semantic Communications," \emph{IEEE J. Sel. Areas Commun., } vol. 40, no. 9, pp. 2584-2597, Sep. 2022.


\bibitem{JingMultimodalSemantic}
J. Gu, X. Zhang, Q. Cui and X. Tao, ``Semantic Communication for Multi-modal Data Transmission," in \emph{Proc. Int. Conf. Wireless Commun. Signal Process. (WCSP),} pp. 208-213, Hangzhou, China, Nov. 2023.


\bibitem{ChaoweiMultimodal}
Wang, Chaowei et al., ``Multimodal semantic communication accelerated bidirectional caching for 6G MEC." \emph{Future Gener. Comput. Syst.,} vol. 140, pp. 225-237, Mar. 2023.

\bibitem{TianYunMultimodal}
Tian, Yun, Jingkai Ying, Zhijin Qin, Ye Jin, and Xiaoming Tao. ``Synchronous Multi-modal Semantic Communication System with Packet-level Coding." \emph{arXiv:2408.04535,} Aug. 2024.

\bibitem{LiYuandiMultimodal}
Li, Yuandi, Zhe Xiang, Fei Yu, Zhangshuang Guan, Hui Ji, Zhiguo Wan, and Cheng Feng. ``Multimodal Trustworthy Semantic Communication for Audio-Visual Event Localization." \emph{arXiv:2411.01991,} Nov. 2024.

\bibitem{WangPenghongMultimodal}
P. Wang, et al., ``Distributed Semantic Communications for Multimodal Audio-Visual Parsing Tasks," \emph{IEEE Trans. Green Commun. Netw.,} vol. 8, no. 4, pp. 1707-1716, Dec. 2024. 


\bibitem{GuoJieMultimodal}
J. Guo et al., ``Distributed Task-Oriented Communication Networks with Multimodal Semantic Relay and Edge Intelligence," \emph{IEEE Commun. Mag.,} vol. 62, no. 6, pp. 82-89, Jun. 2024.



\bibitem{ZhangGuangyiMultimodal}
G. Zhang, Q. Hu, Z. Qin, Y. Cai, G. Yu and X. Tao, ``A Unified Multi-Task Semantic Communication System for Multimodal Data," \emph{IEEE Trans. Commun.,} vol. 72, no. 7, pp. 4101-4116, Jul. 2024.



\bibitem{ZhuYiboMultimodal}
Y. Zhu, H. Gu, J. Nie, J. Tang, J. Jin and Y. Zhang, ``Hashing-Based Multi-Modal Semantic Communication," in \emph{Proc. IEEE Wireless Commun. Netw. Conf.,} pp. 1-6, Dubai, UAE, Apr. 2024.


\bibitem{WangWenjunMultimodal}
W. Wang, M. Liu and M. Chen, ``CA\_DeepSC: Cross-Modal Alignment for Multi-Modal Semantic Communications," in \emph{Proc. IEEE Global Commun. Conf.,} pp. 5871-5876, Kuala Lumpur, Malaysia, Dec. 2023.


\bibitem{GaoYunMultimodal}
Y. Gao, D. Wu, J. Song, L. Zhou, H. Sari and Y. Qian, ``Cross-Modal Semantic Communications Over Wireless Emergency Networks," in \emph{Proc. IEEE Int. Conf. Commun. (ICC),} pp. 1891-1896, CO, USA, Jun. 2024.

\bibitem{HeYangshuoMultimodal}

Y. He, G. Yu and Y. Cai, ``Rate-Adaptive Coding Mechanism for Semantic Communications With Multi-Modal Data," \emph{IEEE Trans. Commun.,} vol. 72, no. 3, pp. 1385-1400, Mar. 2024. 

\bibitem{ZhiyiOneReciver}
Z. Tian, C. Zhang, W. Wang, H. Bogucka and S. Yu, ``ROSE: A Receiver-Oriented Semantic Communication Framework," \emph{IEEE Network,} Early Access, Jul. 2024, DOI: 10.1109/MNET.2024.3435038. 


\bibitem{NguyenOneTransmitterMultiple}
Nguyen, Loc X., et al., ``Semantic Communication Enabled 6G-NTN Framework: A Novel Denoising and Gateway Hop Integration Mechanism." \emph{Preprint arXiv:2409.14726,} Sep. 2024.

\bibitem{ChenMingkaiMultimodal}
M. Chen, M. Liu, W. Wang, H. Dou and L. Wang, ``Cross-modal Semantic Communications in 6G," in \emph{Proc. IEEE/CIC Int. Conf. Commun. (ICCC),} Dalian, China, Aug. 2023.



\bibitem{Yang_semantic}
Y. Yang et al., ``Environment Semantics Aided Wireless Communications: A Case Study of mmWave Beam Prediction and Blockage Prediction." \emph{IEEE J. Sel. Areas Commun.,} vol. 41, no. 7, pp. 2025-2040, Jul. 2023.

\bibitem{Huang_semantic}
Huang, Yinuo, et al., ``Joint Active and Passive Beamforming for RIS-Aided Semantic Communication." \emph{IEEE Trans. Veh. Technol.,} vol. 73, no. 12, pp. 19815-19820, Dec. 2024.

\bibitem{Yang_semantic_Sensing}
Yang, Yinchao, et al., ``Joint Semantic Communication and Target Sensing for 6G Communication System." \emph{arXiv:2401.17108,} Jan. 2024.

\bibitem{Avi_Semantic}
Raha, Avi Deb, et al., ``Advancing Ultra-Reliable 6G: Transformer and Semantic Localization Empowered Robust Beamforming in Millimeter-Wave Communications."  \emph{Preprint arXiv:2406.02000,} Jun. 2024.

\bibitem{Puligheddu_SEM-O-RAN}
C. Puligheddu, et al., ``SEM-O-RAN: Semantic and Flexible O-RAN Slicing for NextG Edge-Assisted Mobile Systems," in \emph{Proc. IEEE Conf. Comput. Commun., (INFOCOM)} NY, USA, May 2023.


\bibitem{Sun_S-RAN}
Y. Sun, L. Zhang, L. Guo, J. Li, D. Niyato and Y. Fang, ``S-RAN: Semantic-Aware Radio Access Networks," \emph{IEEE Commun. Mag.,} Early Access, Nov. 2024, DOI: 10.1109/MCOM.004.2400105.


\bibitem{Zhang_Channel_Semantics}
Y. Zhang et al., ``AI Empowered Channel Semantic Acquisition for 6G Integrated Sensing and Communication Networks," \emph{IEEE Network,} vol. 38, no. 2, pp. 45-53, Mar. 2024.

\bibitem{Lotfi_Semantic}
F. Lotfi, et al., ``Semantic-Aware Collaborative Deep Reinforcement Learning Over Wireless Cellular Networks," in \emph{Proc. IEEE Int. Conf. Commun. (ICC),} pp. 5256-5261, Seoul, South Korea, May 2022.

\bibitem{Loc_SemanticFederated}
L. X. Nguyen et al., ``An Efficient Federated Learning Framework for Training Semantic Communication Systems," \emph{IEEE Trans. Veh. Technol.,} vol. 73, no. 10, pp. 15872-15877, Oct. 2024.


\bibitem{Zhao_Semantic}
B. Zhao et al., ``On Forecasting-Oriented Time Series Transmission: A Federated Semantic Communication System," \emph{IEEE Trans. Mobile Comput.} vol. 23, no. 12, pp. 13728-13744, Dec. 2024.

\bibitem{Biamontequantum}
J. Biamonte et al., “Quantum Machine Learning,” \emph{Nature,} pp. 195–202, vol. 549, no. 7671, Sep. 2017.
\bibitem{Khalidquantum}
U. Khalid, et al., ``Quantum Semantic Communications for Metaverse: Principles and Challenges," \emph{IEEE Wireless Commun.} vol. 30, no. 4, pp. 26-36, Aug. 2023.

\bibitem{Nunavathquantum}
N. Nunavath, et al., ``Quantum Semantic Communications for Graph-Based Models," in \emph{Proc. IEEE Int. Workshop Signal Process. Adv. Wireless Commun. (SPAWC),} pp. 871-875, Lucca, Italy, Sep. 2024.
\bibitem{Chehimiquantum}
M. Chehimi, C. Chaccour, C. K. Thomas and W. Saad, ``Quantum Semantic Communications for Resource-Efficient Quantum Networking," \emph{IEEE Commun. Lett.,} vol. 28, no. 4, pp. 803-807, Apr. 2024.
\bibitem{Tariqquantum}
S. Tariq, U. Khalid, B. E. Arfeto, T. Q. Duong and H. Shin, ``Integrating Sustainable Big AI: Quantum Anonymous Semantic Broadcast," \emph{IEEE Wireless Commun.} vol. 31, no. 3, pp. 86-99, Jun. 2024.



\end{thebibliography}
\end{document}